\documentclass{aa} 
\usepackage{graphicx}
\usepackage{natbib}
\usepackage{amsmath}
\usepackage{amsfonts}
\usepackage{amscd}
\usepackage{booktabs}
\usepackage{longtable}
\usepackage{multirow}
\usepackage{subfigure}
\usepackage{url}
\usepackage[titletoc,toc,title]{appendix}

\usepackage{bm}
\usepackage{color}
\usepackage[usenames,dvipsnames,svgnames,table]{xcolor}
\usepackage{tikz}
\usetikzlibrary{arrows,arrows.meta,shapes,decorations.pathmorphing,positioning,intersections,calc,backgrounds}
\tikzstyle{line} = [draw, -latex']

\usepackage{txfonts}
\usepackage{pdfpages}
\bibpunct{(}{)}{;}{a}{}{,} 
\begin{document}
 \title{The seven sisters DANCe}
   \subtitle{IV. Bayesian hierarchical model}
   \author{J. Olivares
          \inst{1,2,3}
          \and L.M. Sarro
	  \inst{1}
	  \and E. Moraux
	  \inst{2}
	  \and A. Berihuete
          \inst{4}
	  \and H. Bouy
	  \inst{3}
	  \and S. Hernandéz-Jiménez
	  \inst{1}
	  \and E. Bertin
	  \inst{6}
	  \and P.A.B. Galli
	  \inst{7}
	  \and N. Huelamo
	  \inst{5}
	   \and J. Bouvier
	  \inst{2}
	  \and D. Barrado
	  \inst{5}
          }

   \institute{Dpt. de Inteligencia Artificial , UNED, Juan del Rosal,
     16, 28040 Madrid, Spain\\ \email{lsb@dia.uned.es} 
     \and
     Univ. Grenoble Alpes, IPAG, F-38000, Grenoble, France\\
     CNRS, IPAG, F-38000, Grenoble, France
     \and
     Laboratoire d'astrophysique de Bordeaux, Univ. Bordeaux, CNRS, B18N, all\'ee Geoffroy Saint-Hilaire, 33615 Pessac, France.\\ \email{javier.olivares-romero@u-bordeaux.fr}
     \and
     Dpt. Statistics and Operations Research, University of
     C\'adiz, Campus Universitario R\'io San Pedro s/n.  11510 Puerto
     Real, C\'adiz, Spain 
     \and
     Depto. Astrof\'{\i}sica, Centro de Astrobiolog\'{\i}a (INTA-CSIC), ESAC campus,   P.O. Box 78,
     E-28691 Villanueva de la Ca\~nada, Spain
     \and
     Institut d’Astrophysique de Paris, CNRS UMR 7095 and UPMC, 98bis bd Arago, F-75014 Paris, France
     \and
     Instituto de Astronomia, Geof\'isica e Ci\^{e}ncias Atmosf\'ericas, Universidade de S\~ao Paulo, Rua do Mat\~ao, 1226, Cidade Universit\'aria, 05508-900, S\~ao Paulo - SP, Brazil
   }

   \date{}
  \abstract
  {The photometric and astrometric measurements of the Pleiades DANCe DR2 survey \citep{Bouy2015} provide an excellent test case for the benchmarking of statistical tools aiming at the disentanglement and characterisation of nearby young open cluster (NYOC) stellar populations.}
  {We develop, test and characterise of a new statistical tool (intelligent system) for the sifting and analysis of NYOC populations.}
  {Using a Bayesian formalism, this statistical tool is able to obtain the posterior distributions of parameters governing the cluster model. It also uses hierarchical bayesian models to establish weakly informative priors, and incorporates the treatment of missing values and non-homogeneous (heteroscedastic) observational uncertainties. }
 {From simulations, we estimate that this statistical tool renders kinematic (proper motion) and photometric (luminosity) distributions of the cluster population with a contamination rate of $5.8\pm0.2$\%. The luminosity distributions and present day mass function agree with the ones found by \citet{Bouy2015} on the completeness interval of the survey. At the probability threshold of maximum accuracy, the classifier recovers $\approx90$\% of \citet{Bouy2015} candidate members and finds 10\% of new ones. }
  {A new statistical tool for the analysis of NYOC is introduced, tested and characterised. Its comprehensive modelling of the data properties allows it to get rid of the biases present in previous works. In particular, those resulting from the use of only completely observed (non-missing) data and the assumption of homoskedastic uncertainties. Also, its Bayesian framework allows it to properly propagate observational uncertainties into membership probabilities and cluster velocity and luminosity distributions. Our results are in a general agreement with those from the literature, although we provide the most up-to-date and extended list of candidate members of the Pleiades cluster.}
  \keywords{Methods: data analysis, Methods: statistical, Proper motions, Stars: luminosity function, mass function, Galaxy:open clusters and associations: individual: M45}
   \maketitle
%


\section{Introduction}
\label{intro}

    The Pleiades is one of the most studied cluster in history\footnote{Probably just after Orion. The SAO/NASA Astrophysics Data System reports, at date, 2734 entries with keyword "Pleiades" since 1543 CE.}. Its popularity comes from its unique combination of properties. It is young \cite[$125\pm8$ Myr,][]{Stauffer1998}, close to the sun \cite[$134.4^{+2.9}_{-2.8}$ pc,][]{Galli2017}, massive \citep[$870 \pm 35 M_{\odot}$,][] {Converse2008}, has low extinction \cite[$A_v=0.12$,][]{Guthrie1987}, and almost solar metallicity \cite[{[Fe/H]$\approx$0},][]{Takeda2017}. From \citet{Trumpler1921} to date, it continues to yield new and fascinating results. Recently, \citet{Bouy2015} found 812 new candidate members for a total of 2109 down to $\approx 0.025 M_{\odot}$. The discovery of these new candidate members root on their excellent multi-archive data \citep{Bouy2013} and their leading-edge multidimensional statistical tool \citep{Sarro2014}. 

In the past, candidate members were selected using proper motions \cite[e.g.][]{Moraux2001}, or a combination of proper motions and cuts in the photometric space \cite[e.g.][]{Lodieu2012}. Only very recently with the works of \citet{Malo2013} \cite[and later][]{Gagne2014}, \citet{KroneMartins2014} and \citet{Sarro2014} the astrometric and photometric data started to be treated simultaneously and consistently to infer cluster (or moving groups) membership probabilities. Since the early work of \citet{Lodieu2012}, photometric bands have proven to be crucial in the identification of new open cluster members. Therefore we leave aside the discussion of the recent works of \citet{Sampedro2016} and \citet{Riedel2017} in which photometric information is not included in their methodologies.

Briefly, \citet{Malo2013} establish membership probabilities to nearby young moving groups using a naive\footnote{Naive Bayes refers to a parametric classifier in which \emph{all} parameters are assumed independent.} Bayesian classifier (BANYAN). Their classification uses kinematic and photometric models of moving groups and field. They create these models with the previously known \emph{bona fide} members and field objects. Their photometric model in the low mass range is extended using evolutionary models. Later, \citet{Gagne2014} developed BANYAN II, and improved \citet{Malo2013} methodology to identify low-mass stars and brown dwarfs by using redder photometric colours. They also included observational uncertainties, and correlations in the XYZ, and in UVW spaces, separately (via freely-rotated ellipsoid Gaussian models which are mathematically equivalent to two separate 3D multivariate Gaussian models), which reduces the naivety of the classifier, amongst other improvements. \citet{KroneMartins2014} establish cluster membership probabilities in a frequentist approach using an unsupervised and data driven iterative algorithm (UPMASK). Their methodology relies on clustering algorithms and the principal component analysis. Although untested, the authors mention that their methodology is able to incorporate proper motions and deal with missing data and different uncertainty distributions. \citet{Sarro2014} infer posterior membership probabilities using a Bayesian classifier (to which we refer as SBB in the rest of this work). Their cluster model is data driven and, since they model some parametric correlations, it is also less naive than that of \citet{Malo2013} and \citet{Gagne2014}. \citet{Sarro2014} construct the field and cluster proper motions models with clustering algorithms and the photometric one with principal curve analysis. Their treatment of uncertainties and missing values is consistent across observed features. Finally, they infer the best parameter values of their cluster model using a Maximum-Likelihood-Estimator (MLE) algorithm.  

The previous methodologies perform well on their designated tasks and successfully led to the identification of many new high probability members of nearby clusters and associations. However, key aspects still need to be tackled or improved. 

The membership probability of an object to a class (e.g. the cluster or field populations) depends on how the later is defined. In parametric classifiers, the classes are defined by the relations amongst their parameters, and the value of these later. Therefore, the classification is sensitive to the parameters defining the class, with different parameter values resulting in different membership probabilities. The majority of the parametric classifiers from the literature fail to report the sensitivity of the recovered membership probabilities to the chosen value of their parametric models.

Missing values can strongly affect any result based on data containing them. If the missing pattern is completely random, then results based only on the complete (non-missing) data are unbiased. However, the less random the missing pattern is, the most biased the results are. See Chap. 8 of \citet{Gelman2013} for a discussion on missing at random and ignorability of the missing pattern. In astronomy, measurements are strongly affected by the brightness of the source in question. The physical limits of detectors lead to non-random distributions of missing measurements. At the bright end, saturation indeed makes measurements useless, while at the faint end, sources beyond the limit of sensitivity are not detected. In a multi-wavelength data set, the intrinsic stellar colours add another level of correlation between sensitivity limits in different filters. Thus, missing values are not completely random, although they could be in some specific and controlled situations, but most of the time, they occur on measurements of bright and faint sources and depend on the source colours. For these reasons, their treatment is of paramount importance. \citet{Gagne2014} and \citet{Sarro2014}  construct their cluster (or kinematic moving group) and field model using only complete data, and then estimate membership probabilities for objects with missing values \cite[only in parallaxes and radial velocities in the case of][]{Gagne2014}. UPMASK does not include any treatment of missing values, although \citet{KroneMartins2014} mention that, in the future, their methodology will be able to incorporate them.

In general, models consist of relations among their variables (or parameters if they are parametric models) and the data. We call these the underlying relations between modelled and observed data. Traditionally, it is assumed that the underlying relations correspond to those seen in the data, which we call observed relations. For example, \citet{Sarro2014} and \citet{KroneMartins2014} establish as underlying relations in their models, the observed ones that they found after applying the principal curve analysis and the PCA to their data. Another possibility is to assume that the underlying relations come from other models. However, using other models may inherit their possible biases, while the observed relations are strongly affected by observational uncertainties, especially when these are not homogeneous (heteroscedastic). For example, the principal curve and principal components analyses (present in UPMASK and SBB) are biased by individual observational uncertainties \citep{Hong2016}. Some remedies include the so popular $\sigma$-clipping procedure and the variance scaling.
 
Inspired by the previous works and moved to address these three major limitations, we now present a new bayesian methodology aiming at statistically modelling the distributions of the open cluster population. It obtains the cluster membership probabilities as a by-product. Our methodology treats parametric and observational uncertainties consistently in a bayesian framework. In this framework, observational uncertainties propagate into the posterior distributions of the parameters. Objects with missing values are consistently included in all elements of our methodology. Particularly, it allows us to construct our field and cluster models with all objects in the data set, in spite of their missing values. As mentioned in the previous paragraph the observed relation between modelled and observed data is subject of bias. Instead, we aim at the \emph{true}\footnote{\emph{True} refers to that observed in the limit of theoretically perfect noise-free observations.} underlying parametric relations which render the observed data after being convolved with the observational uncertainties. In other words, we aim at deconvolving the \emph{true} cluster distributions \cite[see][for another example of deconvolution]{Bovy2011}.

In a Bayesian framework, priors must be established. To avoid as much as possible the subjectivity of choosing priors, we use the Bayesian Hierarchical Model formalism \cite[ see the works of][for applications of HM in astrophysics]{Jefferys2007,Shkedy2007,Hogg2010,Sale2012, Feeney2013}. On it, the parameters of the prior distributions are given by other distributions in a hierarchical fashion. In the words of \citet{Gelman2006}, Bayesian Hierarchical Models "allow a more 'objective' approach to inference by estimating the parameters of prior distributions from data rather than requiring them to be specified using subjective information". However, it comes at a price: these models are computationally more expensive because they require far more parameters than standard approaches. 

This paper proceeds as follows: In Sect. \ref{sect:methodology} we briefly describe the Pleiades data set and explain our methodology. Section \ref{sect:results} contains the results of applying our methodology to both synthetic and the real Pleiades data. In Sect. \ref{sect:discussion} we discuss our results and compare them with the literature. Finally, Sect. \ref{sect:conclusions} contains our conclusion and future perspectives.

\section{Methodology}
\label{sect:methodology}
This work uses as data set the second release (DR2) of the Pleiades catalogue \cite[see Appendix A of][]{Bouy2015} from the DANCe survey \citep{Bouy2013}. Data from this survey have been successfully used to the characterisation of the Pleiades \citep{Sarro2014,Bouy2015,Barrado2016} and M35 \citep{Bouy2015a} clusters. Although this catalogue contains astrometric (stellar positions and proper motions) and photometric ($ugrizYJHK_s$) measurements for 1,972,245 objects, we use only proper motions and the $iYJHK_s$ bands. Our selection aims at comparing our results to those of \citet{Bouy2015} who used the reduced RF-2 representation space ($\mu_{\alpha},\mu_{\delta},J,H,K_s,i-K_s,Y-J$) of \citet{Sarro2014}. Thus, our representation space comprises the proper motions in right ascension and declination, $\mu_{\alpha},\mu_{\delta}$, and the photometric colours and magnitudes, $i-K_s$,$Y,J,H,K_s$. We model the photometry by means of a set of parametric relations in which the colour index $i-K_s$ ($CI$ hereafter) is the independent parameter. We select the $CI$ from among the possible colour indices because of its discriminant properties. \citet{Goldman2013} remarked the importance of using colour indices with the largest difference in wavelength in order to discriminate Hyades members from the field population. They used the colour indices $g-K_s$, $r-K_s$ and $i-K_s$ to perform their photometric selection of members. This result has been confirmed by \citet{Sarro2014}. Using a random forest classifier, these later authors determined that the colour indices $r-K_s$, $i-K_s$ and $Y-J$ where amongst the most discriminant features with mean decrease of node impurity of 156.0,102.0, and 77.9, respectively (see their Table 3).

As will be explained later in this Section, our parametric model yields the photometric bands as functions (injective by definition) of a \emph{true} colour index. Thus, we proceed to select one colour index from the set of the most discriminant ones. On the one hand, the r band is missing in 1222853 sources of the DANCe DR2 catalogue, which is more than $\sim 50\%$ missing entries than in the $i$ band. On the other hand, we attempted to model the magnitudes of the cluster members as a function of the Y-J colour index, but this resulted in large discrepancies with the observed photometry. This is due to the high and almost vertical slope in cluster CMDs resulting from the Y-J colour index, which prevents our injective functions to correctly reproduce the data. On the contrary, the cluster CMDs slope is less pronounced when using the $i-Ks$ colour index. Therefore, in the following we work with the $i-K_s$ colour index. 

    Since both photometry and proper motions carry crucial information for the disentanglement of the cluster population, we restrict the data set to objects with proper motions and at least two observed values in any of our four CMDs: $Y,J,H,K_s$ vs $CI$. This restriction excludes 22 candidate members of \citet{Bouy2015}, which have only one observed value in the photometry. Furthermore, we restrict the lower limit ($CI =0.8$) of the colour index to the value of the brightest cluster member. We do not expect to find new bluer members in the bright part of the CMDs. We set the upper limit ($CI=8$) of the colour index at one magnitude above the colour index of the reddest known cluster member, thus allowing for new discoveries. Due to the sensitivity limits of the DR2 survey in $i$ and $K_s$ bands, objects with $CI>8$ have $K_s$ magnitudes $\geq 16$ mag. These objects are incompatible with the cluster sequence and therefore we discard them a priori as cluster members.

Our current computational constraints and the costly computations associated to our methodology (described throughout this Sect.), prevent its application to the entire data set. However, since the precision of our methodology, as that of any statistical analysis, increases with the number of independent observations, we find that a size of $10^5$ source for our data is a reasonable compromise. Although a smaller data set produces faster results, it also renders a less precise model of the field (in the area around the cluster) and therefore, a more contaminated model of the cluster. For these reasons, we restrict our data set to the $10^5$ objects with highest membership probabilities according to \citet{Bouy2015}. Of this resulting data set, the majority ($\approx$98\%) are field objects with cluster membership probabilities around zero. Thus, the probability of leaving out a cluster member is negligible. For the remaining of the objects in the Pleiades DANCe DR2, we assign membership probabilities \emph{a posteriori}, once the cluster model is constructed (see Sect. \ref{subsect:MemProb}).

To disentangle the cluster and field population we create parametric and independent models for both populations. These models aim at reproducing the observed astrometric and photometric properties of both populations. We infer the set of model parameters, $\boldsymbol{\theta}$, based on the data, $\boldsymbol{D}={\{\boldsymbol{d}_n\}}_{n=1}^N$ (with $N$ the number of sources and  $\boldsymbol{d}_n=\{\mu_{\alpha,n},\mu_{\delta,n},CI_n,Y_n,J_n,H_n,K_{sn}\}$), and the probabilistic framework established by the Bayes theorem:
\begin{equation}
p(\boldsymbol{\theta} | \boldsymbol{D})=\frac{p(\boldsymbol{D}|\boldsymbol{\theta})\cdot p(\boldsymbol{\theta})}{p(\boldsymbol{D})}.
\label{eq:bayes}
\end{equation}

In this equation, $p(\boldsymbol{\theta} | \boldsymbol{D})$ represents the \emph{posterior} probability of the parameters given the data, this is what we aim to infer. In the right side, $p(\boldsymbol{D}|\boldsymbol{\theta})$ stands for the probability of the data given the parameters, also called the \emph{likelihood}\footnote{Throughout the text, we use likelihood for the probability distribution of the data given the values of the parameters.} of the data, $p(\boldsymbol{\theta})$ represents the \emph{prior} beliefs about the relative probabilities of different parameter values, and $p(\boldsymbol{D})$, also known as \emph{evidence} or \emph{marginal likelihood} (since the parameters have been marginalised over), works as a normalisation constant. Since this last one can be computed by integrating the numerator of Eq. \ref{eq:bayes}, we only focus on the \emph{likelihood} and the \emph{priors}. We describe these terms in more detail throughout the remainder of this Sect..

Assuming that data are independent\footnote{This assumption means that the probability of measuring certain value of an object is independent of the measured value of another object. The DANCe DR2 sample shows no significant correlation amongst the observables it reports, for more details see \citet{Bouy2013}.}, given the parameters, its \emph{likelihood} can be expressed as:
\begin{equation}
\label{gen-model}
p(\boldsymbol{D}|\boldsymbol{\theta})=p({\{\boldsymbol{d}_n\}}_{n=1}^N | \boldsymbol{\theta})=\prod \limits_{n=1}^{N}p(\boldsymbol{d}_n| \boldsymbol{\theta}).
\end{equation}

We call \emph{generative model} to the \emph{likelihood} of one datum, $p(\boldsymbol{d}_n| \boldsymbol{\theta})$, because synthetic data can be drawn from it. Formally, this term must be $p(\boldsymbol{d}_n| \boldsymbol{\theta}, M)$ with $M$ standing for all other information on which the probability distribution depends. This information includes the standard uncertainty of each datum, $\boldsymbol{\epsilon}_n$ (e.g. $\epsilon_{\mu_{\alpha}},\epsilon_{\mu_{\delta}},\epsilon_{CI}$, associated with the common $\pm \sigma$ gaussian uncertainties), and the assumptions we make in the construction of the model. We assume that each datum uncertainty is fixed, which means that the differences between datum $\boldsymbol{d}_n$ and new measurements of the same source will be normally distributed with mean zero and standard deviation $\boldsymbol{\epsilon}_n$. The following section explains the rest of the information $M$, that we use to construct the \emph{generative model}. 

\subsection{The generative model}
\label{subsect:generative-model}
Since we aim at separating the cluster and the field, we model these two overlapping populations separately. Their explicit disentanglement would demand a set of $N$ binary integers to account for the two possible states of each object in our data set: either it belongs to the cluster or to the field population. Since this would be prohibitive for the inference process (in computational terms), we \emph{marginalise}\footnote{Marginalisation is the process by which a parameter is integrated out using a measure or prior.} them using a binomial prior. This marginalisation renders only one parameter, $\pi$, which represents the fraction\footnote{In probability density functions specified as mixtures of other densities the contribution of each of the latter is called its fraction, weight or amplitude. Throughout the text we use them indistinctively.} of field objects in the data set.

Thus, the \emph{generative model} can be expressed as,

\begin{equation}
\label{eq:genmod}
p(\boldsymbol{d}_n | \boldsymbol{\theta}=\{\pi,\boldsymbol{\theta}_f,\boldsymbol{\theta}_c\},\boldsymbol{\epsilon}_n)=\pi \cdot p_f(\boldsymbol{d}_n|\boldsymbol{\theta}_f,\boldsymbol{\epsilon}_n) + (1-\pi)\cdot p_c(\boldsymbol{d}_n| \boldsymbol{\theta}_c,\boldsymbol{\epsilon}_n),
\end{equation}

where $p_f(\boldsymbol{d}_n|\boldsymbol{\theta}_f,\boldsymbol{\epsilon}_n)$ and $p_c(\boldsymbol{d}_n| \boldsymbol{\theta}_c,\boldsymbol{\epsilon}_n)$ are the field and cluster \emph{likelihoods} of the datum $\boldsymbol{d}_n,$ given its standard uncertainty $\boldsymbol{\epsilon}_n$, and the values of  the field and cluster parameters $\boldsymbol{\theta}_f$ and $\boldsymbol{\theta}_c$, respectively. The next two sections explain briefly the \emph{generative models} of the field and cluster. We refer the reader to the Appendix \ref{app:generativemodel} for a more detailed explanation of both models and the relations among their parameters.

In the following, we assume that the observed quantities are drawn from a probability distribution centred in the \emph{true} values. These are then convolved with the probability distribution of the observational uncertainties, which we assume to be a multivariate Gaussian. 

In the DANCe DR2 data set, proper motions and photometric bands are computed independently from each other. Thus, it does not report correlations amongst the observables uncertainties. For this reason, in the multivariate Gaussian describing the observational uncertainties we set the off-diagonal elements to zero, except those corresponding to the $i-K_s$ colour index, which by construction\footnote{Let $A$ be the matrix of transformation from the photometric bands vector $X=\{i,Y,J,H,K_s\}$ to the photometric vector of colour index and bands $Z=\{i-K_s,Y,J,H,K_s\}$. Then $Z=A\cdot X$, and $Cov(Z)=A\cdot Cov(X) \cdot A^T$.} contain the correlation of the $i$ and $K_s$ bands.

Our methodology aims at deconvolving the observational uncertainties to obtain the intrinsic dispersion of the \emph{true} values.This intrinsic dispersion is the convolution of several processes (e.g. unresolved binaries, extinction, variability), which we will attempt to separate in future versions of our model.

Due to its heterogeneous origin, the DANCe Pleiades DR2 has a high fraction of photometric missing values \cite[see Table 1 of][]{Sarro2014}. In our data set, less than 1\% of the objects have values in all photometric bands (and it should be remembered that only these complete sources are used in \citet{Bouy2015} to construct the cluster model which is eventually applied to infer the membership probabilities of the incomplete sources). Therefore, the treatment of objects including missing values is of paramount importance to our methodology. In the following, we deal with the missing values in these objects by setting them as parameters, which we marginalise over all their possible values with the aid of priors. In general we use uniform priors, otherwise, we give specific details.

\subsubsection{The field population model}
\label{sect:field}

We assume that the field  distributions of proper motion and relative photometry are independent, and thus can be factorised. This assumption is not entirely correct since the relative photometry is affected by distance, and the later is correlated with proper motions. Nevertheless, we assume independence amongst proper motion and photometric bands based on the following points: i) the entire DANCe DR2 renders small (<0.1) correlations amongst these observables, and ii) assuming independence reduces the number of free parameters of the field model from 728 to 366. Thus, although this assumption renders a less accurate model, it reduces the complexity of the later by $\sim 50$\%. 

We also assume that the distribution of proper motions and relative photometry are described by Gaussian Mixture Models (GMM). The flexibility of GMM to fit a variety of probability distributions geometries make them a suitable model to describe the density of the heterogeneous data from the DANCe DR2. We fit these GMM to field objects in our data set. We select as field objects those having cluster membership probability lower than 0.75 according to \citet{Bouy2015}, approximately $98,000$ objects. We verify that the number of hypothetically misclassified objects is negligible compared to the size of our data set ($100,000$ objects). With the contamination and true positive rates reported by \citet{Sarro2014}: $\approx 8\%$ and $ \approx96\%$ respectively (at probability threshold $p=0.75$), and the number of candidate cluster members reported by \citet{Bouy2015}, 2109, the number of misclassified objects would be $\approx 258 $, which represents a negligible fraction ($ \lesssim0.26$\%) of our data set. Furthermore, assuming that these misclassified objects would be in the field-cluster "boundary", we can safely assume that they would be spread all over the cluster photometric sequence and over a halo around the proper motion of the cluster. This assumption, together with their negligible fraction, leads us to assume that their contribution to the field parameters is negligible. Thus, we keep the field parameters fixed throughout the inference process. Due to our current computing power, this decision is essential since it diminishes considerably (by 336) the number of free parameters, and therefore the computing time. However, it also leads to posterior distributions of cluster parameters that do not reflect  the uncertainties associated to the field model. 

\begin{figure*}
\begin{center}
\resizebox{\hsize}{!}{\includegraphics[page=1,width=\columnwidth]{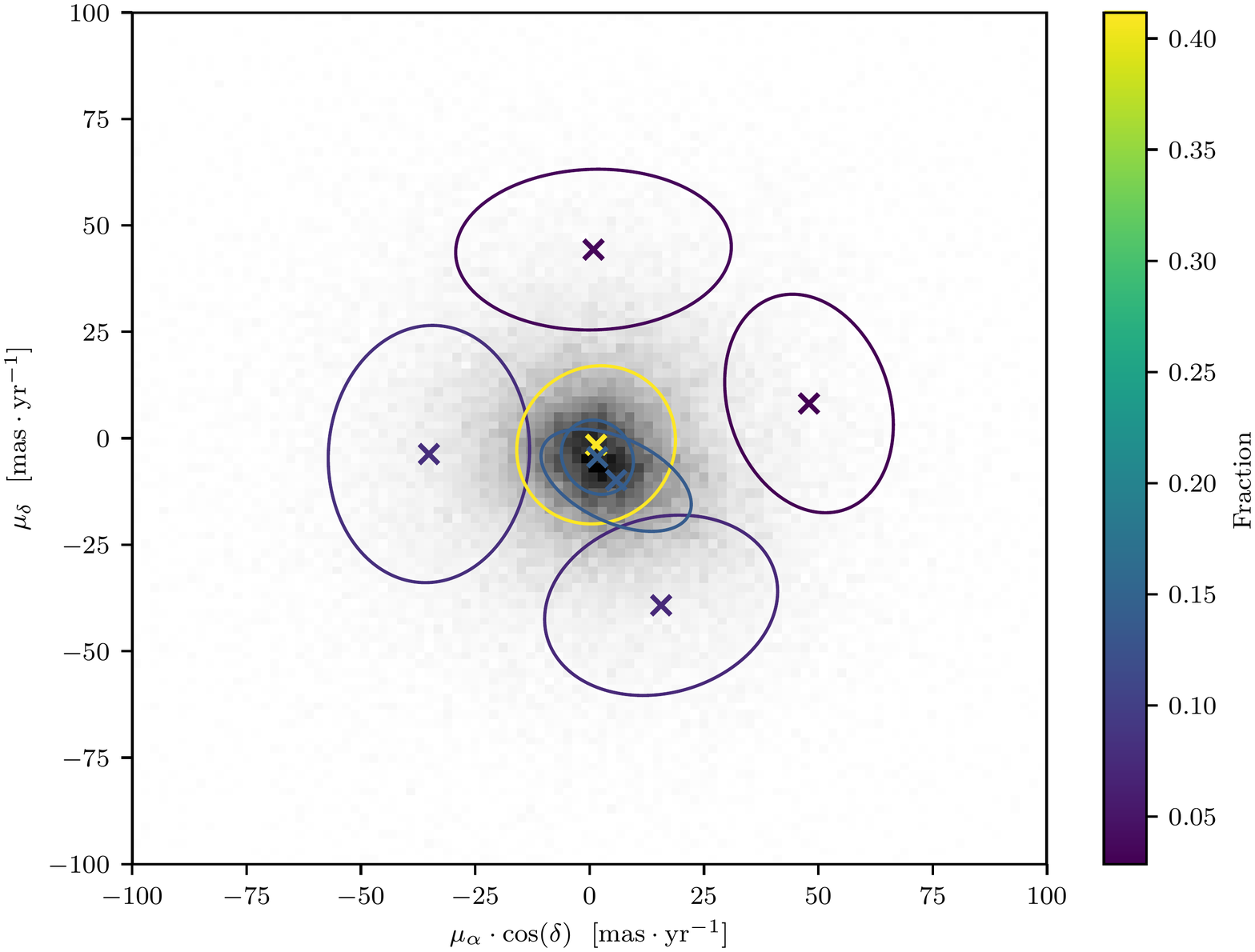}\includegraphics[page=2,width=\columnwidth]{figs/Field_GMM.pdf}}
\caption{Proper motion (left) and $K_s$ vs. $i-K_s$ CMD (right) projections of field models (ellipses and crosses depicting the covariance matrices and means of the GMM) and the density of objects in our dataset (gray pixels in logarithmic scale, shown only to guide the eyes). The colour scale shows the weight of each gaussian in the GMM.}
\label{figure:field_PM-CMD}
\end{center}
\end{figure*}

We determine the number of gaussians in each GMM using the Bayesian Information criterion \cite[BIC,][]{Schwarz1978}. This is a figure-of-merit that combines the likelihood and the number of parameters in the model such that it penalises complex models. Due to the presence of missing values in the photometry, we estimate the parameters of this photometric GMM using the algorithm of \citet{McMichael1996}. This is a generalisation of the Expectation Maximisation (EM) algorithm for GMM in which data with missing values also contribute to estimate the parameters. The number of gaussians suggested by the BIC for this mixture is 14 (amounting to 293 free parameters). The right panel of Fig. \ref{figure:field_PM-CMD} depicts a projection of this multidimensional (5 dimensions) GMM in the subspace of $K_s$ vs $CI$. We notice that, due to the high amount of missing values in the photometry, most of the plotted gaussians in the right panel are empty in this particular projection space.

\begin{figure*}
\begin{center}
\resizebox{\hsize}{!}{\includegraphics[page=1,width=\columnwidth]{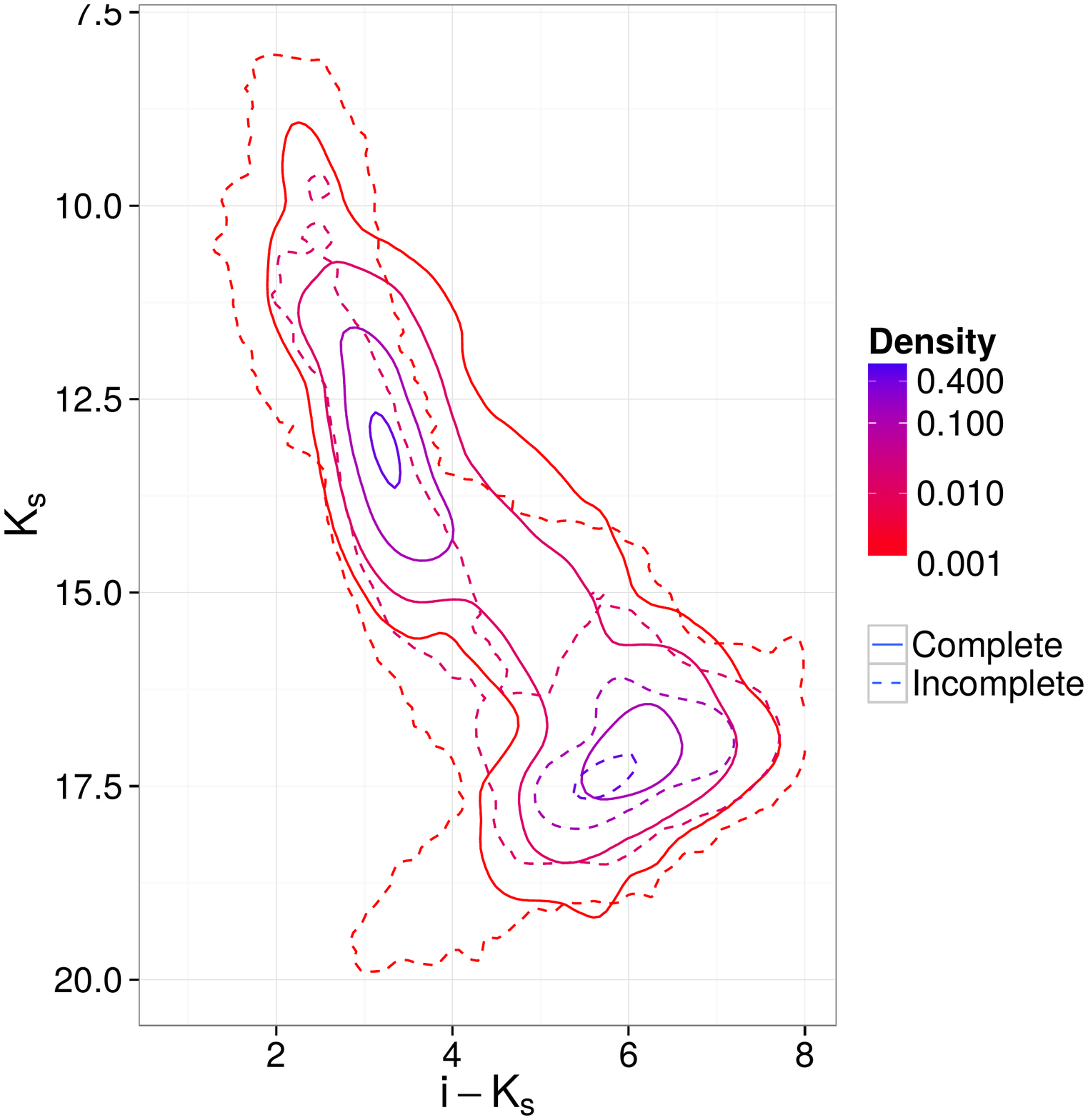}\includegraphics[page=2,width=\columnwidth]{figs/Field-CvsI.pdf}}
\caption{Densities in $K_s$ vs $i-K_s$ (left) and $K_s$ vs. $J-K_s$ (right) CMDs of $10^5$ synthetic sources. We drawn these from two different models: complete (solid line) and incomplete (dashed line). We construct the incomplete model using all objects in our data set, even those with missing values, while for the complete one we use only those object without missing values. The contour values (at $10^{-3},10^{-2},10^{-1},4\times 10^{-1}$) are the same for both complete and incomplete models.}
\label{figure:CvsI}
\end{center}
\end{figure*}

Furthermore, using our entire data set to construct the parameters of the field, allows us to remove biases associated with the use of only the completely observed objects. To illustrate these biases we proceed as follows. First, we take the GMM fitted to the $\approx98,000$ field objects, as described in the previous paragraphs. Since this model takes into account the missing values we call it the incomplete data model. Then, we select only the complete sources in the $\approx98,000$ field data set (which amount up to 1\%) and fit a GMM with the same number of gaussians, 14, as the incomplete data model. We call it the complete data model. Afterwards, for each model, we draw $10^5$ synthetic data points, we call them complete and incomplete, depending on the parent model. In Fig. \ref{figure:CvsI}, we show the associated density of these two synthetic data sets, complete (solid line) and incomplete (dashed line), in the projected $K_s$ vs $i-K_s$ (left) and $K_s$ vs $J-K_s$ (right) CMDs. As this Fig. shows, the complete data model underestimates the density in the faintest regions (where the missing values happen the most), over estimate it in the middle ones ($11<K_s<15$), and shift it at the brightest ones ($K_s\approx 10$ in the $K_s$ vs $J-K_s$ CMD). As this Fig. illustrates, when missing values do not happen at random, the density landscapes of completely observed objects and that of all objects (missing values comprised) differ.

In the case of the proper motions, the BIC favours a model with a large number of gaussians with small weights and large variances distributed all over the observed data space. In order to circumvent this over-complex model, we decided to add a uniform distribution to the GMM. When we apply the BIC to this new mixture of distributions, the modification improves the likelihood and reduces the number of gaussians. The number of gaussians suggested by the BIC for this mixture is 7, plus the uniform distribution (amounting to 42 free parameters). The left panel of Fig. \ref{figure:field_PM-CMD} shows the gaussians of this mixture. As can be seen in this Fig., one of the gaussians in the mixture is centred near the proper motions of the cluster ($\{\mu_{\alpha},\mu_{\delta}\}\sim\{16,-39\}$ $\rm{mas}\cdot \rm{yr}^{-1}$). The weight of this gaussian is small, $0.07$, and only marginally larger than the weight $0.03$ of the gaussian at the upper right corner. Since there is no apparent reason for this gaussian to be coincident with the cluster population, it suggests that within the objects that \citet{Bouy2015} classified as field population, there are some false-negatives with proper motions compatible with those of the cluster population. In future works, we will improve this classification to characterise and minimise possible false-negatives.

\subsubsection{The cluster population model}
\label{sect:cluster}
To model the cluster population, we assume independence between proper motions and photometry. This assumption is not entirely correct since the cluster has a spread in distance, which may introduce a correlation amongst these variables. However, due to the distance to the cluster \cite[$134.4^{+2.9}_{-2.8}$ pc according to][]{Galli2017} we can assume that this spread has a negligible impact in the photometry and proper motions of the cluster members. This correlation and its possible inclusion in the model will be explored in future works. Thus, similarly to the field model, we factorise these two components. \citet{Sarro2014} show evidence of an equal-mass binaries sequence in the Pleiades and model it with a proportion fixed to 20\%. We now model this sequence as a parallel cluster sequence displaced 0.75 magnitudes to the brighter side. Furthermore, since binarity could affect the proper motion of the system, we couple this photometric information to the proper motions by constructing a separate proper motions model for these equal-mass binaries. Additionally, we set the fraction of equal-mass binaries as a free parameter of our model. This will allow us to investigate potential kinematical differences between equal-mass binaries and the rest of the stars, singles and non equal-mass binaries.

\emph{Photometric model of equal-mass binaries and single stars}

To model the cluster sequence in the CMDs we use one truncated series of cubic splines for each of the $YJHK_s$ vs $CI$ CMDs. We choose splines because of their better fitting properties. We tried several polynomial bases (Laguerre, Hermite, Chebyshev) but regardless of their order, they lack the flexibility shown by the splines, particularly in the high slope region around $CI \approx 3$. However, this flexibility comes at a price. Splines require us to set, in addition to the coefficients of the series, a number of points known as knots. Knots are the starting and ending points of each spline section.

The simultaneous inference of spline coefficients and knots, a problem known as \emph{free-knot} splines, introduces multi modality in the parametric space \citep{Lindstrom1999}. To avoid this multi modality, we keep the knots fixed throughout inference. Nevertheless, we apply the \citet{Spiriti2013} methodology\footnote{Implemented in the R package \emph{freeknotsplines}} to the \citet{Bouy2015} members. Doing so, we obtain the best number and position for the knots. These are $CI=\{0.8,3.22,3.22,5.17,8.0\}$. We tested different number of knots, ranging from two to nine, with five the best configuration given by the BIC. 

In \cite{Sarro2014} the cluster sequence was modelled non-parametrically with a principal curve. It had a natural coordinate ($\lambda$) which was not  directly related to any physical parameter. This coordinate no longer holds for the splines model in which now the \emph{true} $CI$ is the independent parameter. Furthermore, as explained in Sect. 1, the principal curve analysis returns the observed relation in the data, not the underlying relation that generates the observations. Instead, splines allow to model the true underlying relation in a parametric way. 

Here, we assume that the observed photometric quantities are drawn from a probability distribution resulting from the convolution of the observed uncertainties, with an \emph{intrinsic} distribution centred at the \emph{true} photometric quantities. We model this \emph{intrinsic} distribution as a multivariate gaussian, whose covariance matrix is the \emph{intrinsic} dispersion, the same all along the cluster sequence. This \emph{intrinsic} dispersion could arise from different astrophysical processes like age, metallicity and distance dispersions, unresolved binaries, transits, variability, etc. Without this dispersion, we will have an over-simplistic model in which the cluster sequence will be an infinitely narrow line, and departures from it would only be explained by the observational uncertainties. In practice, this model would underestimate the posterior membership probabilities of hypothetical good candidates that depart from the ideal cluster sequence. We can have access to this intrinsic dispersion only after deconvolving the observational uncertainties.

The true $CI$ of each object is unknown, even if its observed value is not missing. This means that the true $CI$ of each object is a nuisance parameter which must be marginalised. We show this marginalisation in Equation \ref{eq:examplemarginal}. To marginalise these $CI$ we need a measure. We establish this measure as a truncated ($0.8\leq CI \leq8$) univariate GMM with five components whose parameters are also inferred from the data. 

\begin{align}
 p(\boldsymbol{d}_{ph}| \boldsymbol{\theta}_c,\boldsymbol{\epsilon}_{ph})&=\int p(\boldsymbol{d}_{ph},CI| \boldsymbol{\theta}_c,\boldsymbol{\epsilon}_{ph})\cdot dCI \nonumber \\ 
 &=\int p(\boldsymbol{d}_{ph}|CI, \boldsymbol{\theta}_c,\boldsymbol{\epsilon}_{ph})\cdot p(CI| \boldsymbol{\theta}_c,\boldsymbol{\epsilon}_{ph})\cdot dCI .
 \label{eq:examplemarginal}
\end{align}

In the previous equation, $\boldsymbol{d}_{ph}$, $\boldsymbol{\epsilon}_{ph}$ and, $\boldsymbol{\theta}_c$ correspond to the photometric measurements,   standard photometric uncertainties, and the cluster parameters, respectively. The term $p(\boldsymbol{d}_{ph}|CI, \boldsymbol{\theta}_c,\boldsymbol{\epsilon}_{ph})$ corresponds to the multivariate gaussian associated with the intrinsic dispersion of the cluster. The $CI$ dictates the \emph{true} photometric quantities by means of the splines. The term $p(CI| \boldsymbol{\theta}_c,\boldsymbol{\epsilon}_{ph})$ correspond to the truncated GMM which we use as a measure for the \emph{true} $CI$. Appendix \ref{app:generativemodel} contains more details on this marginalisation and the probability distribution involved on it.

We use the observed $CI$ and magnitudes to reduce the computing time of the marginalisation integral by avoiding regions in which the argument is almost zero (i.e. far away from the measured values). The process is the following: first, we compare the observed photometry to the true one (i.e. the cluster sequence given by the splines) and find the closest point, $p$, using the Mahalanobis metric. This metric uses the sum of the observational uncertainty with the intrinsic dispersion of the cluster sequence as covariance matrix. To define the limits of the marginalisation integral, we use a ball of 3.5 Mahalanobis distances around point $p$. Contributions outside this ball are negligible ($< 4\times10^{-4}$).

Since we model the true photometric quantities of the equal-mass binaries with a parallel sequence displaced 0.75 magnitudes into the bright side (twice the luminosity implies an increase of 0.75 in magnitudes), the only extra parameter needed is the fraction of equal-mass binaries to the total of cluster members.

\emph{Proper motion model of equal-mass binaries and single stars}

We model the proper motions of equal-mass binaries and single stars with a GMM whose parameters are inferred as part of the hierarchical model. The number of gaussians, however, remains fixed throughout inference. Following the BIC criterion, we select four and two gaussians for single and equal-mass binaries, respectively. Furthermore, we also assume that the gaussians in the proper motions GMM share the same mean, one for single stars and one for equal-mass binaries (which need not be equal).

The number of free parameters in our cluster and field models are 84 and 335, respectively. In addition, we use one free parameter, $\pi$ (Eq. \ref{eq:genmod}), to model the fraction of the field in the cluster-field mixture. Thus, our generative model has 420 free parameters. As explained in Sect. \ref{sect:methodology}, due to computational constraints, we use maximum-likelihood techniques to obtain the value of the 335 field parameters. For the remaining ones, we use MCMC to infer their full posterior distribution. In the following section we describe the priors used for the inference of these 85 parameters.

\subsection{Priors}
\label{subsect:priors}
In a Bayesian framework, each parameter in the generative model has a prior, even if it is uniform or improper. The priors we assume are intended to fall in the category of weakly informative priors. A weakly informative prior, following \citet{Gelman2006}, is that in which "the information it does provide is intentionally weaker than whatever actual prior knowledge is available". Although there is no general method for specifying them, a weakly informative prior can be constructed by diminishing the current available information \cite[see for example][]{Gelman2008,Chung2013}. In practice, we construct a weakly informative prior as follows. First, we choose the family distribution and its hyper-parameters such that it resembles the actual prior information. Then, we tune the hyper-parameters such that the statistical variance of the distribution increases with respect to the value found in the first step. In this way, the resulting prior provides less restrictive information than the original one. We choose this kind of priors due to their better properties regarding the regularisation and stability of the posterior computation when compared to reference priors \citep{Simpson2017}, and other non-informative priors \citep{Gelman2006}. We group priors into three main categories, those for fractions, and those for parameters in the proper motion and in the photometrical models. In the following, we explain the kind of distributions we use for the priors. In Appendix \ref{subsect:apppriors}, we give details on the particular parameter values we choose for these distributions.

Fractions are defined for mixtures, which can be GMM or the cluster-field mixture (Eq. \ref{eq:genmod}), and quantify the contribution of each element to the mixture. Thus, they must add to one and be bounded by the $[0,1]$ interval. For priors of fractions we use the multivariate generalisation of the beta distribution: the Dirichlet distribution. This distribution is parametrised by the vector $\boldsymbol{\alpha}$ (where $\{\alpha_i\}_{i=1}^K > 0$, and $K$ is the number of categories) and its support is the set of $K$-dimensional vectors $\rm{x}$ defined in the interval $(0,1)$ and with the property: $||x||=1$ (the sum of their entries equals one). We choose the Dirichlet distribution because it fits perfectly our needs, in addition its variance\footnote{The variance of the Dirichlet distribution, $Dir(\mathbf{X}, \boldsymbol{\alpha})$ is:
\begin{equation}
Var[X_i] = \frac{\alpha_i \cdot(\alpha_0 - \alpha_i)}{\alpha_0^2\cdot(\alpha_0 + 1)} 
\end{equation}
with $\alpha_0=\sum \alpha_i$.
 } can be diminished to tune it as a weakly informative prior. 

We set the priors of means and covariance matrices in the proper motions GMM as bivariate normal and Half--t distributions, respectively. According to \citet{Huang2013}, setting arbitrarily large values of the $\boldsymbol{A}$ parameters in the later distribution leads to arbitrarily weakly informative priors on the corresponding standard deviation terms. Thus, we obtain weakly informative priors by allowing large values of the standard deviations $\boldsymbol{\sigma}$ and $\rm{A}$ parameters, in the bivariate normal and Half--t distributions, respectively. See Appendix \ref{subsect:apppriors} for more details. 

Photometric priors include three categories, those concerning the \emph{true} $CI$, the splines coefficients, and the cluster sequence intrinsic dispersion. For the priors of the means and variances of the true $CI$ GMM, we use the normal and Half--Cauchy distributions, respectively. The later is the recommended choice for a weakly informative prior according to \citet{Gelman2006}. In both distributions we use large values for the variance and $\eta$ parameters (see Appendix \ref{subsect:apppriors}). Thus, both are weakly informative priors.

For the coefficients in the spline series we set the priors as univariate normal distributions. Finally, we use the multivariate Half--t distribution \citep{Huang2013} as a prior for the covariance matrix modelling the intrinsic dispersion of the cluster sequence. Appendix \ref{subsect:apppriors} shows the details on how we tune these distributions to obtain weakly informative priors.

\subsection{Sampling the posterior distribution}
\label{emcee}

There are three possible approaches to obtain the posterior distributions of the parameters in our model: an analytical solution, a grid in parameter space, and the Markov Chain Monte Carlo (MCMC) methods. Given the size of our data set ($10^5$ objects) and the dimension of our inferred model (85 parameters), the analytical solution and the grid approach are discarded a priori.

The MCMC methods offer a feasible alternative to this problem. Briefly, they consist of a particle (or particles) which iteratively moves in the parameter space. Among the many MCMC methods that exist, we select the \emph{stretch} move which is an affine invariant scheme developed by \citet{Goodman2010}. It is implemented to work on parallel in the Python routine \emph{emcee} \citep{Foreman2013}. We choose \emph{emcee} due to the following properties: i) the affine invariance allows a faster convergence over common and skewed distributions \cite[see][for details]{Goodman2010,Foreman2013}, ii) the parallel computation distributes particles over nodes of a computer cluster and thus reduces considerably the computing time, and iii) it requires the hand-tuning of only two constants: the number of particles, and $a$, the parameter of \emph{stretch} distribution \cite[see Eq. 9 of ][]{Goodman2010}. We use 170 particles (twice the number of parameters) and a value of $a=1.3$. These keep the acceptance fraction in the range $0.2 - 0.5$, as suggested by \citet{Foreman2013}.
 
We use \emph{CosmoHammer} \citep{Akeret2013}, a front-end of \emph{emcee}, to control the input and output of data and parameters, as well as the hybrid parallel computing. We run it on a 80 CPUs (cores) computer cluster with 3.5 GHz processors. However, instead of using OpenMP as \citet{Akeret2013} did, we use the \emph{multiprocessing} package of python to distribute the computing of the likelihood among cores in each cluster node. 

Since the evaluation of the likelihood is computationally expensive (it takes approximately 30 days to run in the previously described computer cluster\footnote{For comparison, the methodology of \citet{Sarro2014} takes approximately two days in the same computar cluster.}), we proceed similarly to \citet{Akeret2013}. We provide \emph{emcee} with an optimised set of values of the posterior distribution. These values can be thought of as a ball around the maximum-a-posteriori (MAP) solution. We find them with a modified version of the Charged Particle Swarm Optimiser (PSO) of \citet{Blackwell2002}. It avoids the over-crowding of particles around local best values. The charged version retains the PSO exploratory property by repelling particles that come closer than a certain user specified distance to each other. The repelling force mimics an electrostatic force, thus the name charged PSO. 

The modification that we introduce to the charged PSO relates only to the measuring of distance between particles. The algorithm of \citet{Blackwell2002} computes these distances in the entire parametric space. We find this approach unsuitable for our problem. In it, parameters have different length scales (for example, fractions and proper motions). Therefore, we measure distance between particles and apply the electrostatic force independently in each parameter. Thus, the electrostatic force comes into action only when the relative distance between particles is smaller than $10^{-10}$. We choose this value heuristically.

The PSO does not warrant the finding of the global maximum of the score function \cite[see][and references therein]{Blackwell2002, Clerc2002}. Therefore, we iteratively run PSO and 50 iterations of \emph{emcee} (with the same number of particles as the PSO) until the relative difference between means of consecutive iterations is lower than $10^{-7}$. The iterations of \emph{emcee} guarantee the spreading of the PSO solution without losing the information gained. After convergence of the PSO-\emph{emcee} scheme, we run \emph{emcee} with 175 walkers, until convergence. Neither scheme, PSO alone or PSO-\emph{emcee}, guarantees to find the global maximum and their solution could be biased. However, we use them to obtain a fast estimate of the global maximum, or at least, of points in its vicinity. Nevertheless, the final \emph{emcee} run, during the burning phase, erases any dependance on these initial solutions.

\begin{figure*}
\begin{center}
\resizebox{\hsize}{!}{\includegraphics{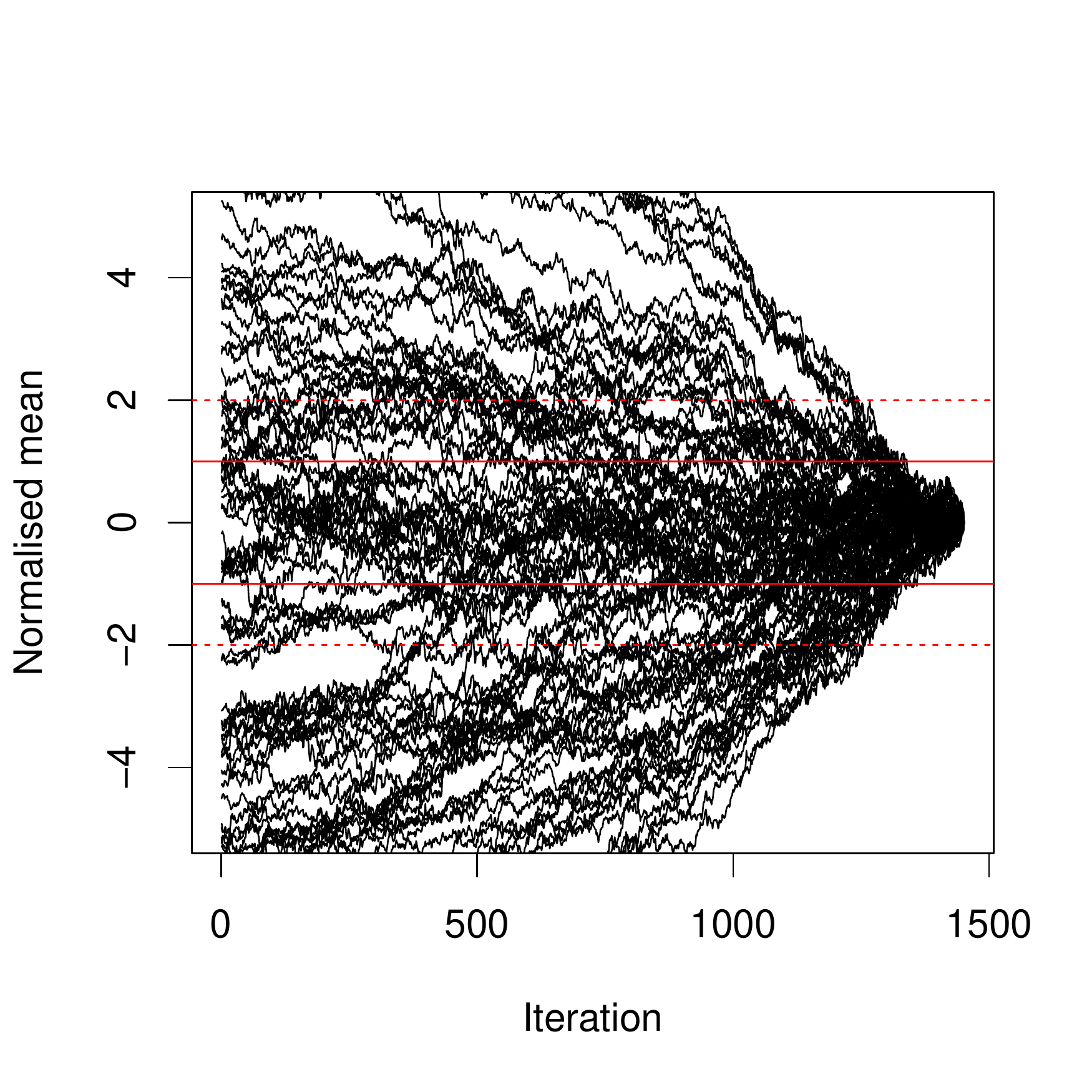}\includegraphics{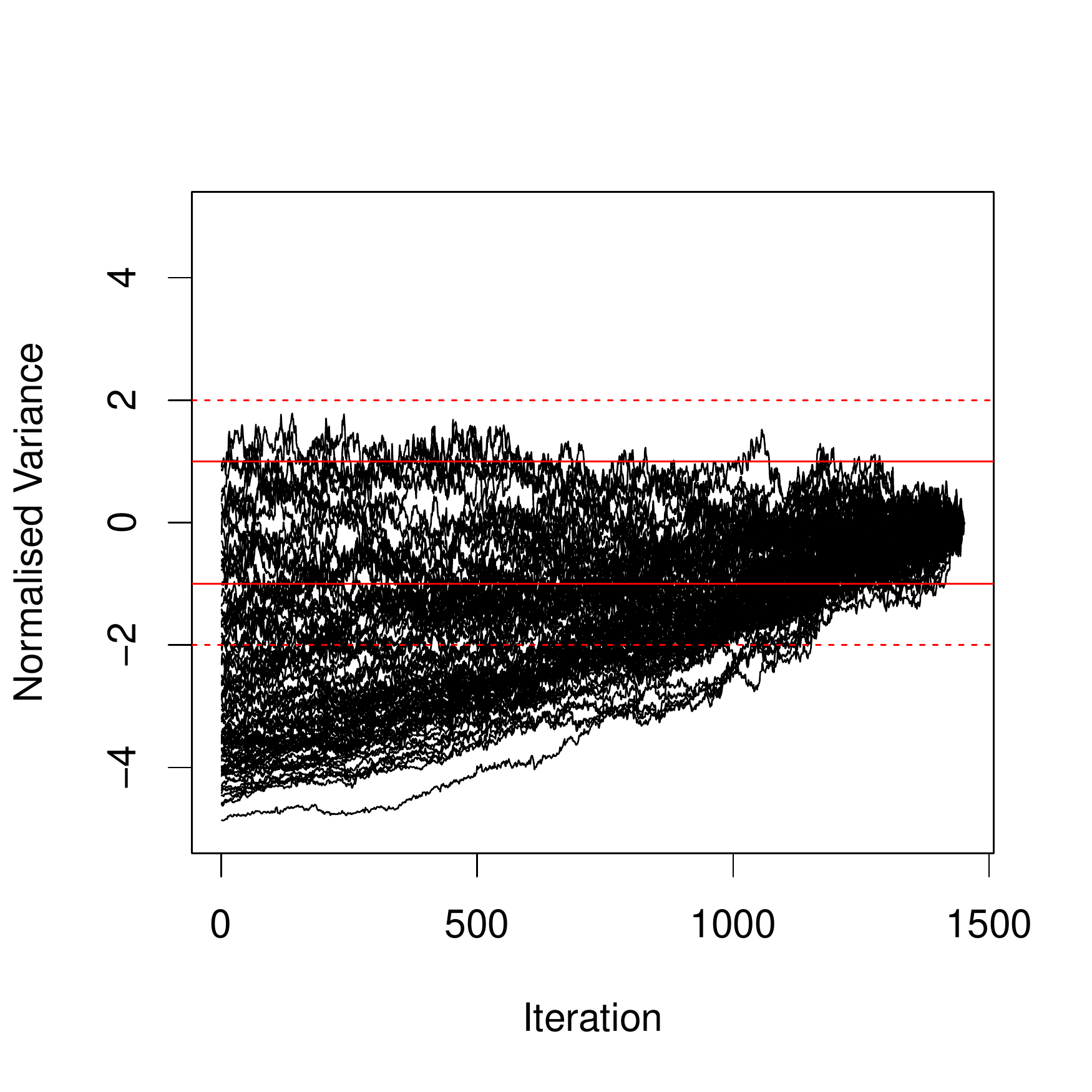}}
\caption{Normalised mean (left panel) and variance (right panel) of  each parameter in our model, given the DANCE DR2 data set as functions of iterations in the MCMC. Each parameter is scaled using the mean and variance of its corresponding ensemble of particles positions at the last iteration. Red lines show one and two sigma levels of these normalisation values. These Figs. depict the evolution of the Markov Chains from the original values provided by the PSO to the convergence. This later shown by the last $\sim 200$ iterations in which the mean and variances are within the two-sigma levels. We notice that some parameters evolve (within the MCMC) in groups, which is related to their correlation.}
\label{convergence}
\end{center}
\end{figure*}

Convergence to the target distribution occurs when each parameter enters into the stationary equilibrium, or normal state. The Central Limit Theorem ensures that this state exists. See \citet{Roberts2004} for guaranteeing conditions and \citet{Goodman2010} for \emph{irreducibility} of the \emph{emcee} stretch move. The stationary or normal state is reached when, in at least 95\% of the iterations, the sample mean is bounded by two standard deviations of the sample, and the variance by the two standard deviation of the variance \footnote{
$sd(\sigma^2)=\sigma^2 \sqrt{\kappa/n + 2/(n-1)}$ with $\kappa$ the kurtosis and $n$ the sample size.
}; see Fig. \ref{convergence}.

Once all parameters have entered the equilibrium state, we stop \emph{emcee} by using the criterion of \citet{Gong2016} \footnote{Implemented in the R package \emph{mcmcse} \citep{mcmcse}}. We choose this criterion because it was developed for high-dimensional problems and tested on Hierarchical Bayesian Models. In this criterion, the MCMC chain stops once its "effective sample size" (ESS, the size that an independent and identically distributed sample must have to provide the same inference) is larger than a minimum sample size computed using the required accuracy, $\epsilon$, for each parameter confidence interval $(1-\delta)$100\%. Our \emph{emcee} run stops once the ESS of the ensemble of walkers is greater than the minimum sample size needed for the required accuracy $\epsilon = 0.05$ on the 68\% confidence interval ($\delta = 0.32$) of each parameter.

\subsection{Membership probabilities}
\label{subsect:MemProb}
The methodology detailed in the previous sections renders the \emph{posterior} distributions of the parameters in the models of cluster and field populations. Cluster membership probabilities are then computed from these distributions by means of Bayes' theorem, (Eq. \ref{eq:bayes}). Applying it to our classification problem, we obtain that the probability of an object with measurement $\boldsymbol{d}_n$, to belong to the cluster population, $C$, is,  
\begin{align}
p(C|\boldsymbol{d}_n) = \frac{p(\boldsymbol{d}_n|C)\cdot p(C) }{p(\boldsymbol{d}_n|C)\cdot p(C) + p(\boldsymbol{d}_n|F)\cdot p(F)},
\label{eq:mem-prob}
\end{align}
where $F$ denotes the field population and, $p(\boldsymbol{d}_n|C)$ and $p(\boldsymbol{d}_n|F)$ are the cluster and field likelihoods, respectively. Probabilities $p(C)$ and $p(F)$ are the prior probabilities of the object to belong to the cluster and field, respectively. For these prior probabilities we use the fraction of field and cluster stars (i.e. the values of $\pi$ and $1-\pi$ in Eq. \ref{eq:genmod}, respectively), which the model infers at each MCMC iteration. The same reasoning is then applied to the probability of an object to be an equal-mass binary. In this case, the two populations are the equal-mass binaries and the stars in the main cluster sequence \footnote{Notice that currently we only give the probability of star to be an equal-mass binary (high mass ratio binary). Our methodology is not yet able to disentangle single stars from binaries of low mass ratio.}.

All terms in Eq. \ref{eq:mem-prob} depend on the model parameters, even the prior probabilities as mentioned before. Thus, each realisation from the joint posterior distribution of the model parameters (i.e. each iteration of the MCMC) results in a value for both cluster and equal-mass binaries membership probabilities. Therefore, upon convergence of the MCMC, sampling the joint posterior distribution of the model parameters results also in the sampling of the cluster and equal-mass binaries membership probabilities of each object.

Once the generative model has been learned from the $10^5$ sample (i.e. the MCMC has converged), we obtained the cluster and equal-mass binaries membership probabilities of all the objects in the DANCe catalogue. Computing 1700 samples of the membership probabilities for each of the $\sim 2$ million stars in the DANCe DR2 takes 4.11 hours. In Table \ref{table:probs} (available entirely at the CDS) we summarise the cluster and equal-mass membership probabilities of the DANCe DR2 objects marginalised over the posterior distribution of the cluster parameters. We also report the sensitivity of these membership probabilities to the cluster parameters by means of the standard deviation of the 1700 samples obtained for each object in the data set.

\section{Results}
\label{sect:results}

In this Sect. we analyse the results obtained by applying our methodology on synthetic and real data. The synthetic data enable us to quantify the reliability of the methodology and evaluate the impact that missing values have on it. This synthetic analysis requires at least three runs: one on the real data (to obtain the best values from which we generate the synthetic data), and two on the synthetic one. These last two runs correspond to data sets with and without missing values. As mentioned before, our methodology is computationally expensive. Therefore, for these three runs we use $10^4$ objects samples. The real data sample contains the objects with the highest membership probabilities as given by \citet{Bouy2015}. These objects are closer to the cluster, in the sense of membership probability, than the remaining 9$\times 10^4$ objects. Therefore, the field probability density in the region occupied by the cluster is higher and more concentrated (around the cluster) than the field density estimated using the larger and distant to the cluster $10^5$ objects sample. Thus, we assume that results obtained on the smaller sample have higher contamination, and lower recovery rates than those obtained on the larger sample, the more distant $10^5$ objects. The higher contamination and lower recovery rates arise from the concentration and higher values of the field probability density around the cluster, respectively. Therefore, results on the next subsection are upper and lower limits to the contamination and recovery rates, respectively.

\subsection{Reliability and impact of missing values}
\label{subsect:analysis}
We measure the performance of our methodology as a classifier (member vs. non member) by means of synthetic data on which members and non-members are known. To generate the synthetic data we draw $10^4$ random samples of the \emph{generative model} (see, Sect. \ref{subsect:generative-model}), whose parameters were found using the $10^4$ sample of real data.

As explained in Sect. \ref{sect:methodology}, our data set has a high fraction of missing values. The pattern of missing values is not random and depends on the magnitudes and colours of the objects. Therefore, we reproduce in each synthetic datum the pattern of missing values of one of its closer neighbours in the real data, closer in the euclidean sense. We found these closer neighbours in each of the CMDs: $\{K_s,J-K_s\},\{J,J-H\},\{K_s,H-K_s\},\{J,Y-J\},\{K_s,i-K_s\}$. These are, in decreasing order, the bands and colours with the fewer missing values. Assigning the missing pattern of the nearest real neighbour results in a biased sample in which objects with complete (non missing) values are underestimated. This bias roots in the fact that euclidean distances are smaller, or at most equal, when measured in subspaces (missing values) compared to those measured in the entire space (non-missing values). To avoid this, for each of the previous CMDs we: i) find the real objects with non-missing values and calculate their fraction, $f_r$, from the total real data, ii) take a sample, from the synthetic data, whose fraction, $f_s$, from the total synthetic data, is equal to $f_r$, iii) assign to the objects in this synthetic sample the pattern of missing values of the nearest neighbour from among the real objects found in (i). In this way, the synthetic data has similar fractions, of missing and non-missing values, to those of the real data.

Uncertainties are assigned as follows. We set the proper motions uncertainties to those of the nearest neighbour in the real data. This scheme, however, cannot be applied in the case of photometry. In the photometric space and due to the presence of missing values, the nearest neighbour scheme returns uncertainties that are biased towards the less precise measurements. Again, the euclidean metric results in the preferential choosing of objects with missing values. Since these missing values occur mostly at the faint end, where uncertainties are larger, it results in a bias towards larger uncertainties. To avoid this, we first fit polynomials (8th degree) to the uncertainties as a function of the magnitudes. Then, we use these polynomials to assign uncertainties to the synthetic photometric data.

To estimate the performance of our classifier to recover cluster members, we apply our methodology to synthetic data sets with and without missing values. In these results, we count the positives (cluster members, TP), negatives (field members, TN), false positives (field members classified as cluster members, FP) and false negatives (cluster members classified as field members,FN). With them we calculate the true positive rate (TPR), contamination rate (CR),  accuracy (ACC) and precision (or positive predictive value, PPV), which are defined as follows. 

\begin{align}
TPR&=\frac{TP}{TP+FN}\nonumber \\
CR&=\frac{FP}{FP+TP}\nonumber\\
PPV&=\frac{TP}{TP+FP}\nonumber\\
ACC&=\frac{TP+TN}{TN+FN+TP+FP}\nonumber
\end{align}

We use the mode to summarise membership probability distributions. To quantify the uncertainties of the previous quantities, we draw five realisations of the synthetic data set with missing values. Since we use the results of the non-missing values data set only for comparisons, we draw it only once. 

\begin{figure*}
\begin{center}
\resizebox{\hsize}{!}{\includegraphics{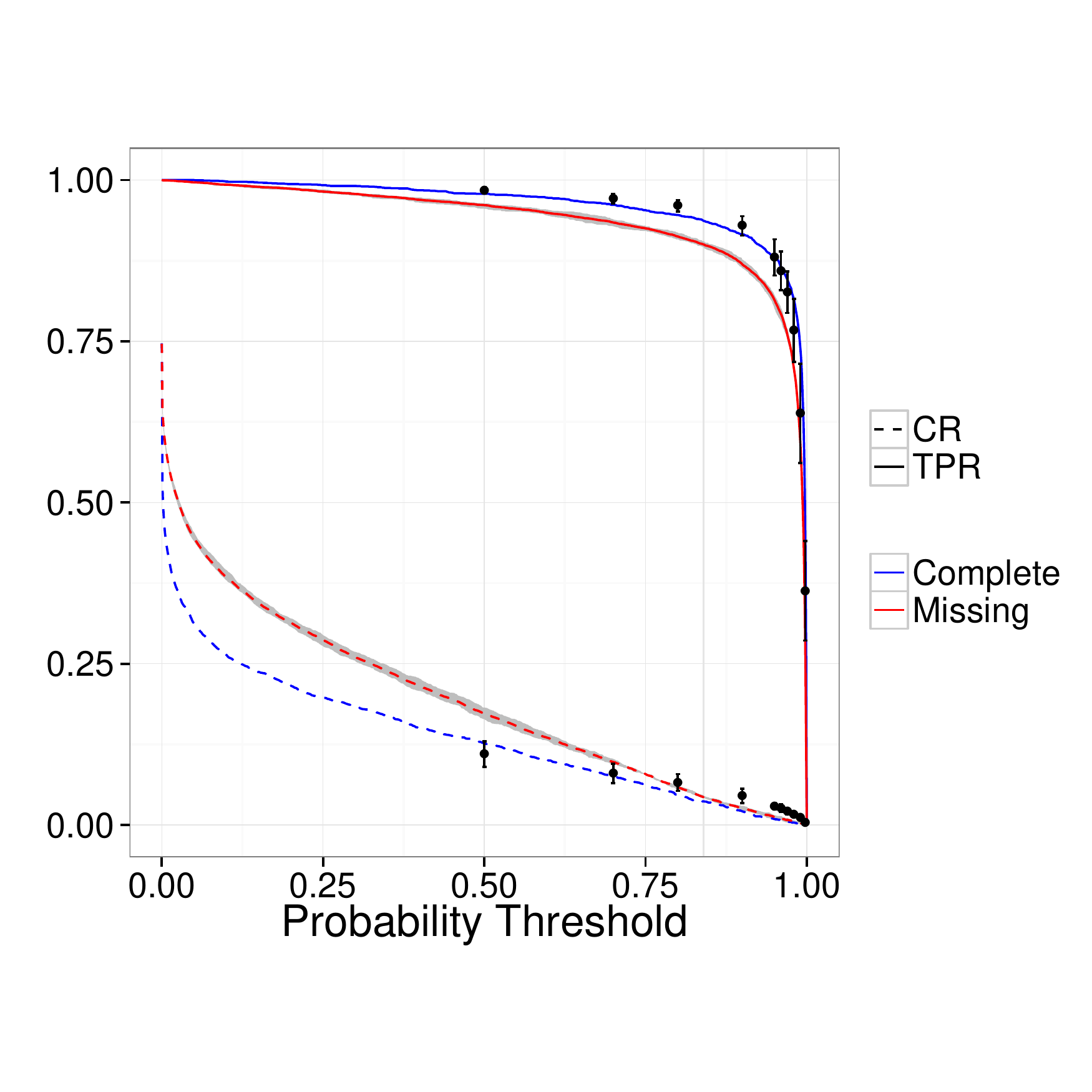}\includegraphics{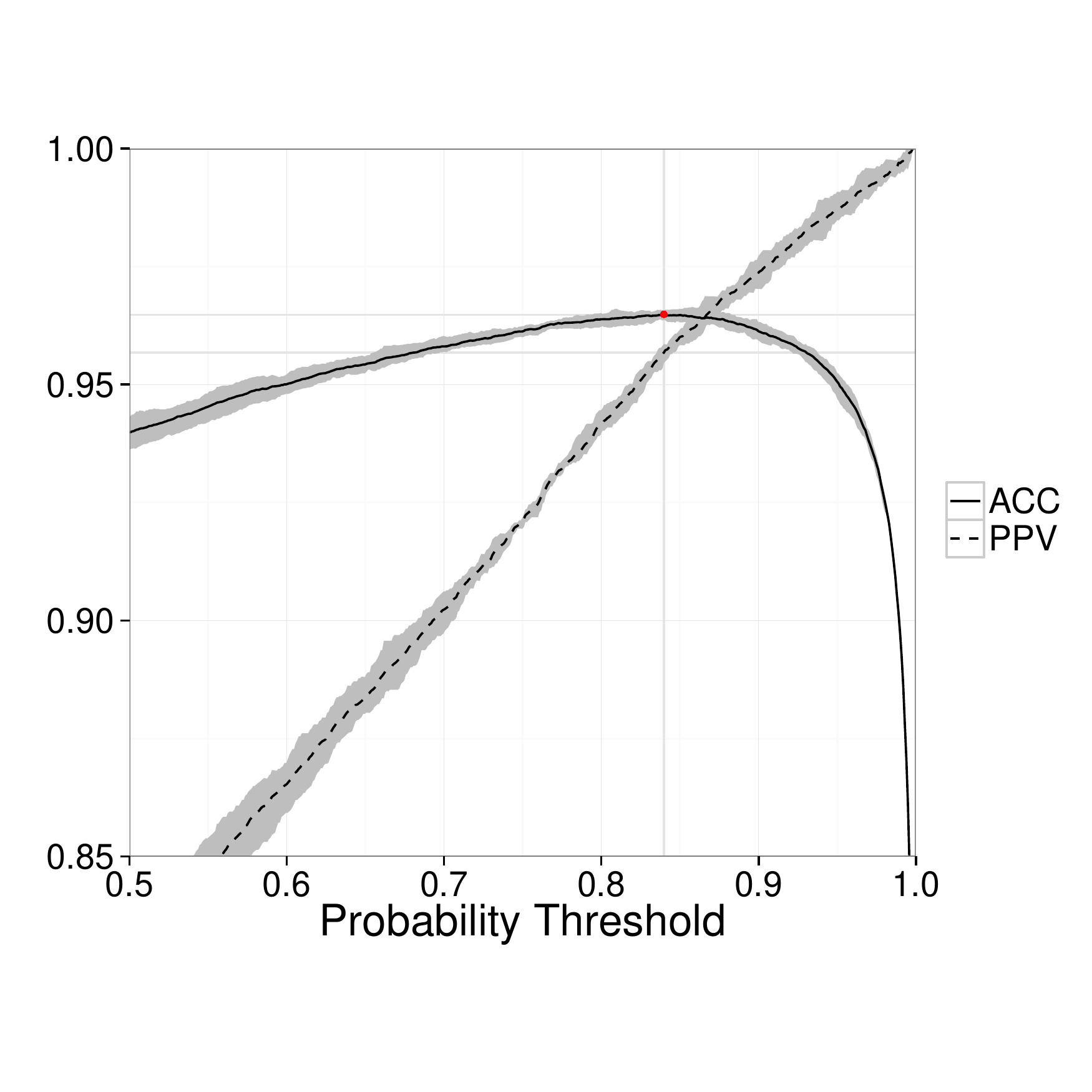}}
\caption{Left: The TPR (solid line) and CR (dashed line) of our methodology  when applied on synthetic data sets with and without missing values (red and blue lines, respectively). In black dots we show the TPR and CR reported by \citet{Sarro2014} for their non-missing values model. Right: Accuracy and precision as a function of probability threshold for our classifier when applied on synthetic data with missing values. The higher accuracy is obtained at $p_t=0.84$ (red dot). In both panels, the grey areas show the maximum deviations from the mean of the results of the five missing-values synthetic data sets.}
\label{figure:tfpr-roc}
\end{center}
\end{figure*}

The left panel of Fig. \ref{figure:tfpr-roc} shows the TPR (solid lines) and CR (dashed lines) in the presence (red lines) and absence (blue lines) of missing values. We measure both quantities as functions of the probability threshold used to define members and non-members. In the missing value case, the lines and the shaded grey regions depict the mean and deviations, respectively, of the results from the five synthetic data sets. As it is shown, the missing values have a negative impact in our classification process by diminishing the TPR  and increasing the CR. Nevertheless, our methodology delivers low ($\lesssim 8\%$) contamination rates above the probability threshold $p \approx 0.75$. In this Fig. and for the sake of comparison, we also show the CR and TPR  (as black dots) reported in Table 4 of \citet{Sarro2014}. This Fig. shows that, the TPR of our methodology measured on data without missing values is similar to that of \citet{Sarro2014}. This is expected since those authors use only completely observed objects to construct their model. However, the TPR we measure on missing values data, at $p_t=0.84$, is $\approx 4\%$ lower than that of \citet{Sarro2014} and the one we measure on non-missing values data. On the other hand, the CR of our methodology above $p=0.8$ outperforms the CR reported by \citet{Sarro2014} in spite of the missing values in our data sets. Nonetheless, we stress the fact that this comparison is not straight forward because of the following reasons. First, \citet{Sarro2014} infer their cluster model using only non-missing-value objects, later they apply it over objects with and without missing values. Second, their synthetic data set and ours are essentially different. They are constructed with different generative models, different number of elements, and different missing value patterns. 

The right panel of Fig. \ref{figure:tfpr-roc} shows the ACC and the PPV of our classifier when applied on synthetic data with missing values. The lines and the grey regions depict the mean and the maximum deviations of the results on the five synthetic data set. As this panel shows, the probability threshold with higher accuracy is $p_t = 0.84$. In what follows, and only for classification purposes, we use it as our cluster membership probability threshold. At this threshold the CR is $4.3\pm0.2$\%, the TPR is $90.0\pm0.05$\%, the ACC is $96.5\pm0.1$\%, and the PPV is $95.6\pm0.2$\%. The quoted uncertainties correspond to the maximal deviations from the mean of results in the five missing-values synthetic data sets.

\begin{figure}
\begin{center}
\resizebox{\hsize}{!}{\includegraphics{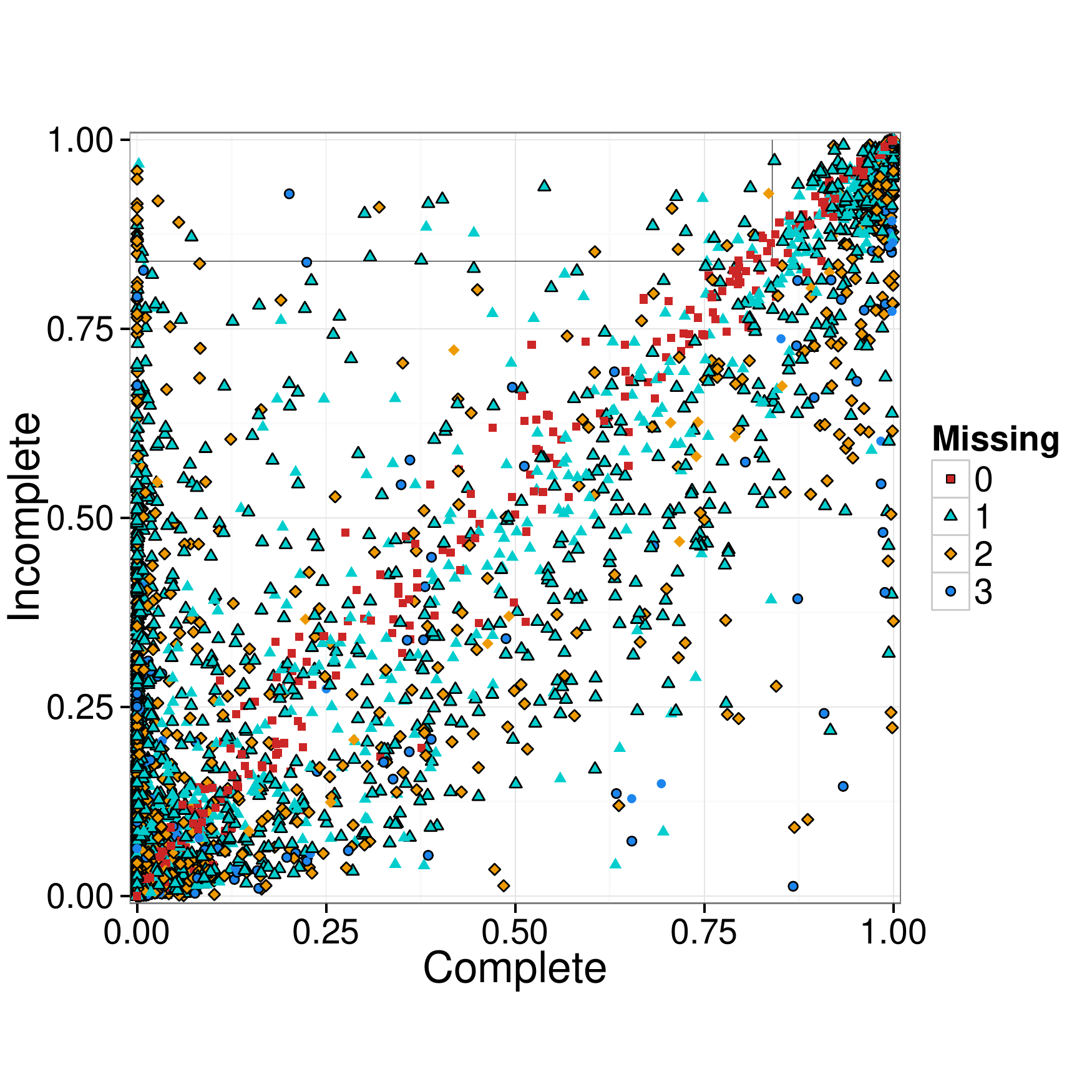}}
\caption{Comparison between the cluster membership probabilities recovered from the synthetic data with missing values (Incomplete) and without them (Complete). The colour and shape indicate the amount of missing values. The symbols enclosed in black indicate a missing $CI$. The top left box contains objects considered as contaminants due to missing values, at the probability threshold $p_t=0.84$. We notice that objects lacking just one observable appear to have larger biases than those lacking two or three. However, this is an artefact of the relative frequencies of their numbers. As explained in the text, the bias, if any, results mainly from a missing $CI$.}
\label{figure:IncVsCom}
\end{center}
\end{figure}

We investigate further on the impact of missing values. In Fig. \ref{figure:IncVsCom} we compare the cluster membership probabilities we recover in the presence of missing values (vertical axis) to those without missing values (horizontal axis). As can be seen in this Fig., the missing values impact our results by spreading the membership probabilities. This spread is expected since in general, decisions are compromised by the loss of information. The box (region above $p_t$) contains the objects which can be considered as the contaminants (at $p=p_t$) resulting from missing values. These objects have a small impact, representing only 1.8\% of the contamination (indicated by the difference between the CR for missing and complete cases in left panel of Fig. \ref{figure:tfpr-roc} at  $p_t$). The most striking difference between both probabilities comes from objects lacking the $CI$ (enclosed in black). Our methodology uses the \emph{true} $CI$ to prescribe the \emph{true} photometry, and the observed $CI$ to constrain the marginalisation integral of the \emph{true} $CI$. Thus, it is expected that a missing $CI$ will produce a probability spread. These missing $CI$ objects show two different behaviours. In one case, there are sources with membership probabilities $p_{complete} \approx0$ which have overestimated probabilities in the incomplete case (vertical axis). In the other case, the sources in the combed area below the line of unit slope have underestimated probabilities in the incomplete case. While the first case contributes to the CR the second one diminishes the TPR. The first case reaches the maximum difference at $p \approx 0$ (difference between red and blue dashed lines in Fig. \ref{figure:tfpr-roc}), thus its impact in our results is marginal. The second case, however, represents the unavoidable (in our model) loss of members due to the missing values (4\% at $p_t=0.84$). In a future version we will try to diminish this breach. In spite of the mentioned behaviours, the root-mean-square (rms) of the difference between membership probabilities of both data sets (with and without missing values) is 0.12, which we consider an small price given the gained improvements due to the inclusion of missing values. This rms drops to only 0.02 for objects with completely observed values (red squares) in both data sets. The previous effects show an overall agreement between results on data sets with and without the missing values, nonetheless, care must be taken when dealing individually with objects lacking this colour index. 

Finally, as explained in Sect. 1, our methodology aims at determining the statistical distributions of the cluster population. Our model returns these distributions without any threshold in cluster membership probabilities. In our methodology, each object contributes to the cluster distributions proportionally to its cluster membership probability. In this sense our results are free of any possible bias introduced by hard cuts in the membership probability. Nevertheless, contamination is still present and must be quantified. To quantify it, we compute the expected value of the CR\footnote{To compute the $\langle CR \rangle$ we use the following  formula:
\begin{equation}
\langle CR \rangle = \int_0^1 CR(p)\cdot p \cdot \rm{d}p,\nonumber
\end{equation}
with $p$ the probability threshold used to obtain the CR.

}. It is $\langle CR \rangle=5.8\pm 0.2$\%. In it, each CR contributes proportionally to the  probability threshold at which it is measured. 

\subsection{Results on the Pleiades}
\label{subsect:results} 

In the previous section we characterised the effectiveness of our methodology, quantified its contamination and found an objective probability threshold based on synthetic data. In this section, we present the results of applying this methodology to the real data set of Sect. \ref{sect:methodology}. First, we give the cluster and equal-mass binaries membership probabilities together with a summary of the probability distributions describing the cluster population. Afterwards, we derive the luminosity functions in the $J,H$ and $K_s$ bands.

The high dimensionality of our results prevents their direct graphical representation. Nevertheless, in what follows we present them projected onto the subspaces of proper motions and the $K_s$ vs. $i-K_s$ CMD. 

\begin{figure*}
\begin{center}
\resizebox{\hsize}{!}{\includegraphics[page=1]{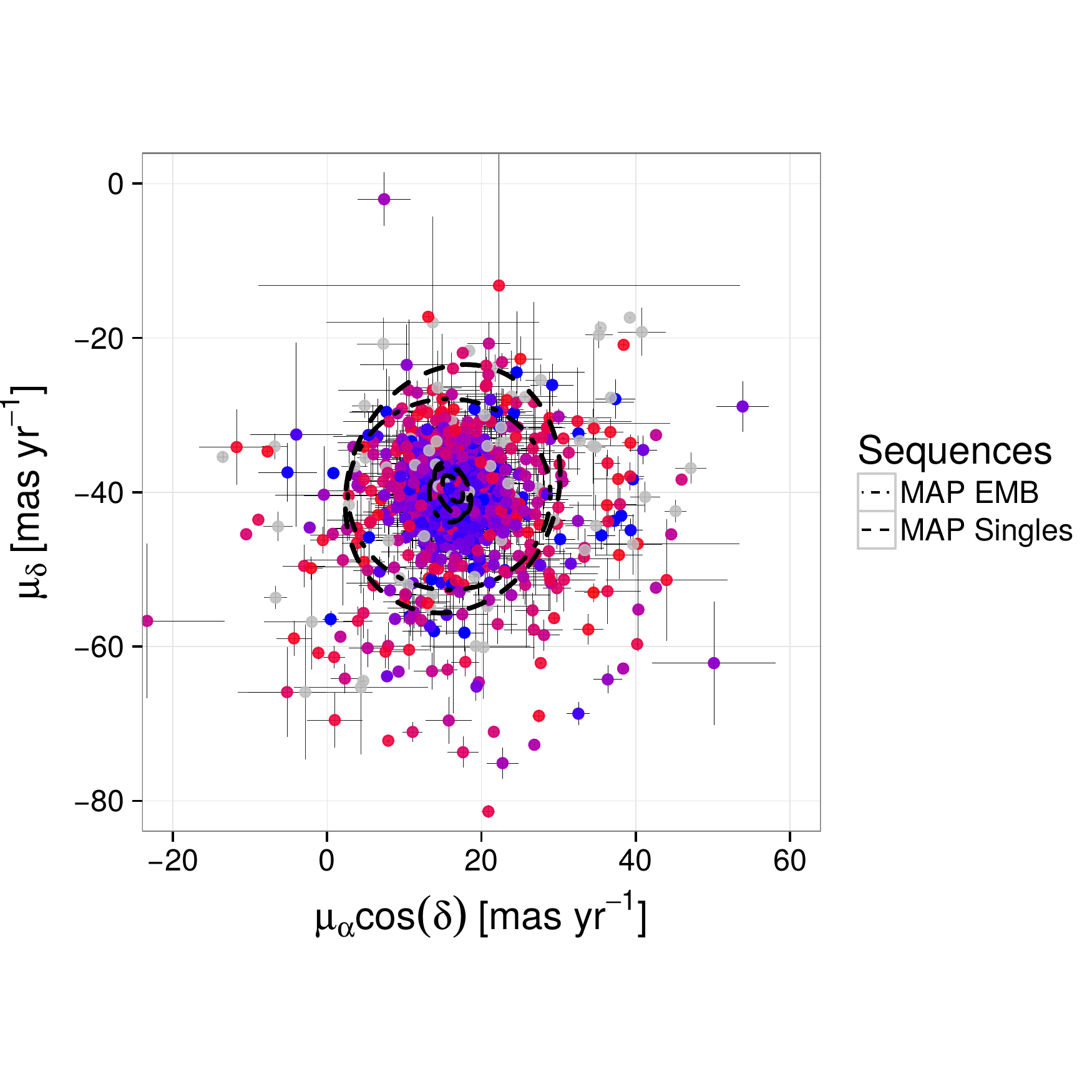}\includegraphics[page=2]{figs/Probability.pdf}}
\caption{Proper motion (left) and  $K_s$ vs. $i-K_s$ CMD (right) of our candidate members (HMPS, see text).  Grey dots depict candidate members whose cluster membership probability is below the probability threshold $p_t$ but are only included because it sensitivity to the cluster parameters reaches the $p_t$. The lines represent the MAP of the parameters in the equal-mass binaries (dot-dashed line labeled as MAP EMB) and single stars (dashed line labeled as MAP Singles) models. Standard uncertainties in photometry are in general smaller than symbols.} 
\label{figure:probabilities}
\end{center}
\end{figure*}

\begin{figure*}
\begin{center}
\resizebox{\hsize}{!}{\includegraphics[page=1]{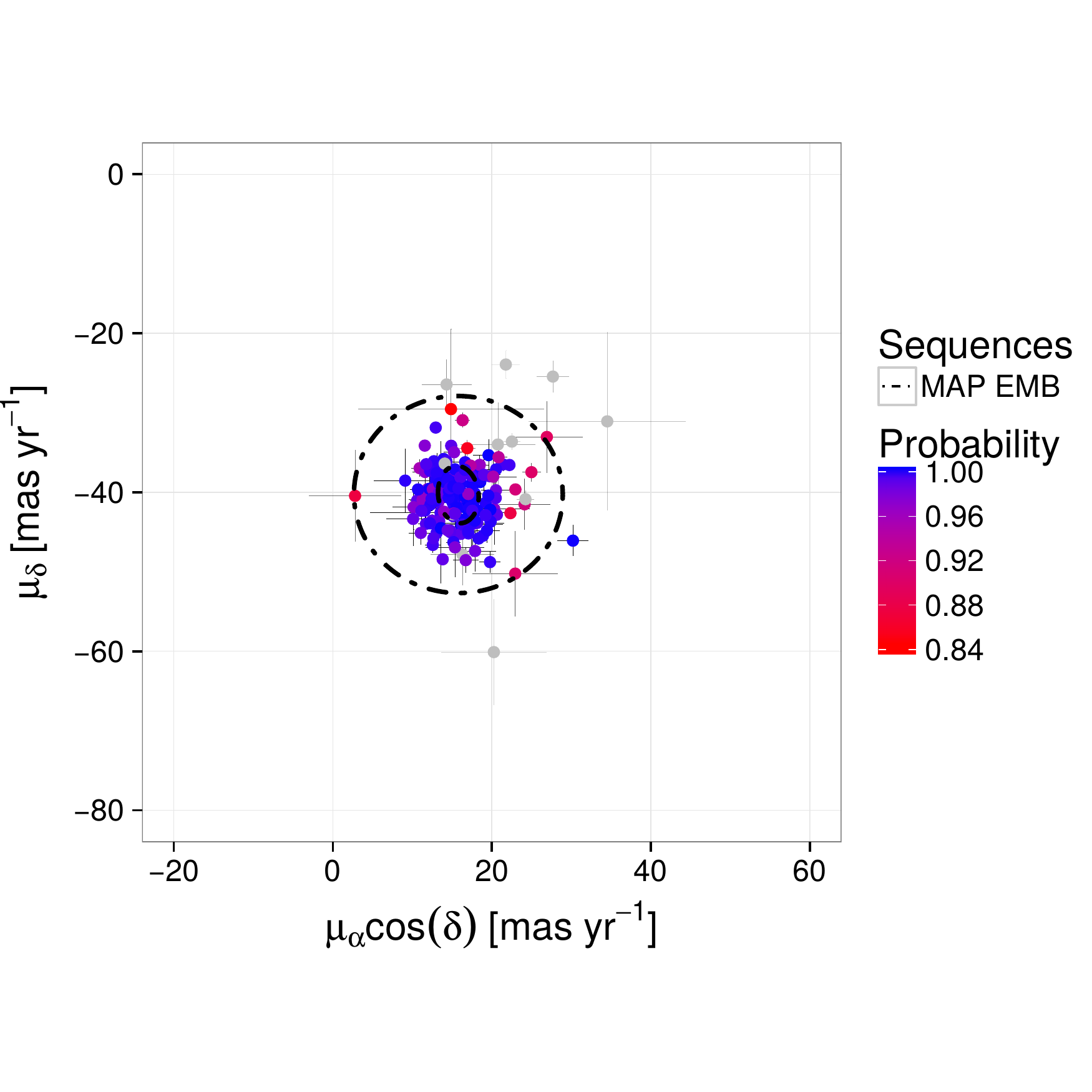}\includegraphics[page=2]{figs/Bs_members.pdf}}
\caption{Proper motion (left) and  $K_s$ vs. $i-K_s$ CMD (right) of our candidate members classified as equal-mass binaries. Captions as in Fig. \ref{figure:probabilities}.} 
\label{figure:probabilities_binaries}
\end{center}
\end{figure*}

\begin{figure}
\begin{center}
\resizebox{\hsize}{!}{\includegraphics[page=1]{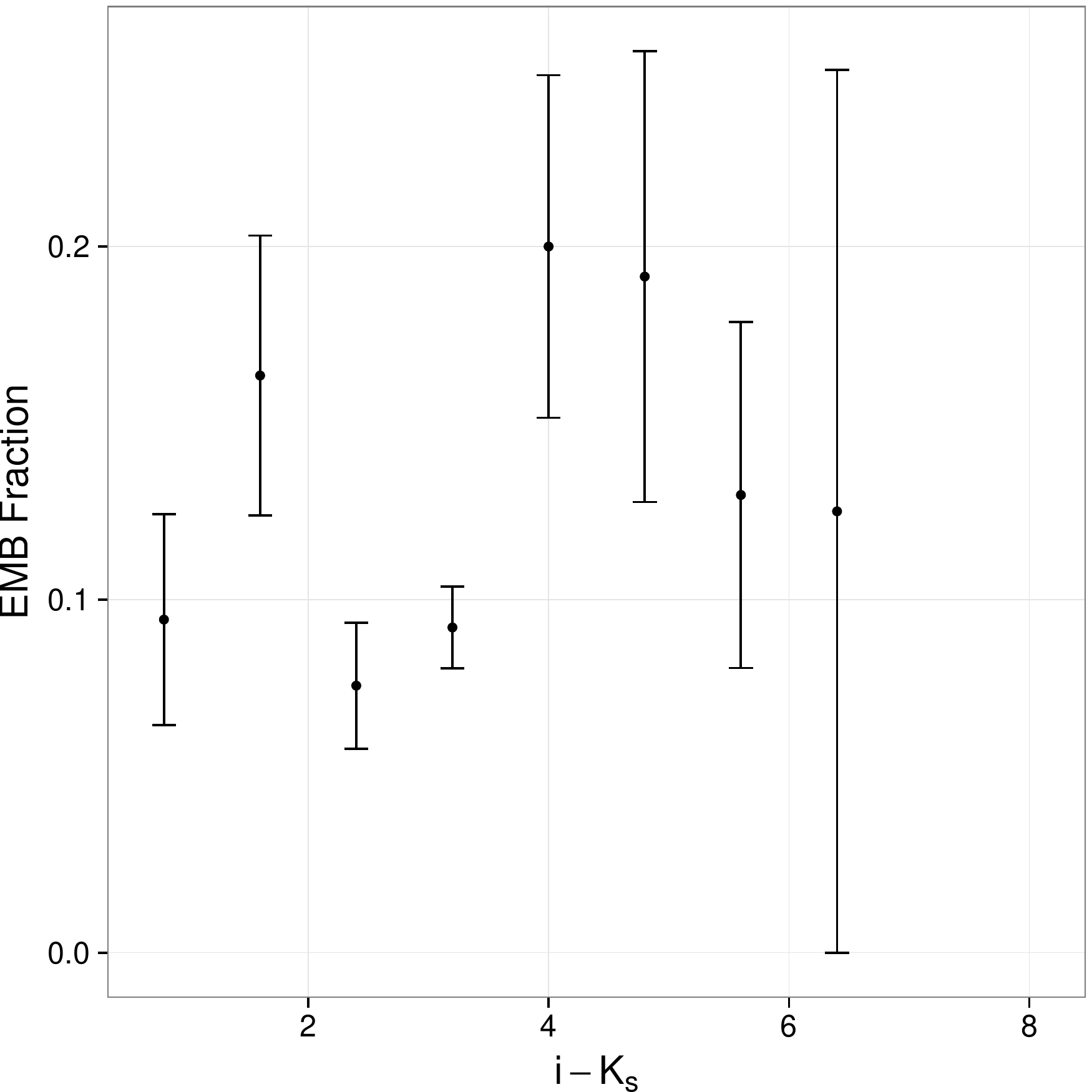}}
\caption{Fraction of candidate members classified as equal-mass binaries as a function of the $CI$ (binned intervals). Uncertainties are Poissonian.} 
\label{fig:BsFraction}
\end{center}
\end{figure}

Once the MCMC converged (see Sect. \ref{emcee}), we used the last 10 iterations (1700 samples of the parameters) to compute the cluster and equal-mass binaries membership probabilities (Eq. \ref{eq:mem-prob}) for the totality of the objects in the DANCe DR2. These membership probabilities are summarised in Table \ref{table:probs}, which also contains a flag indicating if the object has a missing $CI$ (see Sect. \ref{subsect:analysis} for a discussion on the impact of the missing $CI$). In addition, Figs. \ref{figure:probabilities} and \ref{figure:probabilities_binaries} show the cluster and equal-mass binaries membership probabilities for those objects considered as cluster candidate members. The Figures are projected into the subspaces of proper motions and $K_s$ vs. $i-K_s$ CMD, and also show the modes of the posterior distributions for parameters in the cluster and equal-mass binaries models (with dashed and dotted lines, respectively). We consider that an object is a candidate member if its membership probability plus its sensitivity to the cluster parameters ($P_c + \sigma_{P_c}$) is larger than the probability threshold $p_t=0.84$. In the DANCE DR2 data set there are 1973 objects fulfilling this criterion, in the following we refer to them as the High Membership Probability Sample (HMPS). We consider that an object is an equal-mass binary if its equal-mass binary probability is greater than 0.5. Figure \ref{fig:BsFraction} gives the fraction of candidate members, in bins of $CI$, classified as equal-mass binaries by our methodology. Uncertainties are Poissonian.

\begin{table*}[ht]
\centering
\caption{Cluster and equal-mass binaries membership probabilities. Columns one and two show the \citet{Bouy2015} and \citet{Sarro2014} identifiers. Columns three and four show the cluster $P_c$ and equal-mass binaries $P_{EMB}$ membership probabilities. Columns five and six represent the sensitivity of the membership probability to the cluster parameters. Finally, column seven is a flag for a missing CI (see discussion on Sect. \ref{subsect:analysis}.} 
\label{table:probs}
\begin{tabular}{lrrrrrl}
  \hline
Bouy+2015 ID & Sarro+2014 ID & $P_c$ & $P_{EMB}$ & SD $p_c$ & SD $p_{EMB}$ & Missing CI \\ 
  \hline
J035422.48+233812.0 & 5169343 & 0.9995 & 0.0218 & 4.5598e-05 & 4.0915e-03 & FALSE \\ 
  J035437.36+231332.7 & 5053887 & 0.9997 & 0.0797 & 3.2831e-05 & 1.3454e-02 & FALSE \\ 
  J035203.59+250113.5 & 5283439 & 0.9993 & 0.0053 & 6.3711e-05 & 1.0784e-03 & FALSE \\ 
   \hline
\end{tabular}
\end{table*}

\begin{longtab}
\begin{longtable}{llll}
\caption{Mode, 16 and 84 percentiles of each parameter posterior distribution.} \\ 
  \hline
\hline
Parameter & Mode & $p_{16\%}$ & $p_{84\%}$ \\ 
  \hline
\endfirsthead
\caption{continued.}\\
\hline
\hline
Parameter & Mode & $p_{16\%}$ & $p_{84\%}$ \\ 
\hline
\endhead
\hline
\endfoot
Field fraction & 0.967626 & 0.966728 & 0.968706 \\ 
  Cs fraction & 0.901310 & 0.893141 & 0.913867 \\ 
  Cs PM fraction 1 & 0.001348 & 0.001348 & 0.001349 \\ 
  Cs PM fraction 2 & 0.554420 & 0.526243 & 0.564641 \\ 
  Cs PM fraction 3 & 0.226456 & 0.176031 & 0.253682 \\ 
  Bs PM fraction 1 & 0.104137 & 0.103853 & 0.104189 \\ 
  Color fraction 1 & 0.086417 & 0.076100 & 0.096909 \\ 
  Color fraction 2 & 0.535425 & 0.355132 & 0.568704 \\ 
  Color fraction 3 & 0.255825 & 0.237339 & 0.295576 \\ 
  Color fraction 4 & 0.045698 & 0.021080 & 0.202029 \\ 
  Mean color 1 & 1.296733 & 1.239353 & 1.338206 \\ 
  Mean color 2 & 3.286141 & 3.110532 & 3.324626 \\ 
  Mean color 3 & 3.349153 & 3.328641 & 3.363734 \\ 
  Mean color 4 & 3.778554 & 3.665767 & 3.883132 \\ 
  Mean color 5 & 5.672365 & 5.470019 & 5.756913 \\ 
  Variance color 1 & 0.091936 & 0.078186 & 0.173507 \\ 
  Variance color 2 & 0.358799 & 0.301102 & 0.434674 \\ 
  Variance color 3 & 0.026534 & 0.026429 & 0.026586 \\ 
  Variance color 4 & 0.270303 & 0.269459 & 0.271854 \\ 
  Variance color 5 & 0.311564 & 0.276107 & 0.461157 \\ 
  Mean PM Cs[1,1] & 16.271646 & 16.200754 & 16.354460 \\ 
  Mean PM Cs[1,2] & -39.547045 & -39.709392 & -39.450590 \\ 
  Variance Cs[1,1] & 0.000000 & 0.000000 & 0.000000 \\ 
  Variance Cs[1,2] & 0.000000 & -0.000000 & 0.000000 \\ 
  Variance Cs[1,3] & 0.000000 & 0.000000 & 0.000000 \\ 
  Variance Cs[2,1] & 193.906953 & 193.372914 & 194.660560 \\ 
  Variance Cs[2,2] & 17.062073 & 6.833801 & 27.315831 \\ 
  Variance Cs[2,3] & 259.170334 & 258.826603 & 259.569312 \\ 
  Variance Cs[3,1] & 5.611323 & 4.612911 & 6.838203 \\ 
  Variance Cs[3,2] & -2.397476 & -3.634720 & -1.733489 \\ 
  Variance Cs[3,3] & 11.683655 & 11.681949 & 11.686234 \\ 
  Variance Cs[4,1] & 1.745191 & 1.620717 & 1.909607 \\ 
  Variance Cs[4,2] & -0.844115 & -0.853521 & -0.836687 \\ 
  Variance Cs[4,3] & 2.955694 & 2.581374 & 3.487740 \\ 
  Mean PM Bs[1,1] & 15.790953 & 15.594552 & 16.288392 \\ 
  Mean PM Bs[1,2] & -40.284523 & -40.413779 & -40.146048 \\ 
  Variance Bs[1,1] & 172.438086 & 97.796259 & 292.228562 \\ 
  Variance Bs[1,2] & -0.466568 & -0.502744 & -0.426990 \\ 
  Variance Bs[1,3] & 153.522924 & 65.163766 & 323.783775 \\ 
  Variance Bs[2,1] & 6.531003 & 6.493103 & 6.577238 \\ 
  Variance Bs[2,2] & -0.477653 & -2.424205 & -0.030562 \\ 
  Variance Bs[2,3] & 13.029726 & 11.597810 & 13.665141 \\ 
  Coefficient [1,1] & 6.861635 & 6.709436 & 6.997937 \\ 
  Coefficient [1,2] & 12.598796 & 12.575167 & 12.610011 \\ 
  Coefficient [1,3] & 10.646387 & 10.630062 & 10.657488 \\ 
  Coefficient [1,4] & 16.326419 & 16.284197 & 16.333162 \\ 
  Coefficient [1,5] & 16.879828 & 16.803494 & 16.942147 \\ 
  Coefficient [1,6] & 21.089951 & 20.961436 & 21.142862 \\ 
  Coefficient [1,7] & 23.308945 & 23.287670 & 23.329476 \\ 
  Coefficient [2,1] & 7.590568 & 7.581280 & 7.621498 \\ 
  Coefficient [2,2] & 11.632580 & 11.573835 & 11.678375 \\ 
  Coefficient [2,3] & 10.213682 & 10.211374 & 10.215566 \\ 
  Coefficient [2,4] & 15.671901 & 15.637808 & 15.675306 \\ 
  Coefficient [2,5] & 16.133208 & 16.070742 & 16.186593 \\ 
  Coefficient [2,6] & 19.294644 & 19.184446 & 19.329717 \\ 
  Coefficient [2,7] & 22.204061 & 21.851121 & 22.391354 \\ 
  Coefficient [3,1] & 7.555386 & 7.535184 & 7.567120 \\ 
  Coefficient [3,2] & 11.042313 & 11.010642 & 11.119179 \\ 
  Coefficient [3,3] & 9.488983 & 9.481286 & 9.497607 \\ 
  Coefficient [3,4] & 15.177808 & 15.138846 & 15.182508 \\ 
  Coefficient [3,5] & 15.339961 & 15.279761 & 15.397310 \\ 
  Coefficient [3,6] & 18.630147 & 18.560011 & 18.687693 \\ 
  Coefficient [3,7] & 19.520324 & 19.420499 & 19.559654 \\ 
  Coefficient [4,1] & 7.509144 & 7.488793 & 7.518872 \\ 
  Coefficient [4,2] & 10.918408 & 10.895172 & 11.004896 \\ 
  Coefficient [4,3] & 9.342386 & 9.330147 & 9.345062 \\ 
  Coefficient [4,4] & 14.789592 & 14.757252 & 14.795049 \\ 
  Coefficient [4,5] & 14.972319 & 14.917123 & 15.027838 \\ 
  Coefficient [4,6] & 17.582518 & 17.538436 & 17.679884 \\ 
  Coefficient [4,7] & 18.539285 & 18.175453 & 18.825374 \\ 
  Covariance Phot [1] & 0.127928 & 0.127801 & 0.128069 \\ 
  Covariance Phot [2] & 0.030551 & 0.027874 & 0.032170 \\ 
  Covariance Phot [3] & 0.007303 & 0.006981 & 0.007456 \\ 
  Covariance Phot [4] & -0.000587 & -0.001206 & 0.000065 \\ 
  Covariance Phot [5] & -0.012082 & -0.012178 & -0.012055 \\ 
  Covariance Phot [6] & 0.000000 & 0.000000 & 0.000000 \\ 
  Covariance Phot [7] & -0.009338 & -0.009368 & -0.009275 \\ 
  Covariance Phot [8] & -0.025712 & -0.027672 & -0.024934 \\ 
  Covariance Phot [9] & -0.027181 & -0.029058 & -0.025045 \\ 
  Covariance Phot [10] & 0.000027 & 0.000010 & 0.000164 \\ 
  Covariance Phot [11] & -0.001027 & -0.004293 & 0.002920 \\ 
  Covariance Phot [12] & -0.000516 & -0.004143 & 0.002762 \\ 
  Covariance Phot [13] & 0.024515 & 0.024512 & 0.024523 \\ 
  Covariance Phot [14] & 0.022874 & 0.022207 & 0.023354 \\ 
  Covariance Phot [15] & 0.000592 & 0.000587 & 0.000595 \\ 
\hline
\label{table:parameters}
\end{longtable}
\end{longtab}

We summarise the posterior distributions of cluster parameters in Table \ref{table:parameters}. It contains the mode and uncertainty of each parameter in our model. Uncertainty is expressed by the 16 and 84 percentiles of the parameter marginal posterior distribution. In Appendix \ref{app:details} we give details of these parameters and their definition. Briefly, the first six correspond to the fractions of field, cluster sequence ($Cs$) and to the weights in the proper motions GMMs of single stars and equal-mass binaries. The next 14 describe the \emph{true} colour index distribution, (fractions, means and variances). The next 14, from Mean PM Cs[1,1] to Variance Cs[4,3], and eight ,from Mean PM Bs[1,1] to Variance Bs[2,3], describe, respectively, the proper motions GMM of cluster and equal-mass binaries. The next 28 correspond to the coefficients of the cubic splines, with seven coefficients for each band ($Y, J, H$ and $K_s$). The last 15 correspond to the entries of the Cholesky decomposition of the covariance matrix which represents the intrinsic dispersion of the cluster sequence, $\Sigma_{clus}$.

\begin{figure*}
\begin{center}
\resizebox{\hsize}{!}{\includegraphics[page=1]{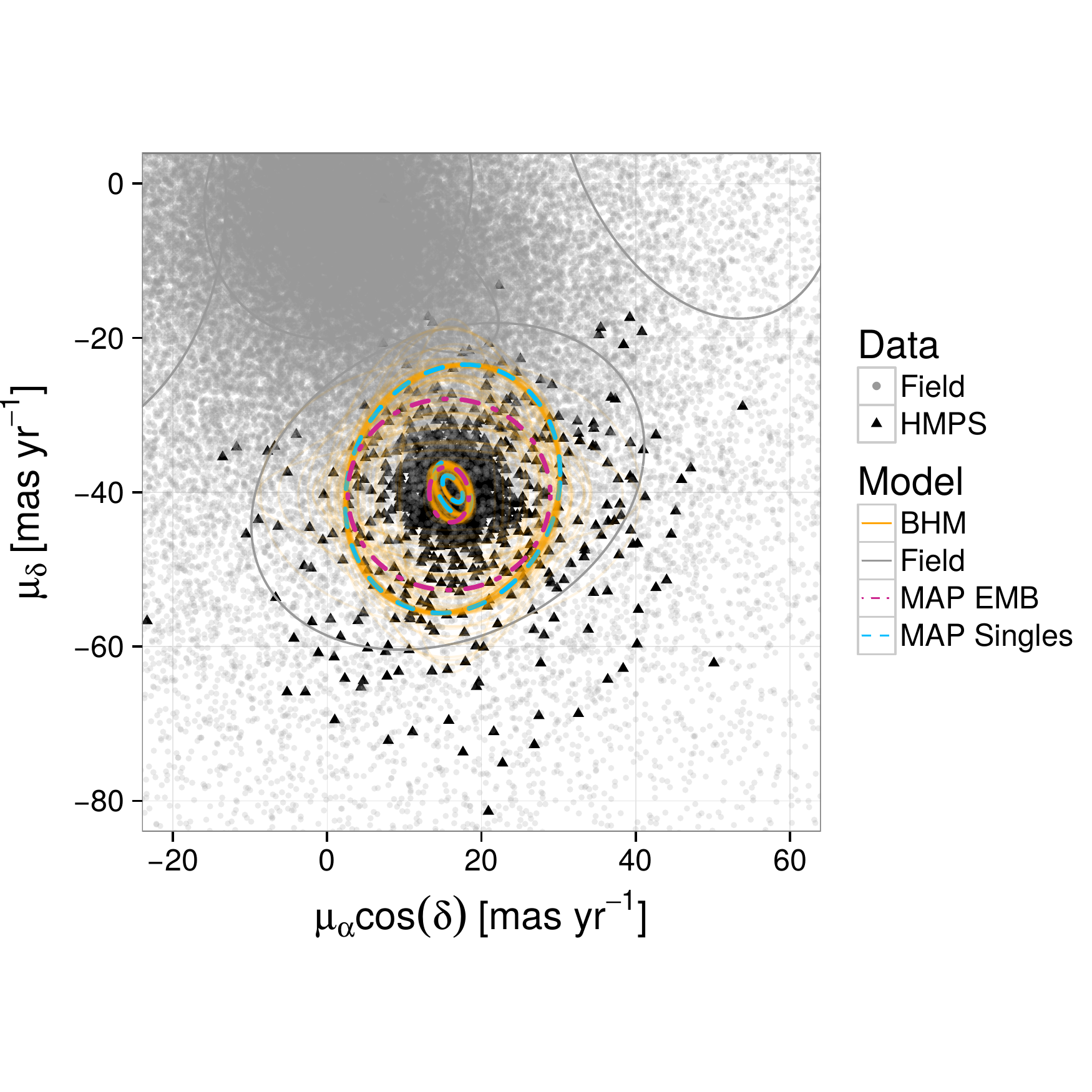}\includegraphics[page=2]{figs/MembersModel.pdf}}
\caption{Proper motion (left) and $K_s$ vs. $i-K_s$ CMD (right) projections of cluster and field models and objects in our dataset. Our candidate members in the HMPS  are in black triangles and the field objects in grey dots. Grey ellipses delineate the covariance matrices of the field GMM. The cluster model is represented by the MAP (dot-dashed and dashed lines labelled as MAP EMB and MAP Singles corresponding to equal-mass binaries and single stars, respectively) and 100 samples (orange lines) of the posterior distributions.}
\label{figure:PM-CMD}
\end{center}
\end{figure*}

In Fig. \ref{figure:PM-CMD} we show some of these distributions. It depicts objects and models in the subspaces of proper motions an $K_s$ vs. $i-K_s$ CMD. The grey ellipses delineate the GMM of the field model. We notice that, due to the high amount of missing values, most of the plotted ellipses are empty. The orange lines portray a sample of 100 realisation of the  posterior distributions of the cluster parameters. We plot, in black triangles and grey dots respectively, those objects that we classify as candidate members and as field population. We draw the mode of the posterior distributions of parameters modelling single stars and equal-mass binaries with dashed blue and dot-dashed maroon, respectively. Although the number of gaussians describing the cluster proper motions for the single stars is four, one of them collapses to fractions and covariances near zero. 

For the sake of clarity, the right panel of Fig. \ref{figure:PM-CMD} does not show the parameters related to the width of the cluster sequence. Thus, this last one appears as a narrow line. Also, and as explained in Sect. \ref{sect:field}, we build the field photometric model using the five photometric dimensions of our data set and, more importantly, we take into account the missing values. For these reasons the grey ellipses of the right panel apparently lack objects inside and in their vicinity.

\subsubsection{Luminosity functions}
\label{sect:luminosity}

We derive the distributions of the apparent magnitudes $J,H$, and $K_s$ using the posterior distributions of the parameters in our photometric model. Briefly, we do this by transforming the true $CI$ distribution into the $J,H,K_s$ distributions using the splines series and the intrinsic dispersion of the cluster sequence. The Appendix \ref{subsect:deriveluminosity} describes in detail how we do this transformation.

Then, we obtain the luminosity distributions using the magnitude distributions, the parallax and extinction of the cluster. We assume that the parallax is normally distributed with mean, 7.44 mas, and standard deviation 0.42 mas \citep{Galli2017}. This parallax distribution is convolved with the magnitude distributions to obtain the absolute magnitude distributions. We notice that this scheme results in a smoother distribution than the hypothetical one resulting from the transformation of relative to absolute magnitudes by means of individual parallaxes; the smoothness results from a lack of information. However, we do so because we do not have individual parallaxes. Finally, we deredenned them employing the canonical value, $A_v=0.12$ \citep{Guthrie1987}, which we transform to the $J,H,K_s$ values using the extinction law of \citet{Cardelli1989}.

\begin{figure}
\begin{center}
\resizebox{\hsize}{!}{\includegraphics{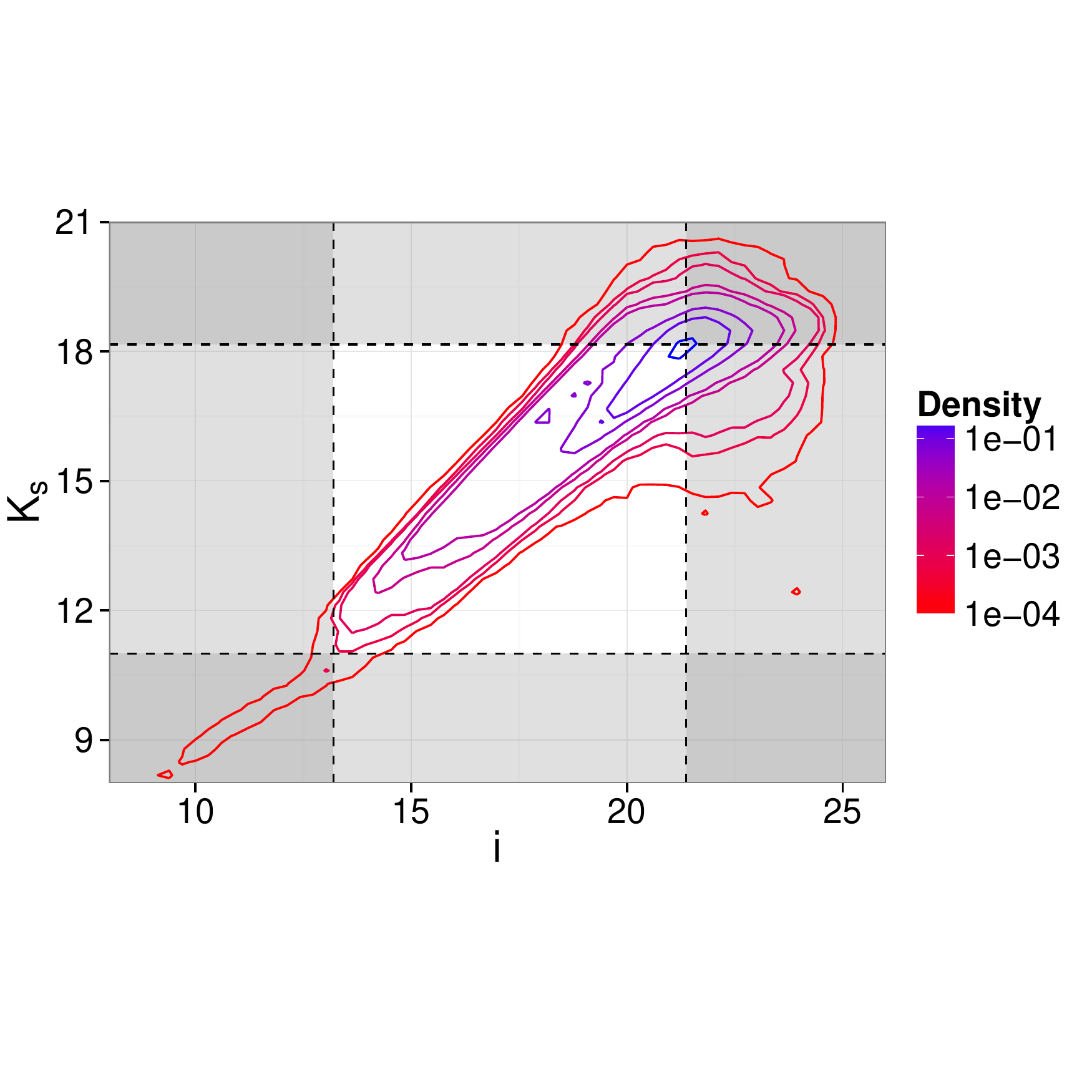}}
\caption{Density of all DANCe DR2 sources in $K_s$ vs $i$ magnitudes. Lines show our completeness limits, $13.2<i<21.4$ mag and $11<K_s<18.1$ mag. The grey area is considered incomplete.}
\label{figure:completeness}
\end{center}
\end{figure}

Since our methodology prescribes the \emph{true} photometric quantities based on the \emph{true} colour index $CI$, therefore, the completeness limits of this $CI$ dictate those of the photometric bands. The upper completeness limits that \citet{Bouy2015} estimate for $i$ and $K_s$ are $i\approx23$ mag and $K_s\approx18$ mag (see their appendix A). As they also mention, due to the heterogeneous origins of the DANCe DR2 survey, its completeness is not homogeneous over its entire area. To overcome this issue, they identified a region, the inner three degrees of the cluster, with homogeneous spatial and depth coverage and restricted their sample to it. Here, instead of restricting the sample, we assume that the UKIDSS survey provides the homogeneous spatial coverage at the faint magnitudes, and quote more conservative completeness limits at the bright end. Figure \ref{figure:completeness} shows the $K_s$ and $i$ density  for all sources in the Pleiades DANCe DR2. The upper completeness limits correspond to the point with maximum density, $i=21.4$ mag, $K_s=18.1$ mag. For the lower completeness limits we choose $i=13.2$ mag and $K_s=11.0$ mag because the density of brighter objects shows a sharp decline, probably due to saturation.
Thus, we define the $CI$ completeness interval as that of all the points, along the cluster sequence in the $K_s$ vs. $i-K_s$ CMD, for which $i$ and $K_s$ are bounded by their upper and lower completeness limits, respectively. This results on $2.7<i-K_s<5.6$ mag. With it and the cluster sequence, we derive the completeness intervals for the $J,H,K_s$. Finally, we transform these intervals to absolute magnitudes and deredden them. 

\begin{figure}
\begin{center}
\resizebox{\hsize}{!}{\includegraphics{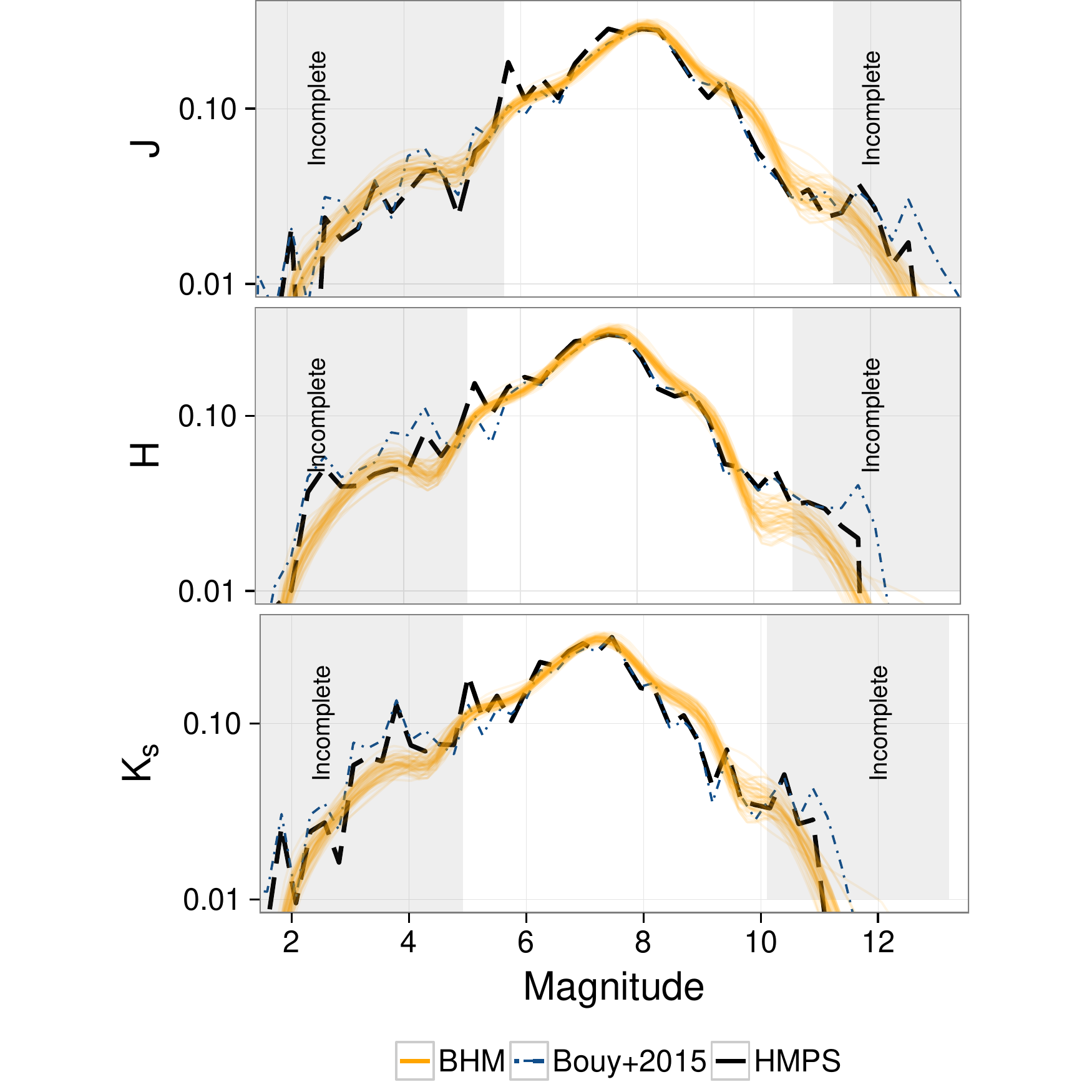}}
\caption{Luminosity functions from $J,H,K_s$ derived from the model(orange lines labelled BHM). Also shown the regions of incompleteness and the luminosity functions computed from: the candidate members of \citet{Bouy2015} (dot-dashed blue line), and our candidate members (HMPS, dashed black line).}
\label{figure:Luminosities}
\end{center}
\end{figure}

The luminosity distributions in $J,H,K_s$ together with their completeness limits are depicted (orange lines, hereafter continuous BHM-Bayesian Hierarchical Model) in Fig. \ref{figure:Luminosities}. For the sake of comparison we also show the luminosity distributions of: i) our candidate members (HMPS, as a black dashed line), and, ii) the candidate members of \citeauthor{Bouy2015} (\citeyear{Bouy2015}, blue dot-dashed line). We impute the missing values of the discrete cases using the nearest euclidean neighbour. The difference between the continuous BHM function and the HMPS comes from the imputed missing values and the objects used to obtain them. The BHM uses all objects proportionally to their cluster membership probability while the HMPS uses only the high probability candidate members. We expect differences since the HMPS is not a random sample of the continuous BHM, therefore their distributions are not exactly alike. The differences between the HMPS and that of \citet{Bouy2015} arise mainly at the bright and faint end ($K_s\approx 4$ mag and $K_s\approx11$ mag). We argue that the origin of these differences lay in our new candidate members and the rejected ones of \citet{Bouy2015} (as discussed in Sect. \ref{sect:discussion}).

\section{Discussion}
\label{sect:discussion}

In this Sect., we focus on the differences between our results and those found by \citet{Bouy2015} on the DANCe DR2 data set. First, we discuss the differences in the cluster membership probabilities, particularly on the new candidate members and the rejected ones. Later, we obtain the present day mass function, compare it with theoretical and empirical ones, and elaborate on the statistical differences that we found.

\subsection{Comparison with previous results} 

\begin{figure}
\begin{center}
\resizebox{\hsize}{!}{\includegraphics{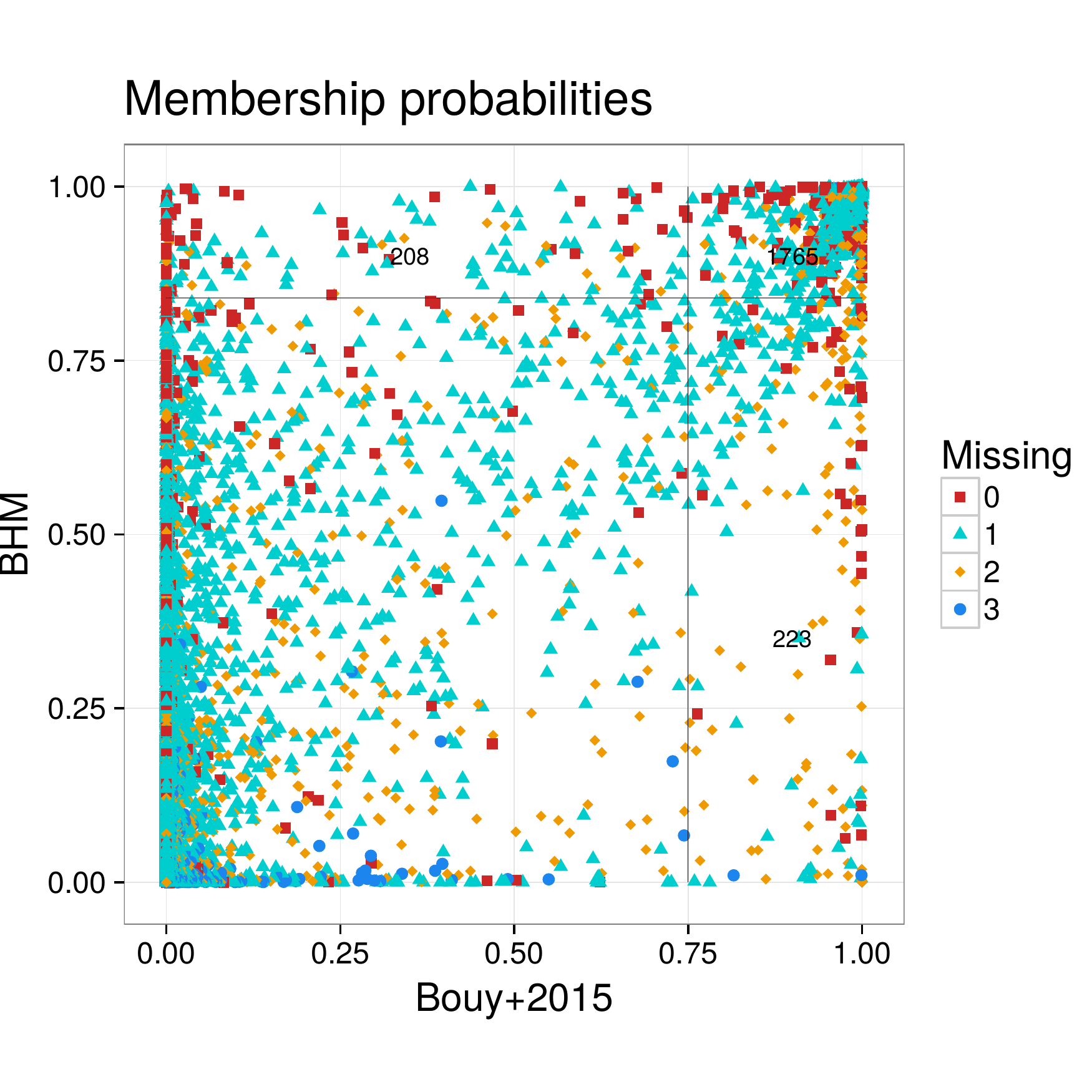}}
\caption{Recovered membership probabilities compared to those of \citet{Bouy2015}. Lines show the 0.75 and $p_t=0.84$ probability thresholds used in both works. The numbers indicate the new candidate members (top left), rejected candidate members (bottom right), and common candidate members (top right).}
\label{figure:HM-SBB}
\end{center}
\end{figure}

The works of \citet{Bouy2015} and ours, although essentially different, have common elements which allow their comparison. In spite of the differences, both agree on $\approx 90$\% of the recovered candidate members (the upper right corner of Fig. \ref{figure:HM-SBB}). In what follows, we detail the differences for individual objects.

In Fig. \ref{figure:HM-SBB} we directly compare, for objects in our data set, the cluster membership probabilities recovered by both works. Although our results on the posterior distributions of the cluster population do not depend on this probability threshold, we use it here only to illustrate differences in the classification processes. As shown in this Fig., there is an overall outstanding, 99.6\% agreement between both methodologies, which is shown by the upper right and lower left boxes of Fig. \ref{figure:HM-SBB}. Nonetheless, the differences are worthy of discussion.

\begin{figure*}
\begin{center}
\resizebox{\hsize}{!}{\includegraphics[page=1]{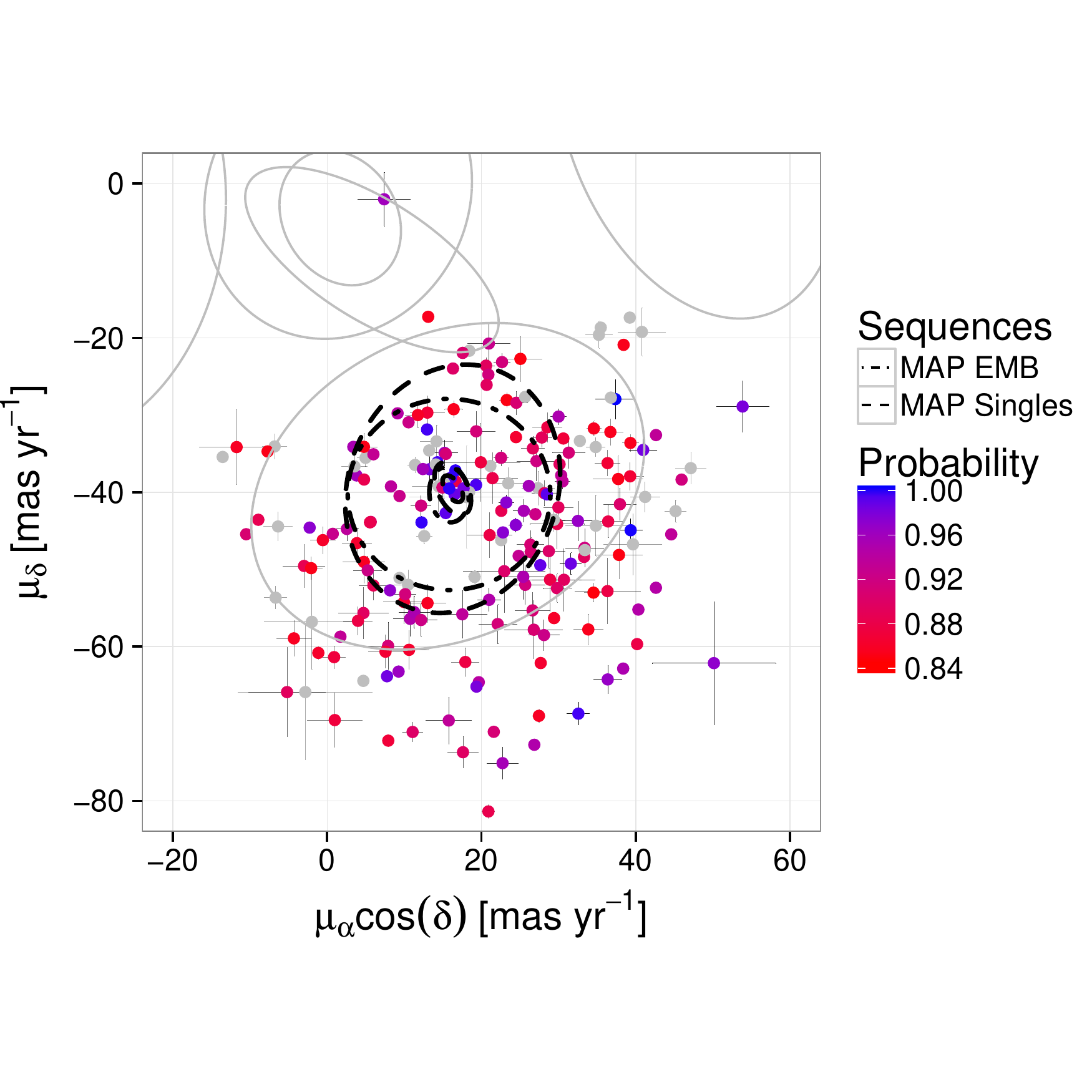}\includegraphics[page=5]{figs/NewOnes.pdf}}
\caption{Proper motion (left) and $K_s$ vs. $i-K_s$ CMD (right) showing the new candidate members found in this work. Captions as in Fig. \ref{figure:probabilities}.}
\label{figure:newones}
\end{center}
\end{figure*}

\begin{figure*}
\begin{center}
\resizebox{\hsize}{!}{\includegraphics[page=1]{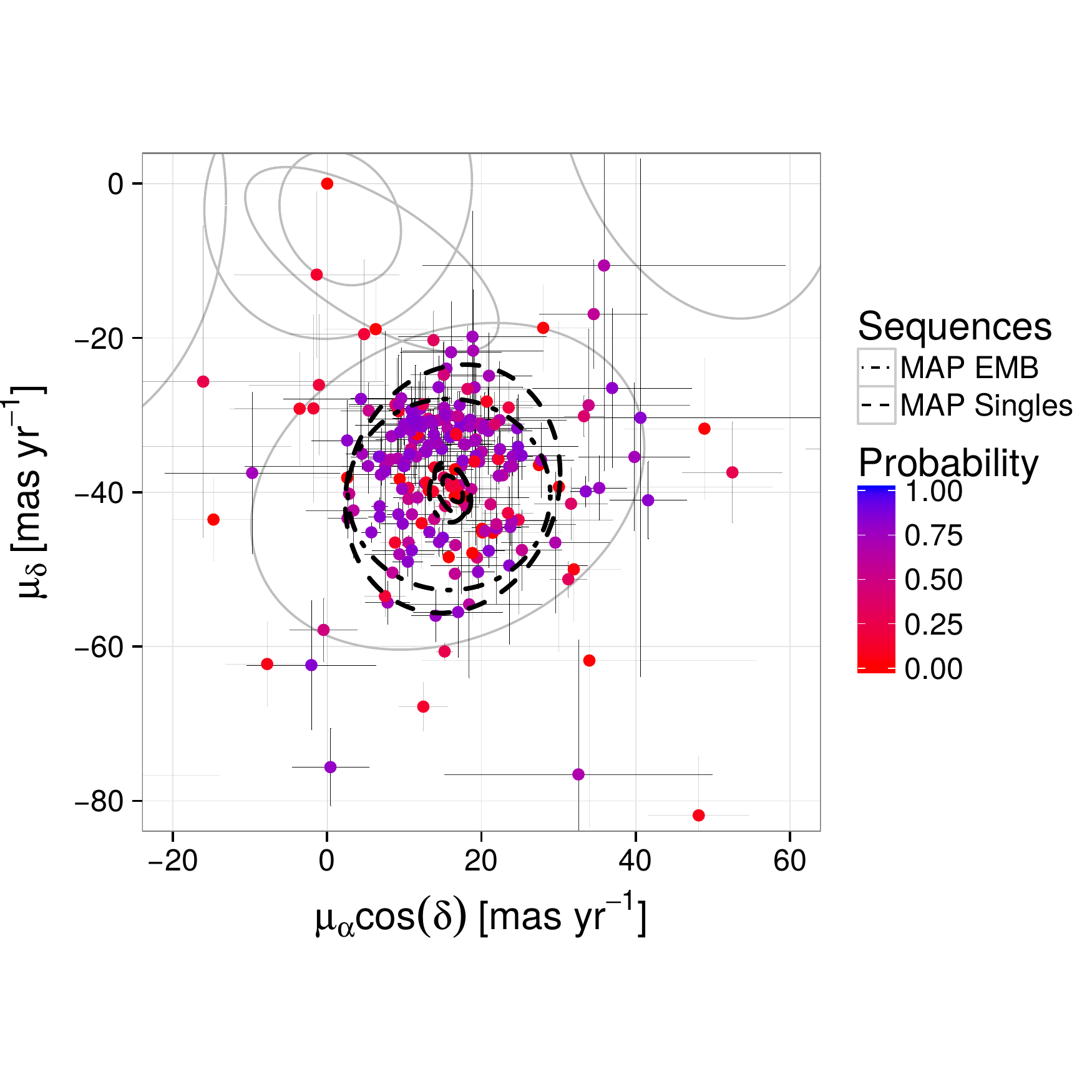}\includegraphics[page=2]{figs/Rejecteds.pdf}}
\caption{Proper motion (left) and $K_s$ vs. $i-K_s$ CMD (right) showing the rejected candidate members of \citet{Bouy2015}. Captions as in Fig. \ref{figure:probabilities}.}
\label{figure:rejecteds}
\end{center}
\end{figure*}

The rejected candidates of \citeauthor{Bouy2015} (\citeyear{Bouy2015}, at the lower right box of Fig. \ref{figure:HM-SBB}) amount to 12\% of their candidate members. This value is higher than the contamination rate reported by \citet{Sarro2014}, $7.3\pm1.4$\%. Also, the fraction of our new candidates (upper left box), 10\%, is higher than the $4.3\pm0.2$\% CR reported on Sect. \ref{subsect:analysis}. We plot the new candidates and the rejected ones of \citet{Bouy2015} in Figs. \ref{figure:newones} and \ref{figure:rejecteds}, respectively. In what follows we address these differences.

The new candidate members have proper motions uncertainties (median $\overline{\mu_{\alpha,\delta}}=\{1.37,1.37\} \rm mas\cdot yr^{-1}$) two times larger than those of the candidate members in common (median $\overline{\mu_{\alpha,\delta}}=\{0.68,0.68\} \rm mas\cdot yr^{-1}$). Also, as shown by Fig. \ref{figure:newones}, the majority of them (171) have probabilities lower than 0.95, are located in a halo around the locus of the cluster proper motions and on top of the cluster sequence in the $K_s$ vs $i-K_s$ CMD. On the contrary, the new candidates with probabilities higher than 0.95 (37), lay in the centre of the cluster proper motions and fall above the cluster sequence in the $K_s$ vs $i-K_s$ CMD. Thus, we hypothesise that, i) objects with photometry compatible with the cluster sequence but in the proper motions halo, have higher membership probabilities in our methodology due to the increased flexibility of the cluster proper motions model (four gaussians instead of the two of \citealp{Bouy2015}), and ii) objects at the centre of the cluster proper motions but above the cluster sequence are multiple systems \cite[probably triple systems which can amount to 4\% of the population][]{Duquennoy1991} with an increased membership probability due to our more flexible photometric model of the cluster and equal-mass binaries sequences.

\begin{figure*}
\begin{center}
\resizebox{\hsize}{!}{\includegraphics[page=1]{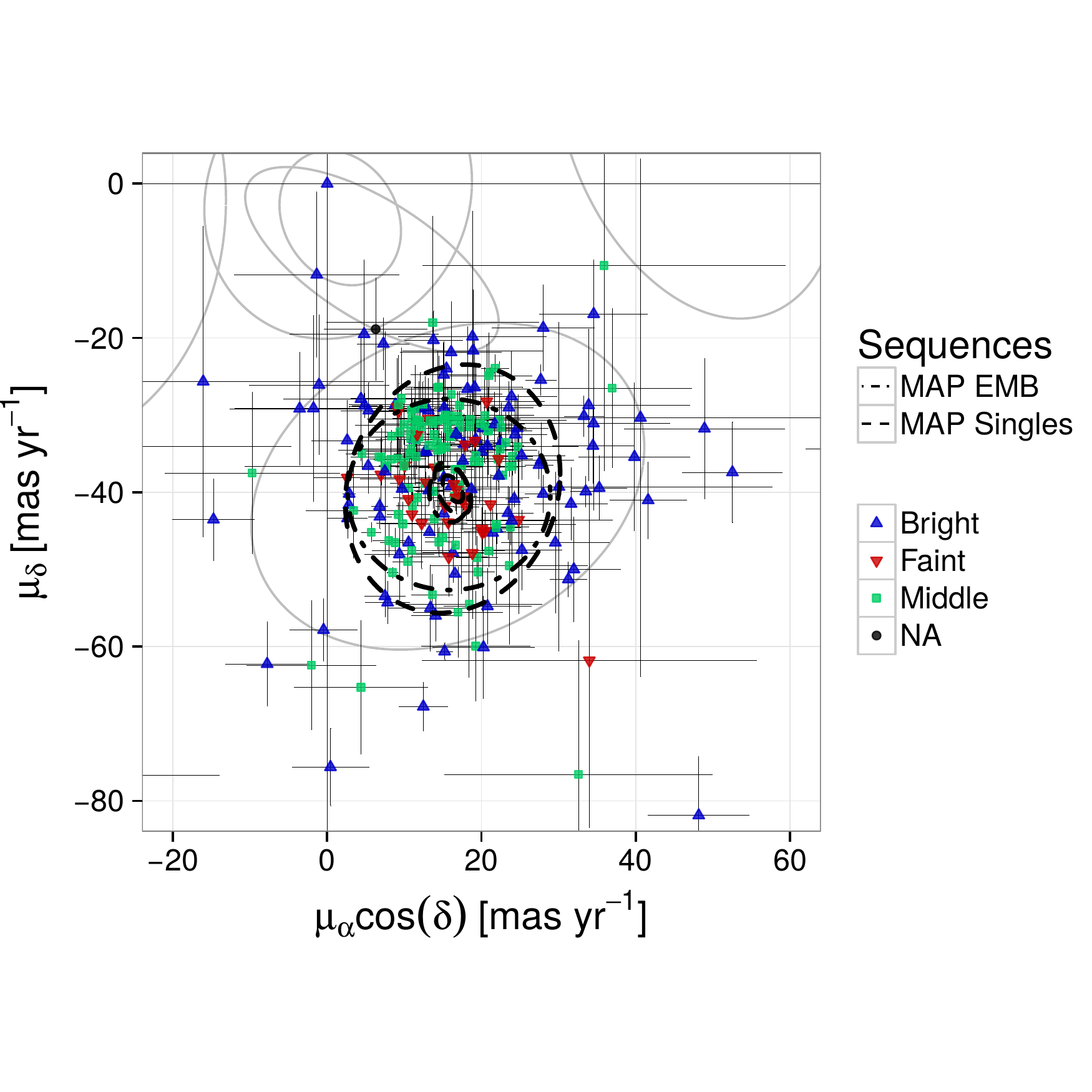}\includegraphics[page=2]{figs/RejectedsCOLORS.pdf}}
\caption{Proper motion (left) and $K_s$ vs. $i-K_s$ CMD (right) showing the rejected candidate members of \citet{Bouy2015}. The colours and shapes are a proxy for their $K_s$ magnitude. The dot-dashed and dashed lines labelled MAP EMB and MAP Singles correspond to the MAP of the equal-mass binaries and single star models, respectively.}
\label{figure:rejectedsCOLORS}
\end{center}
\end{figure*}

The rejected candidates of \citet{Bouy2015}, as it is shown in Figs. \ref{figure:rejecteds} and \ref{figure:rejectedsCOLORS}, have proper motions uncertainties (median $\overline{\mu_{\alpha,\delta}}={\{3.19,3.20\}} \rm mas\cdot yr^{-1}$) more than four times larger than those of the candidates in common and are distributed along the cluster sequence. The relatively high membership probability among these objects occurs at the middle of the cluster sequence (green squares of Fig. \ref{figure:rejectedsCOLORS}) while the lowest probabilities occur at the bright and faint ends (blue and red triangles of Fig. \ref{figure:rejectedsCOLORS}, respectively), where the missing values happen the most. We stress the fact that \citet{Bouy2015} construct their field model using a sample of $\approx 20,000$ objects without missing values. Proceeding in that way underestimates the photometric field density in the regions where missing values happen (see Fig. \ref{figure:CvsI}). Underestimating the photometric field likelihood leads to an increase in the cluster field likelihood ratio, and therefore it increases the cluster membership probabilities. Furthermore, the proper motions uncertainties of objects at the bright end (median $\overline{\mu_{\alpha,\delta}}=\{4.10,4.21\}\rm mas\cdot yr^{-1}$ and depicted as blue triangles), faint end (median $\overline{\mu_{\alpha,\delta}}=\{3.4,3.4\} \rm mas\cdot yr^{-1}$ depicted as red triangles), and at the middle magnitudes (median $\overline{\mu_{\alpha,\delta}}=\{2.6,2.6\} \rm mas\cdot yr^{-1}$ depicted as green squares) are approximately 6, 5 and 4 times larger than those of the candidates in common. Thus, we hypothesise that higher proper motion uncertainties and field likelihoods are responsible for our lower membership probabilities of  \citet{Bouy2015} rejected candidates. However, we stress the fact that, although the probability threshold $p_t=0.84$ returns the maximum accuracy of our methodology, at this value the TPR is just $90.0\pm0.05$\%. Thus, the rejected candidate members of \citet{Bouy2015} cannot be discarded as potential members. To solve this discrepancy it is necessary to have lower proper motion uncertainties and fewer missing values. Future steps will be taken to try to solve this issue.

Finally, the discrepancies in the individual membership probabilities of both works, \citet{Bouy2015} and ours, arise from the subtle but important differences between them. The inclusion of missing values in our methodology have two main consequences. First, the use of missing values in the field photometric model leads to lower membership probabilities than those of \citet{Bouy2015} in the regions where missing values happen the most. Second, the use of missing values in the construction of the cluster model allow us to include the information of good candidate members that were otherwise discarded a priori. This last point, together with the higher flexibility of our cluster model allow us to rise the membership probability of the previously discarded candidates. Furthermore, as shown by the red squares in the upper left corner of Fig. \ref{figure:HM-SBB}, the higher flexibility of our cluster model allow us to include as new candidate members previously rejected objects with complete (non-missing) values.  

\subsection{Present day mass function}

Now, we proceed to compare the photometric distributions of the cluster population to those present in the literature. 
First, we compute the present day system mass function (PDSMF) and compare it to the Initial Mass Functions (IMF) of \citet{Chabrier2005} and \citet{Thies2007}. Then, we analyse and discuss the differences between the Pleiades PDSMF and those of the Trapezium and Hyades clusters. 

We obtain the PDSMF, independently in the $J,H,K_s$ bands, by transforming the continuous luminosity distributions obtained in Sect. \ref{sect:luminosity}. These are transformed into system mass functions using the mass-luminosity relations given by the BT-Settl models of \citeauthor{Allard2012} (\citeyear{Allard2012}, the grid CIFIST2011bc for the 2MASS Vega photometric system, and $i$ band from the SDSS AB sytem), whit exactly the same grid used by \citeauthor{Bouy2015}(\citeyear{Bouy2015}, private communication). Since the luminosity functions of Sect. \ref{sect:luminosity} correspond to the luminosity of systems (single and binary stars) and the mass-luminosity relation is not a linear one, we derive the PDSMF by adding the mass distributions of single stars to those of the equal-mass binaries taking into account the proportion of equal-mass binaries inferred by the model. We notice that: i) working with only two mass ratios, 0 and 1, is an over-simplistic assumption that we plan to remedy in future versions of our methodology, and, ii) the PDSMF is derived from the luminosity functions, which in turn are derived from the posterior distributions of the cluster parameters. Thus, the uncertainties in the PDSMFs result from the propagation of those of the cluster parameters\footnote{In the literature it is customary to derive the mass distribution from individual masses of stars or systems, and then assign Poisson uncertainties accordingly. The methodology introduced in this work is conceptually different. The uncertainties in our mass distribution result from those of the posterior distribution of the cluster parameters, which in turn are propagated from those of the data and the model itself.}. 

We assume an age of 120 Myr for the Pleiades together with solar metallicity. We notice that, due to the uncertainty in the age \cite[$125\pm8$ Myr][]{Stauffer1998} and metallicity of the Pleiades, the previous assumptions are over-simplistic. The mass-luminosity relation must incorporate all sources of uncertainty (e.g. models, metallicities, ages). However, the analysis of these uncertainties and their impact in the PDMF is outside the scope of the present work. Here, we make these simplistic assumptions to directly compare our results with those of \citet{Bouy2015}. The transformation from luminosities to masses is proportional to the derivative of the mass-luminosity relation, and indeed very sensitive to it \cite[see][ for some words of caution]{DAntona1998}. Therefore, we decide to fit the BT-Settl grid with splines, and obtain the derivatives from this fit. 

\begin{figure}[htbp]
\begin{center}
\resizebox{\hsize}{!}{\includegraphics[page=1]{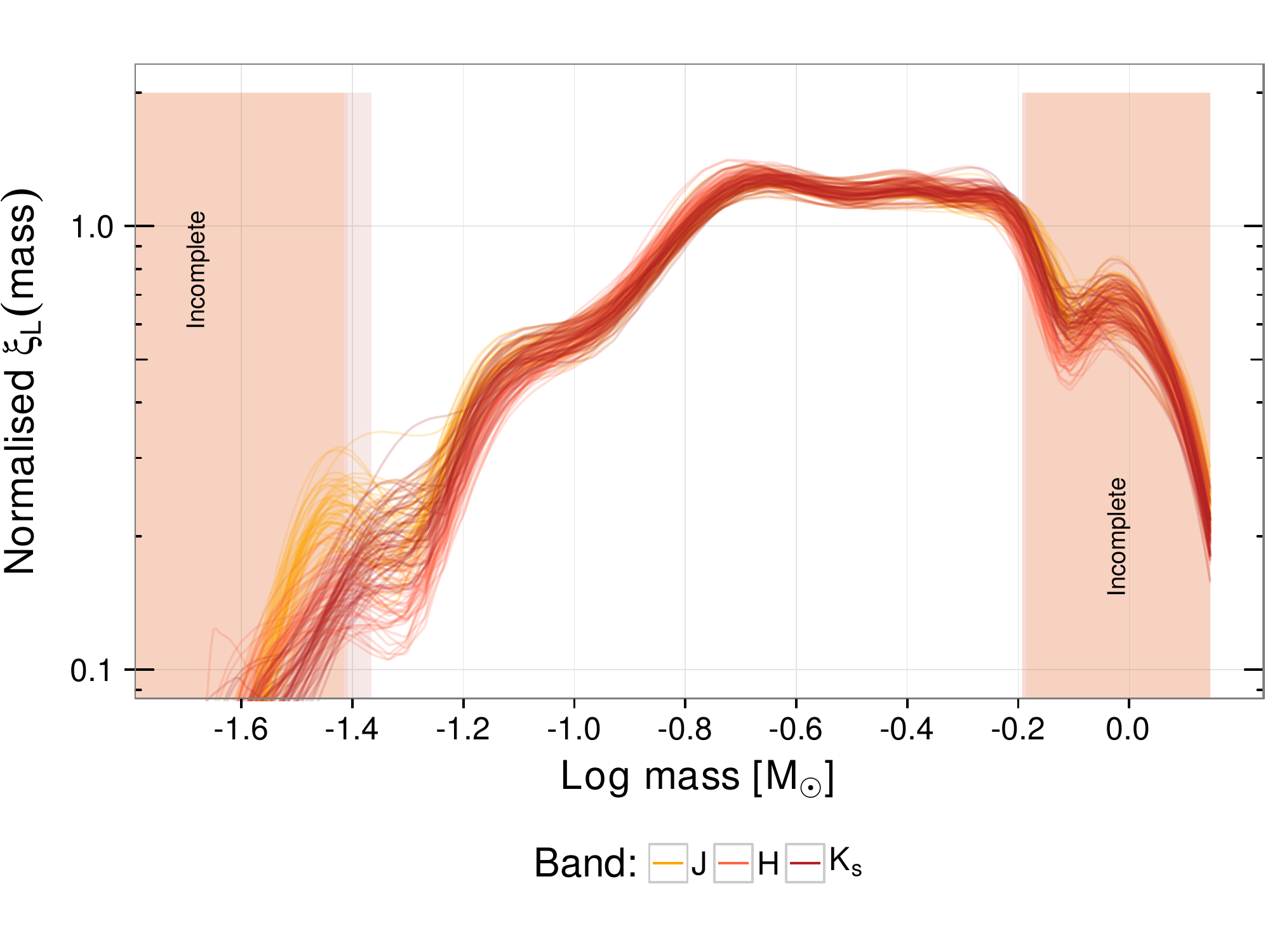}}
\caption{Normalised PDSMF in $J,H,K_s$ bands.}
\label{fig:MassFunction}
\end{center}
\end{figure}

\begin{figure}[htbp]
\begin{center}
\resizebox{\hsize}{!}{\includegraphics[page=1]{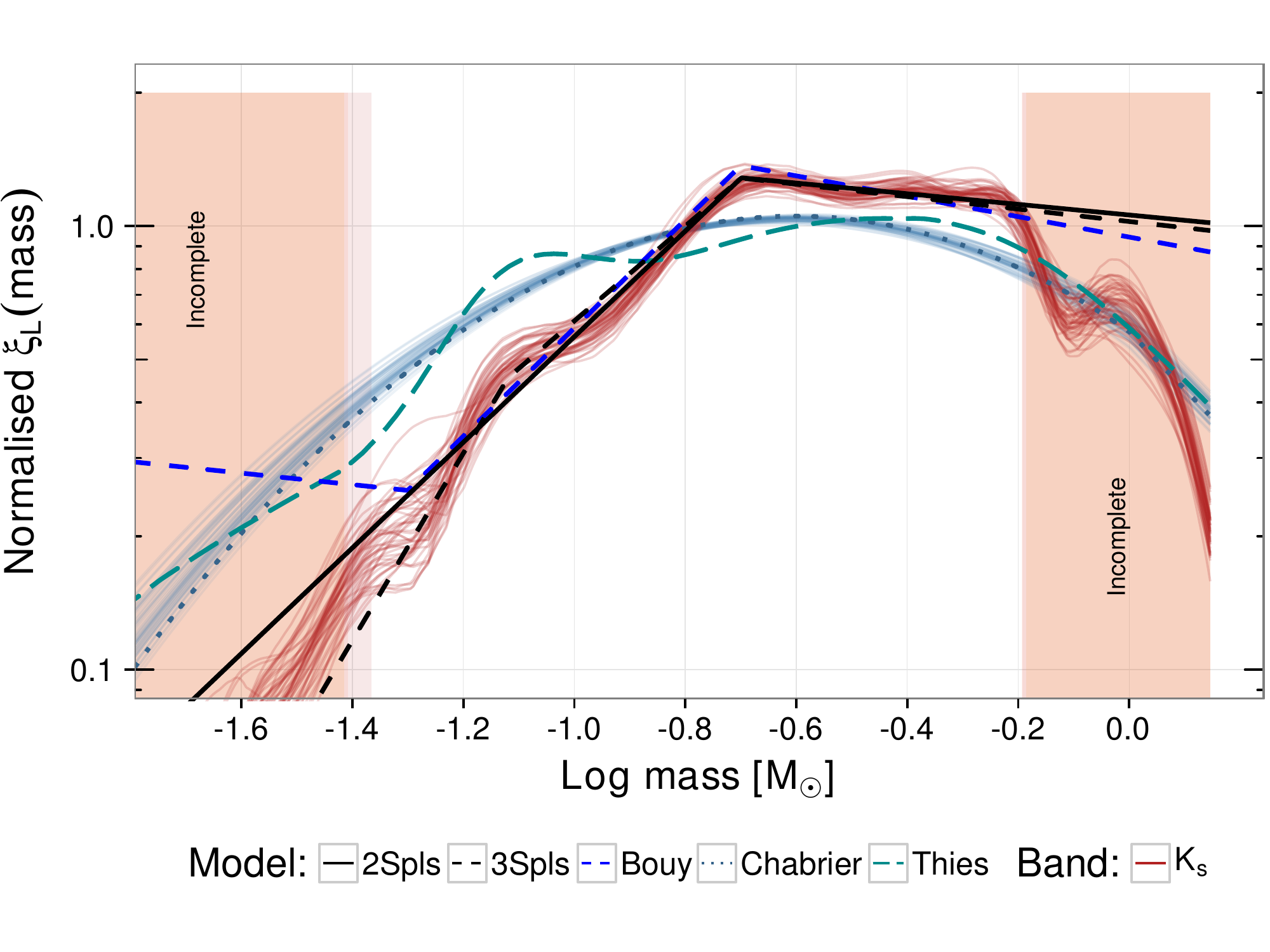}}
\caption{Normalised PDSMF in $K_s$ band together with the IMFs of \citet{Chabrier2005, Thies2007} and fits to the PDSMF found by us and \citet{Bouy2015}.}
\label{fig:ModelsMassFunction}
\end{center}
\end{figure}

Figure \ref{fig:MassFunction} shows the logarithmic PDSMF ($\xi_L$) for the $J,H,K_s$ bands normalised on the completeness limits of the survey (see Sect. \ref{sect:luminosity}). Figure \ref{fig:ModelsMassFunction} shows the PDSMF in the $K_s$ band, superimposed to the PDSMF proposed by  \citet{Bouy2015} and, the IMFs of \citet{Thies2007} and \citet{Chabrier2005}. For this last one, we show its standard uncertainty \cite[taken from][]{Chabrier2003} as a sample of blue lines. As shown in this Fig., the PDSMFs of this work compare well with each others, and, in the overlap interval, with the one proposed by \citet{Bouy2015}. However, the difference that the PDSMFs show above $0.3 M_{\odot} (-0.5 < \log M/M_{\odot})$ may have its origin on the new and rejected candidate members, which are preferentially M stars (with masses in the range $0.075 - 0.6 M_{\odot}$ or $-1.12 < \log M/M_{\odot} < -0.22$). Also, similarly to what \citet{Bouy2015} pointed out, there is a possible flattening in the PDSMF below $50 M_{Jup}$ ($\log M/M_{\odot} < -1.3$). However, due to the level of uncertainty in this region we have not enough evidence to claim it.

Using \emph{PyMultiNest}\footnote{PyMultiNest is a Python implementation of the MultiNest code \citep{2009MNRAS.398.1601F}. MultiNest is a multimodal Nested Sampling algorithm that computes the evidence, with its uncertainty, and posterior samples of possible multimodal distributions.} \citep{Buchner2014}, we fit three models to synthetic samples of our $K_s$ band PDSMF in the completeness interval: a log-normal distribution and two power-law distributions, $m^{-\alpha}$, with two and three segments. Table \ref{tab:fitPDSMF} show the parameters of these models together with their evidence. In addition and for comparison,  this Table also includes the BIC value for the power-law models of \citet{Bouy2015} and \citet{2001MNRAS.322..231K}, and the log-normal model of \citet{Chabrier2005}. Judging by the evidence in our models the best fits are the two and three segment power-laws (black lines in Fig. \ref{fig:ModelsMassFunction}, labelled as 2Spls and 3Spls, respectively). Judging by the BIC values the best model is the power-law, particularly the two and three slopes models of this work and the one of \citet{Bouy2015}. However, given the uncertainties of both the evidence and BIC, the three previous models are equally good at describing our data set. Under this similarity of evidences, the prejudice of simplicity can be use to choose the two-slopes power-law over the three-slopes power-law. This distributions is similar to that found by \citet{Bouy2015}, except for the flat part in the low-mass range and the less step slope in the high mass range. The best models found here are in clear discrepancy with the IMFs of \citet{Chabrier2005} and of \citet{Thies2007}. We notice that this apparent discrepancy may have its origin on the not yet established uncertainties in the mass-luminosity relationships, on dynamical effects associated with age, or most probably in a combination of both of them.

\begin{table*}[ht!]
\caption{Parameters, evidences and BIC values of models fitted to the ten synthetic samples of the PDSMF. For comparison, we also show the parameters of the power-laws and log normal functions of \citet{Bouy2015,2001MNRAS.322..231K} and \citet{Chabrier2005}.}
\begin{center}
\begin{tabular}{lllll}
Reference &Model    &Parameters           & Log Evidence        & BIC\\
\hline
This work &LogNormal&$m_c=0.35\pm0.03$    &                     &  \\
          &         &$\sigma=0.45\pm0.04$ & $-155 \pm 31$      &  $311\pm 60$ \\
\hline
This work & Power law     &$\alpha_0=-0.19\pm0.11$ \ \ $m \in [0.04,0.20\pm0.02]$ &                & \\ 
          & Two segments  &$\alpha_1=1.12\pm0.08$ \ \ $m \in [0.20\pm0.02,0.56]$   &$1950\pm20$ & $-3899\pm42 $ \\
\hline
This work & Power law       &$\alpha_0=-1.11\pm0.47$ \ \ $m \in [0.04,0.076\pm0.05]$       &                & \\
          &  Three segments &$\alpha_1=-0.07\pm0.51$ \ \ $m \in [0.076\pm0.05,0.20\pm0.13]$ &                & \\ 
          &                 &$\alpha_2=1.14\pm0.45$ \ \ $m \in [0.20\pm0.13,0.56]$        &$1950\pm20$ & $-3897\pm43$ \\
\hline
\citet{Bouy2015} & Power law     &$\alpha_0=1.13\pm0.6$ \ \ $m \in [0.035,0.05]$        &                & \\
                 & Four segments &$\alpha_1=0.22\pm0.1$ \ \ $m \in [0.05,20]$   &                & \\ 
                 &               &$\alpha_2=1.23\pm0.1$ \ \ $m \in [0.20,0.6]$          &                & \\
                 &               &$\alpha_3=3.56\pm0.1$ \ \ $m > 1.58 $          &                & $-3891\pm47$ \\
\hline
\citet{2001MNRAS.322..231K} & Power law     & $\alpha_0=1.3$ \ \ $m < 0.5$   &                & \\
                   & Two segments  & $\alpha_1=2.3$ \ \ $m > 0.5$  &                & $-2289\pm124$ \\
\hline
\citet{Chabrier2005} & Log Normal & $m_c=0.25_{-0.016}^{+0.021}$   &                & \\
                     &            & $\sigma=0.55_{-0.01}^{+0.05}$  &                & $566\pm25$\\
\hline
\end{tabular}
\tablefoot{The BIC and evidence values and its uncertainties correspond to the mean and the sample standard deviation obtained from the ten realisations of the synthetic samples, each containing the expected number of cluster members (3290). The uncertainty of the evidence also contains, added in quadrature, the value reported by the MultiNest algorithm \citep{2009MNRAS.398.1601F}. }
\end{center}
\label{tab:fitPDSMF}
\end{table*}%

Our PDSMF allows us to give a lower limit to the mass of the cluster. The average and mode mass of the cluster members (computed within the completeness limits) are $0.24 \pm 0.01 M_{\odot}$ and $0.26 \pm 0.09 M_{\odot}$, respectively. We compute the expected number of cluster members as the integral, over the whole range of membership probabilities, of number of objects at each membership probability, and its value is $3290 \pm 140$ objects. The product of the mean mass times the expected number of members is $795^{+40}_{-28} M_{\odot}$. Since we still lack the high mass range of the PDSMF, this value is a lower limit to the mass of the cluster. However, we cannot make any further claim based on our results because the quoted uncertainties are probably underestimated. They do not take into account the uncertainties in the mass-luminosity relations, which are yet to be established.

\subsection{Comparison with empirical mass functions.}
 
\begin{figure*}[htp]
\begin{center}
\resizebox{\hsize}{!}{\includegraphics{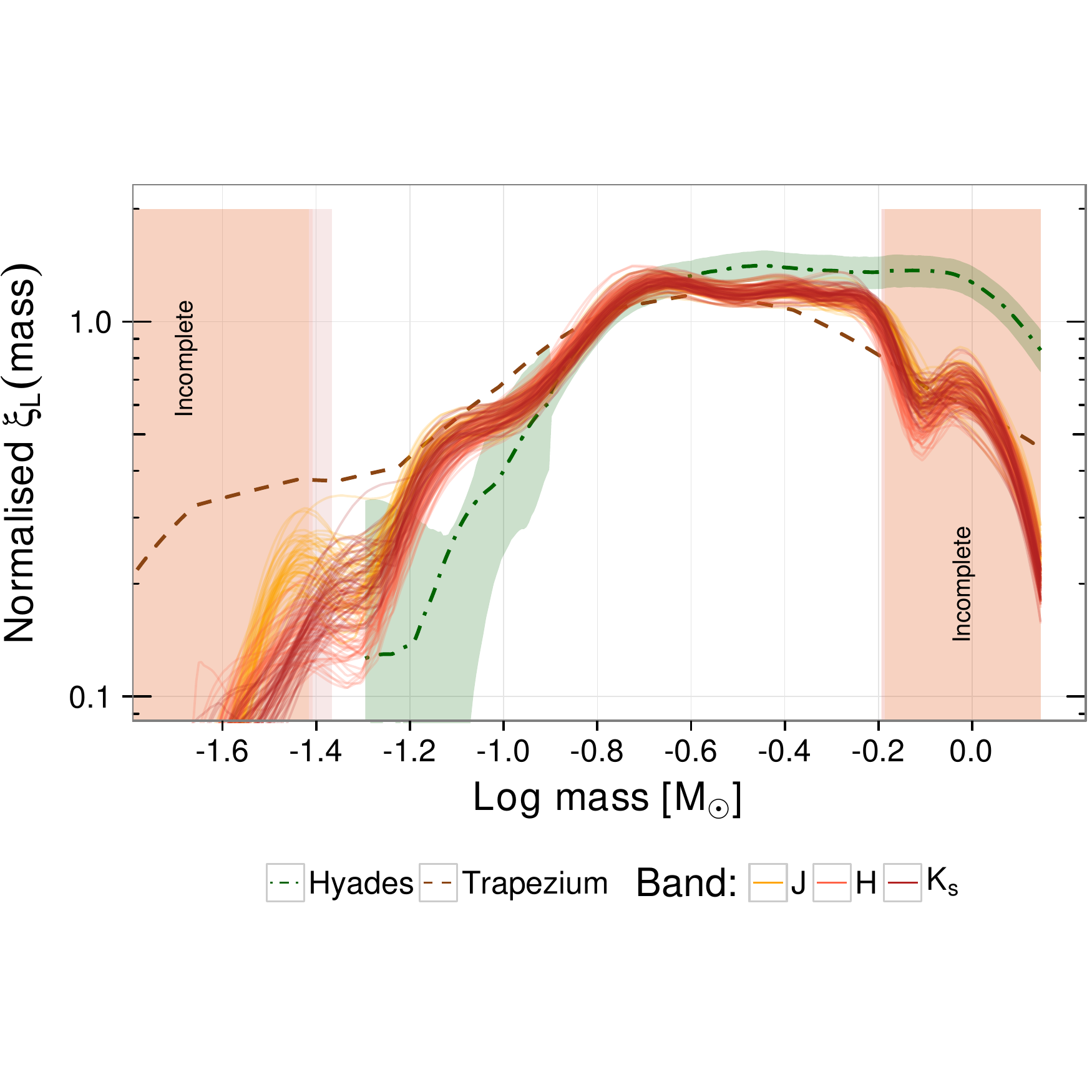}\includegraphics{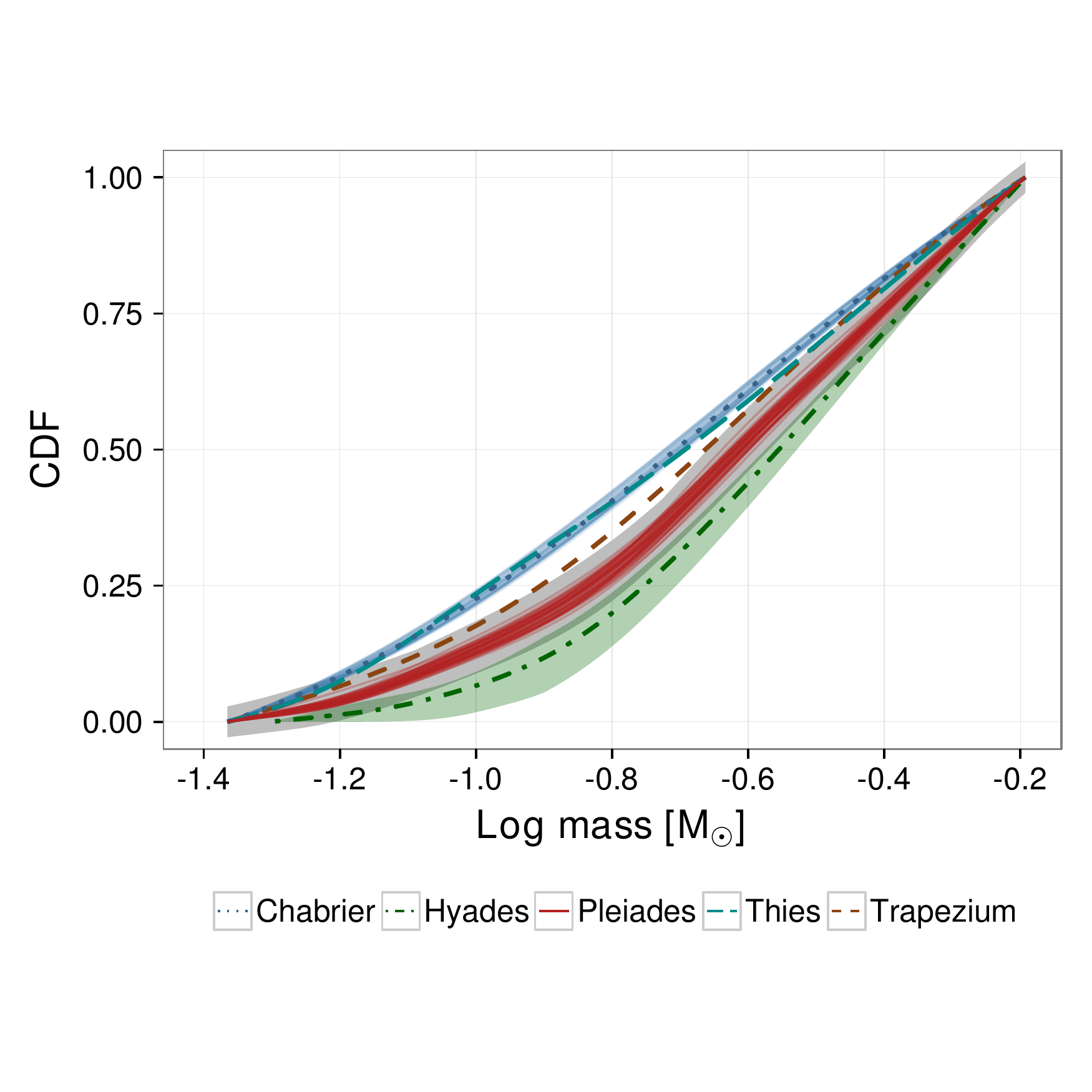}}
\caption{Left: PDSMFs of the Pleiades (derived here for $J,H,K_s$ bands), Trapezium, and Hyades, from \citet{Bouy2015}. They are normalised in the interval of completeness. Right: Cumulative distribution functions (CDF) of the PDSMFs from left panel and that of \citet{Chabrier2005} and \citet{Thies2007} system initial mass function (normalised also in the interval of completeness). The Pleiades CDF shown is just from $K_s$ band. The grey area depicts the area in which the null hypothesis of same PDSMF as that of the Pleiades can not be rejected (at $\alpha=0.01$).}
\label{fig:PDSMFcomparison}
\end{center}
\end{figure*}

Dynamical effects may have an impact on the cluster mass function. Figure \ref{fig:PDSMFcomparison} (left panel) compares the PDSMF from the Pleiades \cite[$120\pm8$ Myr][]{Stauffer1998} derived here, to those of the younger \cite[0.2 to 1.4 Myr,][]{Muench2002} and farther \cite[$414\pm7$ pc,][]{2007A&A...474..515M} Trapezium, and to the older \cite[$648\pm45$ Myr,][]{2009ApJ...696...12D} and closer \cite[$47.5\pm3.6$ pc,][]{2011AJ....141..172M} Hyades clusters. These PDSMFs correspond to those of  Fig. 11 of \citet{Bouy2015} (private communication). As mentioned by \citet{Bouy2015}, the abundance of low-mass stars and brown dwarfs in the range $0.03 - 0.1 \ \ M_{\odot}$($\log M/M_{\odot} \approx \{-1, -1.4\}$) seems to diminish with time (since the PDSMF is normalised, this produces a relative increase of low-mass stars in the range $-0.4 < \log M/M_{\odot} < -0.2$). This effect is consistent with the classical scenario in which low-mass stars and brown dwarfs are ejected as the cluster relaxes \cite[see for example][]{2004A&A...426...75M,1987MNRAS.224..193T}. To test the validity of this scenario, at least the statistical significance of the observed differences among the PDSMF of this three clusters, we test the null hypothesis that the Trapezium and the Hyades have the same PDSMF as the Pleiades. Since we just have the cluster model of the Pleiades, we are not able to perform model comparison in a bayesian fashion. Thus, to do the statistical comparison of these three clusters PDSMF we use the Kolmogorov-Smirnov and Anderson-Darling tests. 

The right panel of Fig. \ref{fig:PDSMFcomparison} shows the cumulative distribution functions (CDFs) of  the Trapezium, Pleiades (only in $K_s$ band) and Hyades PDSMFs. Also and for comparison,  we show the CDFs of \citet{Chabrier2005} and \citet{Thies2007} IMFs. The grey area around the Pleiades CDF shows the hypothesis test in which we compare each CDF with that of the Pleiades. The null hypothesis is that each compared CDF is exactly that of the Pleiades. We use the Kolmogorov-Smirnov statistic \citep{Pearson1954} with the alpha value $\alpha = 0.01$, to compute the maximum vertical distance $d_{\alpha}$ from the Pleiades CDF (shown as the grey region in Fig. \ref{fig:PDSMFcomparison}). The null hypothesis is rejected only if the tested CDF lies entirely outside the grey region around the Pleiades CDF. As can be seen, neither the IMFs nor the PDSMF of the Trapezium and Hyades lay entirely within the grey area, thus we can reject the null hypothesis that they share the same PDSMF of the Pleiades. Furthermore, since the Kolmogorov-Smirnov test uses only the maximum distance between CDFs, we also applied the more robust Anderson-Darling test \citep{Anderson1952}. In this test, the distance is computed placing more weight in the observations at the tails of the distribution. To transform this distance into a probability we use the statistic and critical values given by \citet{Scholz1987}\footnote{In the AD test, the distance between two distributions $F(x)$ and $G(x)$ is computed as:
\begin{equation}
A^{2}=\frac{m\cdot n}{N}\cdot \int _{-\infty }^{\infty }{\frac{(F(x)-G(x))^{2}}{H(x)\cdot(1-H(x))}\rm{d}H(x)}\nonumber
\end{equation}
 with $n$,$m$ and $N$ the samples of $F$, $G$ and the total of them. The $A^2$ distance is transformed to the $T_{2N}$ statistic following the formula 
 \begin{equation}
T_{2N}=\frac{A^{2}-1}{\sigma_N}\nonumber
\end{equation}
 where the $\sigma_N$ and the critical values of $T_{2N}$ are given in Eq. 4 and Table 1 of \citet{Scholz1987}, respectively.}. This test also rejects the null hypotheses, with probabilities $p < 0.004$, that the Trapezium and Hyades PDSMFs, and the \citet{Chabrier2005} and \citet{Thies2007} IMFs have the same CDF as the Pleiades.

Furthermore, we performed the Anderson-Darling test but only in the low-mass regime ($M/M_{\odot} < 0.1$). This test rejects the null hypothesis that the PDSMF of the Pleiades, in this low-mass regime, was drawn from the IMFs of \citet{Thies2007} and \citet{Chabrier2005} with $p< 0.05$ and $p<0.004$, respectively. However, this test does not entirely reject the null hypothesis that the PDSMFs of Hyades, Trapezium and Pleiades are drawn from the same distribution in the low-mass regime. 
The maximum probabilities render by the uncertainties in the Pleiades PDSMF are $p < 0.13$ and $p < 0.1$ for the Hyades and the Trapezium, respectively.
Nevertheless, these results must be analysed in the light of better constrained uncertainties. In particular those concerning the Hyades and the Trapezium PDSMFs.

The previous tests show that there is mild evidence to claim for differences among the PDSMFs of these three clusters and from IMFs and Pleiades PDSMF. Thus suggesting that these differences may have an origin on dynamical effects associated with age and relaxation. Nevertheless, to claim for reliable evidence supporting these differences the census of the Trapezium and Hyades must be done using the same methods. Also, the uncertainties must be properly established both for the other PDSMFs and for the mass-luminosity relation from which all these PDSMF are derived. 

\section{Conclusions and perspectives}
\label{sect:conclusions}

In this work we create a methodology which models the photometric and astrometric data of the heterogeneous multi-archive DANCe survey \citep{Bouy2013}. We model these data with most of its inherent characteristics: missing values, non-homogeneous observational uncertainties (heteroscedasticity) and correlations (whenever available). This enables us not just to dramatically increase the number of objects used to construct the cluster and field models \cite[$10^5$ compared to the $2\times10^4$ and 1662 of][for their field and cluster model, respectively]{Sarro2014}, but also to obtain results that minimise the biases associated to the lack of treatment for missing values and non-uniform uncertainties. The Bayesian framework used here, together with the MCMC techniques, enables us to accomplish our first objective: sample the posterior distribution of the parameters in the cluster model. This sampling also resulted in the accomplishment of our second objective: the cluster and equal-mass binaries membership probabilities. Finally, we compare our results to those of previous works and found a general agreement. Since our luminosity functions and PDSMF do not use any probability threshold, they are free of any possible bias associated to it. We also provide the list of candidate members which is the most complete up to date. We estimate that at the probability threshold of $p_t=0.84$ (at which our methodology performs the best as a classifier) the contamination and true positive rates are $4.3\pm0.2$\% and $90.0\pm0.05$\%, respectively. We stress the fact that at this probability threshold, up to 10\% of true cluster members may still have membership probabilities below it.

The main limitations of our methodology are: 
\begin{itemize}
\item The computing power it demands.
\item The accuracy of our PDSMF and mass dependent results. The later are now limited by the accuracy of the mass-luminosity relationship, which still needs to be confirmed and calibrated at low masses and very young ages. 
\item The lack of uncertainty in the field model, which remains fixed at the MLE value.
\item The lack of treatment of correlations amongst photometric and proper motions observables, which assumes that cluster members are located in a tight distribution of distances.
\end{itemize}

The methodology presented here represents the ground upon which we will continue improving our data and cluster modelling. In terms of data modelling, future steps will aim at including the spatial, radial velocities and parallaxes distributions, together with their correlations. This will allows us to deal with more complicated configurations, like two (or more) supper imposed clusters, and to apply our methodology to data from other surveys. Regarding the cluster modelling, in future works we will include: extended photometrical and kinematical treatments of binaries (regardless their mass ratio), multiple systems, white dwarfs, and the treatment of extinction. In particular, this last one will enable us to apply our methodology to even younger and embedded star forming regions. 

\begin{acknowledgements}
We are grateful to the anonymous referee for his/her kind, precise and important comments, which considerably improved the quality of this work.
J. Olivares acknowledges founding from the Ministère de l'Enseignement Supérieur et de la Recherche, and thanks Sergio Su\'arez for his kind support in the management of the computer cluster at CAB.
Some of the computations presented in this paper were performed using the Froggy platform of the CIMENT infrastructure (https://ciment.ujf-grenoble.fr), which is supported by the Rhône-Alpes region (GRANT CPER07-13 CIRA), the OSUG@2020 labex (reference ANR10 LABX56) and the Equip@Meso project (reference ANR-10-EQPX-29-01) of the programme Investissements d'Avenir supervised by the Agence Nationale pour la Recherche.
We are grateful to  S. Spiriti, P. Smith and P. Lecuyer for their R package \emph{freeknotsplines}. This study has received financial support from the French State in the framework of the Investments for the Future Program, IdEx Bordeaux, reference ANR-10-IDEX-03-02. This project has received funding from the European Research Council (ERC) under the European Union’s Horizon 2020 research and innovation programme (grant agreement No 682903 COSMIC-DANCE). We are grateful to I. Thies and P. Kroupa for providing the electronic table of their Pleiades mass function predictions. We are grateful to L. Hillebrand and G. Muench for providing an electronic version of the mass and luminosity functions of the Trapezium cluster presented in their respective articles. P.A.B. Galli  acknowledges financial support from FAPESP. This research has been funded by Spanish grants AYA2012-38897-C02-01 and AYA2010-21161-C02-02. E. Moraux acknowledges funding from the Agence Nationale pour la Recherche program ANR 2010 JCJC 0501 1 “DESC (Dynamical Evolution of Stellar Clusters)”. J. Bouvier acknowledges funding form the Agence Nationale pour la Recherche program ANR 2011 Blanc SIMI 5-6 020 01 (“Toupies”). This publication makes use of VOSA, developed under the Spanish Virtual Observatory project supported from the Spanish MICINN through grant AyA2008-02156. Based on observations obtained with MegaPrime/MegaCam, a joint project of CFHT and CEA/DAPNIA, at the Canada-France-Hawaii Telescope (CFHT), which is operated by the National Research Council (NRC) of Canada, the Institut National des Science de l’Univers of the Centre National de la Recherche Scientifique (CNRS) of France, and the University of Hawaii. Based on observations obtained with WIRCam, a joint project of CFHT, Taiwan, Korea, Canada, and France, at the Canada-France-Hawaii Telescope (CFHT), which is operated by the National Research Council (NRC) of Canada, the Institute National des Sciences de l’Univers of the Centre National de la Recherche Scientifique of France, and the University of Hawaii. This paper makes use of data obtained from the Isaac Newton Group Archive, which is maintained as part of the CASU Astronomical Data Centre at the Institute of Astronomy, Cambridge. The data was made publicly available through the Isaac Newton Group’s Wide Field Camera Survey Programme. The Isaac Newton Telescope is operated on the island of La Palma by the Isaac Newton Group in the Spanish Observatorio del Roque de los Muchachos of the Instituto de Astrof\'isica de Canarias. This research used the facilities of the Canadian Astronomy Data Centre operated by the National Research Council of Canada with the support of the Canadian Space Agency. This research draws upon data provided by C. Brice\~no as distributed by the NOAO Science Archive. NOAO is operated by the Association of Universities for Research in Astronomy (AURA) under cooperative agreement with the National Science Foundation. This publication makes use of data products from the Two Micron All Sky Survey, which is a joint project of the University of Massachusetts and the Infrared Processing and Analysis Center/California Institute of Technology, funded by the National Aeronautics and Space Administration and the National Science Foundation. This work is based in part on data obtained as part of the UKIRT Infrared Deep Sky Survey. This research has made use of the VizieR and Aladin images and catalog access tools and of the SIM- BAD database, operated at the CDS, Strasbourg, France. We are grateful to the Isaac Newton Group for the service observations with LIRIS at the WHT. The William Herschel Telescope and its service program are operated on the island of La Palma by the Isaac Newton Group in the Spanish Observatorio del Roque de los Muchachos of the Instituto de Astrof\'isica de Canarias. This publication makes use of data products from the Wide-field Infrared Survey Explorer, which is a joint project of the University of California, Los Angeles, and the Jet Propulsion Laboratory/California Institute of Technology, funded by the National Aeronautics and Space Administration. This work is based [in part] on archival data obtained with the Spitzer Space Telescope, which is operated by the Jet Propulsion Laboratory, California Institute of Technology under a contract with NASA. Support for this work was provided by NASA.
\end{acknowledgements}

\begin{appendices}
\section{Details of the methodology}
\label{app:details}

This Appendix gives specific details of the field and cluster generative models explained in Sect. \ref{subsect:generative-model}, the priors introduced in Sect. \ref{subsect:priors}, and the transformations of colour into magnitude distributions mentioned in Sect. \ref{sect:luminosity}. Also, in Table \ref{table:symbols_parameters} we summarise the parameters in our model, their symbols and the priors we use for them. Additionally, at the end of this Appendix, we schematically represent the relations among parameters in our model by means of probabilistic graphical models. 

\subsection{Details of the generative model}
\label{app:generativemodel} 

In what follows, the subscripts $pm$ and $ph$ stand for proper motion and photometry, respectively. Thus, an object with measurements $\boldsymbol{d}$ has proper motions $\boldsymbol{d}_{pm}$ and photometry $\boldsymbol{d}_{ph}$. Also, we represent the standard uncertainties as the associated covariance matrix $\boldsymbol{\epsilon}$.  In it, almost all off-diagonal elements are zero, with the exception of those of the colour index $i-K_s$ and $K_s$. Thus, $\boldsymbol{\epsilon}_{pm}$ and $\boldsymbol{\epsilon}_{ph}$ refer to the covariance matrices of proper motions and photometric standard uncertainties.
\subsubsection{The field population model}
\label{app:fieldmodel}
As explained in Sect. \ref{sect:field}, we model the field photometry and proper motions as independent distributions. We use mixtures of distributions for both these models. A GMM for the photometric model and a mixture of Gaussians and uniform distributions for the the proper motions model.
Thus, the field likelihood of an object with measurements $\boldsymbol{d}$ and standard uncertainties $\boldsymbol{\epsilon}$ is:
\begin{align}
p_f(\boldsymbol{d}|\boldsymbol{\theta}_f,\boldsymbol{\epsilon})=&\left[ c\cdot\pi_{f,pm,0} + \nonumber \right. \\
&\left. \sum \limits_{i=1}^{M_{pm}}\pi_{f,pm,i}\cdot \mathcal{N}(\boldsymbol{d}_{pm} | \boldsymbol{\mu}_{f,pm,i},\boldsymbol{\Sigma}_{f,pm,i}+\boldsymbol{\epsilon}_{pm})\right] \nonumber \\ 
&\cdot \left[ \sum \limits_{i=1}^{M_{ph}}\pi_{f,ph,i}\cdot \mathcal{N}(\boldsymbol{d}_{ph} | \boldsymbol{\mu}_{f,ph,i},\boldsymbol{\Sigma}_{f,ph,i}+\boldsymbol{\epsilon}_{ph})\right].
\label{eq:field}
\end{align}
In this equation, $\boldsymbol{\theta}_f$ refers to the set of field parameters, with $\boldsymbol{\pi}_f,\boldsymbol{\mu}_f,\boldsymbol{\Sigma}_f$ standing for the fractions, means and covariance matrices, respectively. The first and second brackets contain the models of proper motions and photometry, respectively. The first term of the proper motion model corresponds to the uniform distribution. In it, $c$ is a constant determined by the inverse of the product of the proper motions ranges, and $\pi_{f,pm,0}$ is the fraction of this uniform distribution. The second term in the same bracket is the mixture of $M_{pm}$ gaussians with means $\boldsymbol{\mu}_{f,pm}$ and covariance matrices $\boldsymbol{\Sigma}_{f,pm} +\boldsymbol{\epsilon}_{pm}$. The parameters of the photometric GMM, in the second bracket, are similar to those in the proper motion model, except for the uniform distribution.

\subsubsection{The cluster population model}
\label{app:clustermodel}
In the cluster, we also assume independence between the photometric and proper motions models. The photometric model is a mixture of cluster (subindex $Cs$) and the equal-mass binaries (subindex $Bs$). We model each element of this mixture with multivariate normal distributions, where the means are given by the \emph{true} photometric quantities both of cluster, $\boldsymbol{t}_{ph;Cs}$, and equal-mass binaries, $\boldsymbol{t}_{ph;Bs}$. The covariance matrices of these multivariate normal distributions result from the addition of the covariance matrix of the standard photometric uncertainties, $\boldsymbol{\epsilon}_{ph}$, with the modelled covariance matrix $\Sigma_{clus}$. We assume that this last one describes the intrinsic dispersion of both  cluster and equal-mass binaries sequences. Since by definition, covariance matrices are symmetric and positive semi-definite, then the Cholesky decomposition allows us to describe $\Sigma_{clus}$ with only 15 independent parameters which we infer from the data. 

\emph{The photometric model of the cluster}
In the photometric model, we prescribe the \emph{true} photometric quantities both for the cluster sequence, $\boldsymbol{t}_{ph;Cs}= \{CI,Y,J,H,K_s\}$, and the equal-mass binaries, $\boldsymbol{t}_{ph;Bs}=\{CI,Y-0.75,J-0.75,H-0.75,K_s-0.75\}$, by means of the cubic spline series, $\mathcal{S}$. These series specify the \emph{true} photometric quantities by means of the colour index $CI$, the knots and seven coefficients for each magnitude, $\beta_{Y,J,H,K_s}$. Thus, $Y =\mathcal{S}_Y(CI,\beta_Y), J =\mathcal{S}_J(CI,\beta_J),  H =\mathcal{S}_H(CI,\beta_H),$ and $ K_s =\mathcal{S}_{K_s}(CI,\beta_{K_s})$. We denote the coefficients of all the splines as the 4x7 matrix, $\boldsymbol{\beta}$. Since the \emph{true} photometry of the equal-mass binaries is a linear transformation, $T_{Bs}$, of the \emph{true} photometry of cluster sequence, no extra parameters are required. Therefore, 

\begin{align}
\boldsymbol{t}_{ph;Cs} &= \boldsymbol{\mathcal{S}}(CI, \boldsymbol{\beta}) \label{eq:trueph_Cs}\\
\boldsymbol{t}_{ph;Bs} &=T_{Bs}( \boldsymbol{\mathcal{S}}(CI, \boldsymbol{\beta})).
\label{eq:trueph_Bs}
\end{align}

Thus, the cluster and equal-mass binaries likelihoods of an object with photometric measurements $\boldsymbol{d}_{ph}$ and standard uncertainties $\boldsymbol{\epsilon}_{ph}$ are:
\begin{align}
\label{eq:lik-seq}
 p_{Cs}(\boldsymbol{d}_{ph}| CI, \boldsymbol{\beta},\Sigma_{clus},\boldsymbol{\epsilon}_{ph})={\mathcal{N}}(\boldsymbol{d}_{ph}|\boldsymbol{t}_{ph;Cs}, \boldsymbol{\epsilon}_{ph}+\Sigma_{clus}),\nonumber \\
p_{Bs}(\boldsymbol{d}_{ph}| CI, \boldsymbol{\beta},\Sigma_{clus}, \boldsymbol{\epsilon}_{ph})={\mathcal{N}}(\boldsymbol{d}_{ph}|\boldsymbol{t}_{ph;Bs}, \boldsymbol{\epsilon}_{ph}+\Sigma_{clus}),
\end{align}
where $\boldsymbol{t}_{ph;Cs}$ and $\boldsymbol{t}_{ph;Bs}$ are given by Equations \ref{eq:trueph_Cs} and \ref{eq:trueph_Bs}, respectively.

Modelling the \emph{true} $CI$ for each object in our data set demands a computing power that we currently lack. Instead, we marginalise them with the aid of a truncated GMM whose fractions ($\boldsymbol{\pi}_{CI}$), means ($\boldsymbol{\mu}_{CI}$) and variances ($\boldsymbol{\sigma}_{CI}$) we also infer from the data. This GMM is 

\begin{equation}
\label{eq:colordist}
p_{CI}(CI|\boldsymbol{\pi}_{CI},\boldsymbol{\mu}_{CI},\boldsymbol{\sigma}_{CI})= \sum_{i=1}^5 \pi_{CI,i} \cdot \mathcal{N}_t(CI| \mu_{CI,i},\sigma_{CI,i}).
\end{equation}

In this last Equation, the symbol $\mathcal{N}_t$ stands for the truncated ($0.8<CI<8$) univariate normal distribution.

Then, the marginalisation of $CI$ runs as follows:
\begin{align}
\label{eq:clmarginal}
 p_{Cs}(\boldsymbol{d}_{ph}| \boldsymbol{\theta}_c,\boldsymbol{\epsilon}_{ph})&=\int p_{Cs}(\boldsymbol{d}_{ph},CI| \boldsymbol{\theta}_c,\boldsymbol{\epsilon}_{ph}) \cdot dCI \nonumber \\
 &=\int p_{Cs}(\boldsymbol{d}_{ph}|CI, \boldsymbol{\theta}_c,\boldsymbol{\epsilon}_{ph}) \cdot p_{Cs}(CI| \boldsymbol{\theta}_c,\boldsymbol{\epsilon}_{ph})\cdot dCI \nonumber \\
p_{Bs}(\boldsymbol{d}_{ph}| \boldsymbol{\theta}_c,\boldsymbol{\epsilon}_{ph})&=\int p_{Bs}(\boldsymbol{d}_{ph},CI| \boldsymbol{\theta}_c,\boldsymbol{\epsilon}_{ph})\cdot dCI \nonumber \\
 &=\int p_{Bs}(\boldsymbol{d}_{ph}|CI, \boldsymbol{\theta}_c,\boldsymbol{\epsilon}_{ph})\cdot p_{Bs}(CI| \boldsymbol{\theta}_c,\boldsymbol{\epsilon}_{ph})\cdot dCI.
\end{align}
In these Equations, $\theta_c$ stands for all cluster parameters, and the first and second terms of the integrals in the last equalities correspond to Equations \ref{eq:lik-seq} and \ref{eq:colordist}, respectively. Since $CI$ depends only on  $\boldsymbol{\pi}_{CI},\boldsymbol{\mu}_{CI},\boldsymbol{\sigma}_{CI}$, thus, the cluster and equal-mass binaries likelihoods of datum $\boldsymbol{d}_{ph}$ are 
\begin{align}
\label{eq:lik-seq2}
 &p_{Cs}(\boldsymbol{d}_{ph}|\boldsymbol{\pi}_{CI},\boldsymbol{\mu}_{CI},\boldsymbol{\sigma}_{CI},\boldsymbol{\beta},\Sigma_{clus},\boldsymbol{\epsilon}_{ph}) \nonumber \\
 &=\int{\mathcal{N}}(\boldsymbol{d}_{ph}|\boldsymbol{\mathcal{S}}(CI, \boldsymbol{\beta}), \boldsymbol{\epsilon}_{ph}+\Sigma_{clus}) \nonumber \\
 &\cdot \sum_{i=1}^5 \pi_{CI,i}\cdot \mathcal{N}_t(CI| \mu_{CI,i},\sigma_{CI,i}) \cdot dCI\nonumber \\
&p_{Bs}(\boldsymbol{d}_{ph}|\boldsymbol{\pi}_{CI},\boldsymbol{\mu}_{CI},\boldsymbol{\sigma}_{CI}, \boldsymbol{\beta},\Sigma_{clus}, \boldsymbol{\epsilon}_{ph})\nonumber \\
&=\int{\mathcal{N}}(\boldsymbol{d}_{ph}|T_{Bs}( \boldsymbol{\mathcal{S}}(CI, \boldsymbol{\beta})), \boldsymbol{\epsilon}_{ph}+\Sigma_{clus}) \nonumber \\ &\cdot \sum_{i=1}^5 \pi_{CI,i}\cdot \mathcal{N}_t(CI| \mu_{CI,i},\sigma_{CI,i}) \cdot dCI.
\end{align}

\emph{The proper motions model of the cluster}
For the proper motions models of both cluster and equal-mass binaries we use GMM with 4 and 2 gaussians, respectively. Each with its own fractions, $\boldsymbol{\pi}$, means, $\boldsymbol{\mu}$ and covariance matrices, $\boldsymbol{\Sigma}$. However, gaussians within each GMM share the mean. Since covariance matrices are symmetric, only three independent parameters are needed to describe them. Thus, the cluster and equal-mass binaries likelihoods of object with measurements $\boldsymbol{d}_{pm}$ and uncertainties $\boldsymbol{\epsilon}_{pm}$ are

\begin{align}
p_{Cs}(\boldsymbol{d}_{pm}| \boldsymbol{\pi}_{Cs}, \boldsymbol{\mu}_{Cs},\boldsymbol{\Sigma}_{Cs},\boldsymbol{\epsilon}_{pm})
&= \sum_{i=1}^4\pi_{Cs,i}\cdot \mathcal{N}(\boldsymbol{d}_{pm} | \boldsymbol{\mu}_{Cs},\Sigma_{Cs,i}+\boldsymbol{\epsilon}_{pm}) \nonumber\\
p_{Bs}(\boldsymbol{d}_{pm}| \boldsymbol{\pi}_{Bs}, \boldsymbol{\mu}_{Bs},\boldsymbol{\Sigma}_{Bs},\boldsymbol{\epsilon}_{pm})
&= \sum_{i=1}^2\pi_{Bs,i}\cdot \mathcal{N}(\boldsymbol{d}_{pm} | \boldsymbol{\mu}_{Bs},\Sigma_{Bs,i}+\boldsymbol{\epsilon}_{pm}).
\label{eq:lik-pm}
\end{align}

Finally, the total cluster likelihood of an object with measurement $\boldsymbol{d}$ and uncertainties $\boldsymbol{\epsilon}$ is

\begin{align}
p_c(\boldsymbol{d}|\boldsymbol{\theta}_c,\boldsymbol{\epsilon})&=\pi_{CB}\cdot p_{Cs}(\boldsymbol{d}_{pm}| \boldsymbol{\pi}_{Cs}, \boldsymbol{\mu}_{Cs},\boldsymbol{\Sigma}_{Cs},\boldsymbol{\epsilon}_{pm}) \nonumber \\ &\cdot  p_{Cs}(\boldsymbol{d}_{ph}|\boldsymbol{\pi}_{CI},\boldsymbol{\mu}_{CI},\boldsymbol{\sigma}_{CI},\boldsymbol{\beta},\Sigma_{clus},\boldsymbol{\epsilon}_{ph})\nonumber\\
&+(1-\pi_{CB})\cdot p_{Bs}(\boldsymbol{d}_{pm}| \boldsymbol{\pi}_{Bs}, \boldsymbol{\mu}_{Bs},\boldsymbol{\Sigma}_{Bs},\boldsymbol{\epsilon}_{pm}) \nonumber \\
&\cdot  p_{Bs}(\boldsymbol{d}_{ph}|\boldsymbol{\pi}_{CI},\boldsymbol{\mu}_{CI},\boldsymbol{\sigma}_{CI}, \boldsymbol{\beta},\Sigma_{clus}, \boldsymbol{\epsilon}_{ph}),
\end{align}

where $\pi_{CB}$ is the parameter representing the fraction of single cluster sequence stars $Cs$ (the non equal-mass binaries in the cluster). The photometric and proper motions likelihoods are given by Equations \ref{eq:lik-seq2} and \ref{eq:lik-pm}, respectively.

\subsection{Details of the priors}
\label{subsect:apppriors}

In Sect. \ref{subsect:priors} we give the kind of distributions we use for setting our prior beliefs. Here, we give details on the parameter values we choose for these distributions. In the Hierarchical Bayesian model formalism, the parameters of the distributions governing the priors are called hyper-parameters, here, we stick to that convention. We classify the priors of our parameters into three categories: fractions, means, and covariance matrices. 

As mentioned in Sect.\ref{subsect:priors}, we use the Dirichlet distribution to set the priors of the fractions. For the field-cluster mixture we set the hyper-parameters to $\boldsymbol{\alpha}=\{98,2\}$. The means of the field and cluster fractions distributions resulting from these hyper-parameters, correspond to the fraction of objects in our data set,  that \citet{Bouy2015} classified as field an candidate members, respectively. For the cluster-equal-mass binaries mixtures, we use as hyper-parameter values, $\boldsymbol{\alpha}_{CB}=\{8,2\}$, this induce a  distribution for the fraction of equal-mass binaries whose mean is at 20\%, as suggested by \citet{Bouy2015}. For fractions in the cluster and equal-mass binaries proper motions we set their hyper-parameters to $\boldsymbol{\alpha}_{Cs}=\{1,1,1,1\}$ and $\boldsymbol{\alpha}_{Bs}=\{1.2,8.8\}$. The first values result in equal priors to all components while the second one induce similar means to those recovered after fitting a GMM to the \citet{Bouy2015} candidate members. Since the gaussians in proper motion model of single stars could be interchanged and we observe that a posteriori the fraction of one of them goes to zero, we adopt an even less informative prior for  these fractions and set all $\alpha_{Cs}$ to one. Despite this less informative prior, one of the gaussians still has a negligible contribution in the posterior solution (see Table \ref{table:parameters}).

\begin{figure*}[htbp]
\begin{center}
\includegraphics[page=1,width=8cm]{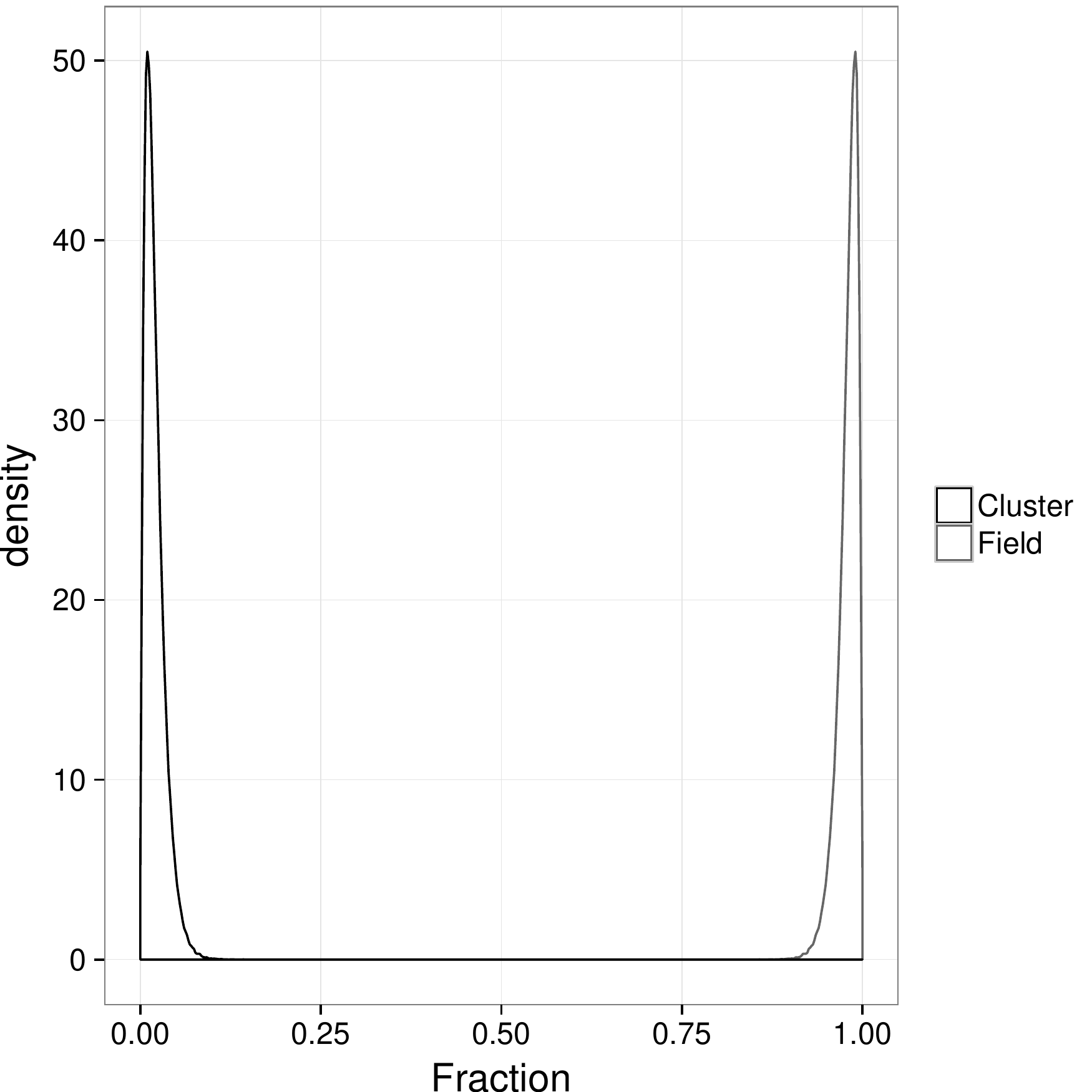}
\includegraphics[page=2,width=8cm]{figs/priors.pdf}\\
\includegraphics[page=3,width=8cm]{figs/priors.pdf}
\includegraphics[page=4,width=8cm]{figs/priors.pdf}
\caption{Prior distribution of fraction parameters. From top left to bottom right, the distributions of field fraction ($\pi$), equal-mass binaries fraction ($1-\pi_{CB}$), and the cluster ($\pi_{Cs}$) and equal-mass binaries ($\pi_{Bs}$) fractions in their proper motion GMM, respectively.}
\label{figure:priors}
\end{center}
\end{figure*}

Although the means of the la distributions correspond to what \citet{Bouy2015} found, the variances of these priors allow us to explore wide ranges of values (except for the fraction of field objects), as it is shown in Fig. \ref{figure:priors}. However, the narrow variance in the cluster-field mixture correspond to our prior belief about the fraction candidate members within our large data set which we expect to be very small $\leq 2\%$. For the fraction in the GMM of the $CI$ distribution, we set all the hyper-parameter values to 1, ($\boldsymbol{\alpha}_{CI}=\{1,1,1,1,1\}$), which results in equal means and large variances for all of them alike.

We select the bivariate normal distribution as prior for the parameters representing the means of the GMM, both for cluster and equal-mass binaries. We set the hyper-parameters of this bivariate normal as those found after fitting a bivariate normal to the candidate members of \citet{Bouy2015}, both cluster and equal-mass binaries together. These values are $\boldsymbol{\mu}_{\mu_{pm}}=\{16.30,-39.62\}$ and $\Sigma_{\mu_{pm}}=\{\{36.84,1.18\},\{1.18,40.71\}\}$. 

We use the Half--t($\nu,\boldsymbol{A}$) distribution as prior for the covariance matrices in the proper motions GMM. We set the hyper-parameter to $\nu=3$ and $\boldsymbol{A}_{pm}=\{10^5,10^5\}$. According to \citet{Huang2013}, a value of $\nu=3$ leads to marginal squared root distributions for all correlation terms of the covariance matrix. And also, arbitrarily large values of $\boldsymbol{A}$ lead to arbitrarily weakly informative priors on the corresponding standard deviation terms.

For the means and variances in the GMM of the $CI$, we use uniform and Half--Cauchy$(0,\eta)$ distributions, respectively. The uniform distribution is defined over the $CI$ span ($0.8<CI<8$) while for the scale of the Half--Cauchy we use a value of $\eta=100$.

Again, we make use of the Half--t$(\nu,\boldsymbol{A})$ distribution to establish the prior for the intrinsic dispersion of the cluster sequence, $\Sigma_{clus}$. $\nu$ is again 3. However, we use  $\boldsymbol{A}_{ph}=\{10,10,10,10,10\}$, which are large values compared to those of the photometric uncertainties.

\begin{figure}[htbp]
\begin{center}
\resizebox{\hsize}{!}{\includegraphics[page=1]{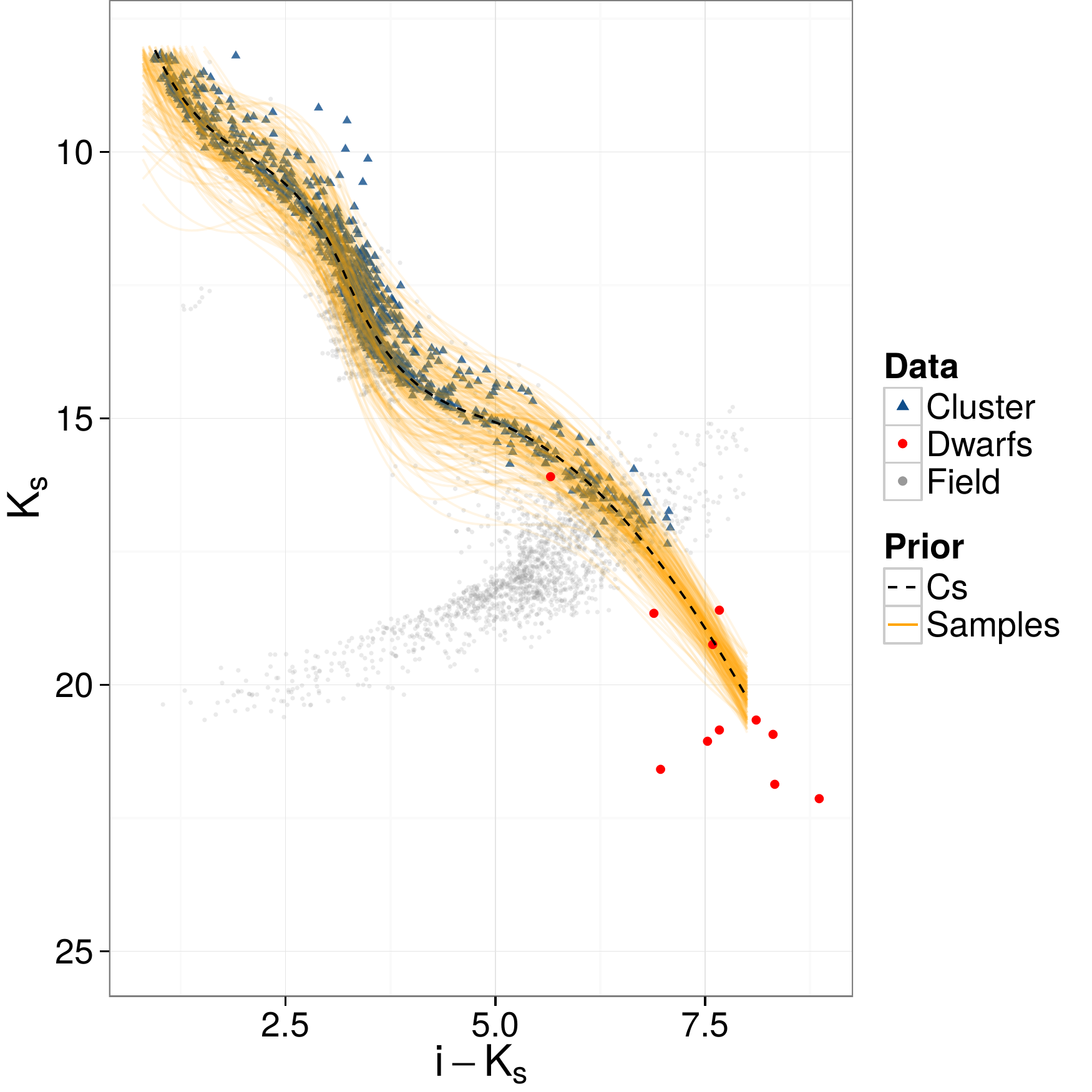}}
\caption{CMD $K_s$ vs. $i-K_s$ showing a sample of the prior for the coefficients in the splines series. Also shown the brown-dwarfs we add from \citet{Faherty2012} sample, and the cluster sequence (dashed line) found after fitting the splines to the brown-dwarfs and candidate members below the equal-mass binaries sequence.}
\label{figure:priorcoefs}
\end{center}
\end{figure}

We use univariate normal distributions to establish the priors for the coefficients in the spline series. To find the values of the hyper-parameters, we proceed as follows. First, we discard equal-mass binaries from \citet{Bouy2015} candidate members. To do this, we iteratively fit the cluster sequence and remove objects above 0.75 magnitudes. Then, to obtain an empirical prior in the region were no members have been found, we complement our list with the brown-dwarfs from the \citet{Faherty2012} sample that have the same bands as our data set. Finally, we fit the splines, and use the coefficients of this fit as means, $\mu_{\beta}$ of the univariate normal distributions. We set the standard deviation to $\sigma_{\beta}=\{1,1,1,1,1,0.5,0.1\}$. These values provide a reasonable compromise between cluster sequences compatible with the previously known candidates and those far away or with exotic shapes. We show a sample of this priors in Fig. \ref{figure:priorcoefs}. This Fig. also shows the brown-dwarfs from \citet{Faherty2012} and the sequence (dashed line) we use to provide the means of the univariate normal distributions.

Finally, in Table \ref{table:hyperparameters}, we summarise all the hyper-parameter values of our Bayesian Hierarchical Model.

\begin{table}[!htp]
\centering
\begin{tabular}{ll}
  \hline
Hp. & Value \\ 
  \hline
$\alpha_{Cs}$      & $\{1,1,1,1\}$\\
$\alpha_{Bs}$      & $\{1.2,8.8\}$\\
$A_{pm}$            & $\{10^5,10^5\}$\\
$\mu_{\mu_{pm}}$    & $\{16.30,-39.62\}$\\
$\Sigma_{\mu_{pm}}$ & $\{36.84,1.18,40.71\}$\\
\hline
$\alpha_{CI}$      & $\{1,1,1,1,1\}$\\
$rg_{CI}$           & $\{0.8,8\}$\\
$\eta$              & 100\\
$\mu_{\beta_Y}$     & $\{7.65,11.47,10.66,16.33,16.49,21.44,22.49\}$\\
$\mu_{\beta_J}$     & $\{7.61, 11.52, 10.20, 15.66, 15.58, 19.88, 21.16\}$\\
$\mu_{\beta_H}$     & $\{7.63, 10.88,  9.50, 15.19, 15.04, 18.68, 20.64\}$\\ 
$\mu_{\beta_K}$     & $\{7.55, 10.81,  9.32, 14.79, 14.62, 17.63, 20.24\}$\\
$\sigma_{\beta}$    & $\{1,1,1,1,1,0.5,0.1\}$\\
$A_{ph}$            & $\{10,10,10,10,10\}$\\
\hline
$\alpha$            & $\{98,2\}$\\
$\alpha_{CB}$       & $\{8,2\}$ \\
$\nu$               & 3\\
   \hline
\end{tabular}
\caption{Hyper-parameters for different blocks of the Hierarchical Bayesian Model. The upper and middle blocks correspond, respectively, to hyper-parameters of the proper motions and photometric models. The lower block to hyper-parameters shared by both last models.}
\label{table:hyperparameters}
\end{table}

\begin{table*}[ht]
\caption{Parameters names, symbols, and priors.}
\centering
\begin{tabular}{lcc|lcc}
  \hline
 Name & Symbol & Prior & Name & Symbol & Prior\\ 
  \hline
  Field fraction	& $\pi$        & Dirichlet($\alpha$)       & Coefficient [1,1]	& $\beta_{Y,1}$     & Normal($\mu_{\beta},\sigma_{\beta}$)\\
  Cs fraction 		& $\pi_{CB}$   & Dirichlet($\alpha_{Cs}$)  & Coefficient [1,2] 	& $\beta_{Y,2}$     & Normal($\mu_{\beta},\sigma_{\beta}$)\\
  Cs PM fraction 1 	& $\pi_{Cs,1}$ & Dirichlet($\alpha_{Cs}$)  & Coefficient [1,3] 	& $\beta_{Y,3}$     & Normal($\mu_{\beta},\sigma_{\beta}$)\\
  Cs PM fraction 2 	& $\pi_{Cs,2}$ & Dirichlet($\alpha_{Cs}$)  & Coefficient [1,4] 	& $\beta_{Y,4}$     & Normal($\mu_{\beta},\sigma_{\beta}$)\\
  Cs PM fraction 3 	& $\pi_{Cs,3}$ & Dirichlet($\alpha_{Cs}$)  & Coefficient [1,5] 	& $\beta_{Y,5}$     & Normal($\mu_{\beta},\sigma_{\beta}$)\\
  Bs PM fraction 1 	& $\pi_{Bs,1}$ & Dirichlet($\alpha_{Bs}$)  & Coefficient [1,6] 	& $\beta_{Y,6}$     & Normal($\mu_{\beta},\sigma_{\beta}$)\\
  Color fraction 1 	& $\pi_{CI,1}$ & Dirichlet($\alpha_{CI}$)  & Coefficient [1,7] 	& $\beta_{Y,7}$     & Normal($\mu_{\beta},\sigma_{\beta}$)\\
  Color fraction 2 	& $\pi_{CI,2}$ & Dirichlet($\alpha_{CI}$)  & Coefficient [2,1] 	& $\beta_{J,1}$     & Normal($\mu_{\beta},\sigma_{\beta}$)\\
  Color fraction 3 	& $\pi_{CI,3}$ & Dirichlet($\alpha_{CI}$)  & Coefficient [2,2] 	& $\beta_{J,2}$     & Normal($\mu_{\beta},\sigma_{\beta}$)\\
  Color fraction 4 	& $\pi_{CI,4}$ & Dirichlet($\alpha_{CI}$)  & Coefficient [2,3] 	& $\beta_{J,3}$     & Normal($\mu_{\beta},\sigma_{\beta}$)\\
  Mean color 1 		& $\mu_{CI,1}$ & Unifrom($rg_{CI}$)        & Coefficient [2,4] 	& $\beta_{J,4}$     & Normal($\mu_{\beta},\sigma_{\beta}$)\\
  Mean color 2 		& $\mu_{CI,2}$ & Unifrom($rg_{CI}$)        & Coefficient [2,5] 	& $\beta_{J,5}$     & Normal($\mu_{\beta},\sigma_{\beta}$)\\
  Mean color 3 		& $\mu_{CI,3}$ & Unifrom($rg_{CI}$)        & Coefficient [2,6] 	& $\beta_{J,6}$     & Normal($\mu_{\beta},\sigma_{\beta}$)\\
  Mean color 4 		& $\mu_{CI,4}$ & Unifrom($rg_{CI}$)        & Coefficient [2,7] 	& $\beta_{J,7}$     & Normal($\mu_{\beta},\sigma_{\beta}$)\\
  Mean color 5 		& $\mu_{CI,5}$ & Unifrom($rg_{CI}$)        & Coefficient [3,1] 	& $\beta_{H,1}$     & Normal($\mu_{\beta},\sigma_{\beta}$)\\ 
  Variance color 1 	& $\sigma_{CI,1}$ &HalfCauchy(0,$\eta$)    & Coefficient [3,2] 	& $\beta_{H,2}$     & Normal($\mu_{\beta},\sigma_{\beta}$)\\
  Variance color 2 	& $\sigma_{CI,2}$ &HalfCauchy(0,$\eta$)    & Coefficient [3,3] 	& $\beta_{H,3}$     & Normal($\mu_{\beta},\sigma_{\beta}$)\\
  Variance color 3 	& $\sigma_{CI,3}$ &HalfCauchy(0,$\eta$)    & Coefficient [3,4] 	& $\beta_{H,4}$     & Normal($\mu_{\beta},\sigma_{\beta}$)\\
  Variance color 4 	& $\sigma_{CI,4}$ &HalfCauchy(0,$\eta$)    & Coefficient [3,5] 	& $\beta_{H,5}$     & Normal($\mu_{\beta},\sigma_{\beta}$)\\
  Variance color 5 	& $\sigma_{CI,5}$ &HalfCauchy(0,$\eta$)    & Coefficient [3,6] 	& $\beta_{H,6}$     & Normal($\mu_{\beta},\sigma_{\beta}$)\\
  Mean PM Cs[1,1] 	& $\mu_{Cs,1}$ & Normal($\mu_{\mu_{pm}},\Sigma_{\mu_{pm}}$) &Coefficient [3,7] 	& $\beta_{H,7}$     & Normal($\mu_{\beta},\sigma_{\beta}$)\\ 
  Mean PM Cs[1,2] 	& $\mu_{Cs,2}$ & Normal($\mu_{\mu_{pm}},\Sigma_{\mu_{pm}}$) &Coefficient [4,1] 	& $\beta_{K,1}$     & Normal($\mu_{\beta},\sigma_{\beta}$)\\
  Variance Cs[1,1] 	& $\Sigma_{Cs,1,1}$ & Half-t($\nu,A_{pm}$)  & Coefficient [4,2] 	& $\beta_{K,2}$     & Normal($\mu_{\beta},\sigma_{\beta}$)\\
  Variance Cs[1,2] 	& $\Sigma_{Cs,1,2}$ & Half-t($\nu,A_{pm}$)  & Coefficient [4,3] 	& $\beta_{K,3}$     & Normal($\mu_{\beta},\sigma_{\beta}$)\\
  Variance Cs[1,3] 	& $\Sigma_{Cs,1,3}$ & Half-t($\nu,A_{pm}$)  & Coefficient [4,4] 	& $\beta_{K,4}$     & Normal($\mu_{\beta},\sigma_{\beta}$)\\ 
  Variance Cs[2,1] 	& $\Sigma_{Cs,2,1}$ & Half-t($\nu,A_{pm}$)  & Coefficient [4,5] 	& $\beta_{K,5}$     & Normal($\mu_{\beta},\sigma_{\beta}$)\\
  Variance Cs[2,2] 	& $\Sigma_{Cs,2,2}$ & Half-t($\nu,A_{pm}$)  & Coefficient [4,6] 	& $\beta_{K,6}$     & Normal($\mu_{\beta},\sigma_{\beta}$)\\
  Variance Cs[2,3] 	& $\Sigma_{Cs,2,3}$ & Half-t($\nu,A_{pm}$)  & Coefficient [4,7] 	& $\beta_{K,7}$     & Normal($\mu_{\beta},\sigma_{\beta}$)\\
  Variance Cs[3,1] 	& $\Sigma_{Cs,3,1}$ & Half-t($\nu,A_{pm}$)  &Covariance Phot[1] 		& $\Sigma_{clus}[1]$ & Half-t($\nu,A_{ph}$)  \\ 
  Variance Cs[3,2] 	& $\Sigma_{Cs,3,2}$ & Half-t($\nu,A_{pm}$)  &Covariance Phot[2] 		& $\Sigma_{clus}[2]$ & Half-t($\nu,A_{ph}$)  \\  
  Variance Cs[3,3] 	& $\Sigma_{Cs,3,3}$ & Half-t($\nu,A_{pm}$)  &Covariance Phot[3] 		& $\Sigma_{clus}[3]$ & Half-t($\nu,A_{ph}$)  \\ 
  Variance Cs[4,1] 	& $\Sigma_{Cs,4,1}$ & Half-t($\nu,A_{pm}$)  &Covariance Phot[4] 		& $\Sigma_{clus}[4]$ & Half-t($\nu,A_{ph}$)  \\ 
  Variance Cs[4,2] 	& $\Sigma_{Cs,4,2}$ & Half-t($\nu,A_{pm}$)  &Covariance Phot[5] 		& $\Sigma_{clus}[5]$ & Half-t($\nu,A_{ph}$)  \\  
  Variance Cs[4,3] 	& $\Sigma_{Cs,4,3}$ & Half-t($\nu,A_{pm}$)  &Covariance Phot[6] 		& $\Sigma_{clus}[6]$ & Half-t($\nu,A_{ph}$)  \\  
  Mean PM Bs[1,1] 	& $\mu_{Bs,1}$ & Normal($\mu_{\mu_{pm}},\Sigma_{\mu_{pm}}$)& Covariance Phot[7]& $\Sigma_{clus}[7]$ & Half-t($\nu,A_{ph}$)  \\ 
  Mean PM Bs[1,2] 	& $\mu_{Bs,2}$ & Normal($\mu_{\mu_{pm}},\Sigma_{\mu_{pm}}$)& Covariance Phot[8]& $\Sigma_{clus}[8]$ & Half-t($\nu,A_{ph}$)  \\ 
  Variance Bs[1,1] 	& $\Sigma_{Bs,1,1}$ & Half-t($\nu,A_{pm}$)  &Covariance Phot[9] 		& $\Sigma_{clus}[9]$ & Half-t($\nu,A_{ph}$)  \\ 
  Variance Bs[1,2] 	& $\Sigma_{Bs,1,2}$ & Half-t($\nu,A_{pm}$)  &Covariance Phot[10] 	& $\Sigma_{clus}[10]$ & Half-t($\nu,A_{ph}$)  \\  
  Variance Bs[1,3] 	& $\Sigma_{Bs,1,3}$ & Half-t($\nu,A_{pm}$)  &Covariance Phot[11] 	& $\Sigma_{clus}[11]$ & Half-t($\nu,A_{ph}$)  \\ 
  Variance Bs[2,1] 	& $\Sigma_{Bs,2,1}$ & Half-t($\nu,A_{pm}$)  &Covariance Phot[12] 	& $\Sigma_{clus}[12]$ & Half-t($\nu,A_{ph}$)  \\  
  Variance Bs[2,2] 	& $\Sigma_{Bs,2,2}$ & Half-t($\nu,A_{pm}$)  &Covariance Phot[13] 	& $\Sigma_{clus}[13]$ & Half-t($\nu,A_{ph}$)  \\  
  Variance Bs[2,3] 	& $\Sigma_{Bs,2,3}$ & Half-t($\nu,A_{pm}$)  &Covariance Phot[14] 	& $\Sigma_{clus}[14]$ & Half-t($\nu,A_{ph}$)  \\  
                        &                   &                       &Covariance Phot[15] 	& $\Sigma_{clus}[15]$ & Half-t($\nu,A_{ph}$)  \\ 
   \hline
\end{tabular}
\label{table:symbols_parameters}
\end{table*}

\subsection{Derivation of the magnitude distributions}
\label{subsect:deriveluminosity}
To derive the $J,H,K_s$ magnitude distributions, we use the distribution of the colour index, $CI$, and the cluster and equal-mass binaries photometric sequences (the spline series). We exemplify this derivation on the $K_s$ band, but similar transformations apply to the rest of the bands. To obtain the distribution of $K_s$ for the cluster objects, we introduce the colour index $CI$, as a nuisance parameter and then we marginalise it. Thus, 

\begin{align}
p(K_s | \boldsymbol{\theta}_c) & = \int p(K_s,CI | \boldsymbol{\theta}_c) \cdot dCI =  \int p(K_s | CI ,\boldsymbol{\theta}_c) \cdot p(CI|\boldsymbol{\theta}_c)\cdot dCI. \nonumber
\end{align}
The term $p(K_s | CI ,\boldsymbol{\theta}_c)$ corresponds to the GMM modelling the distribution of $CI$ (Eq. \ref{eq:colordist}), while $p(K_s | CI ,\boldsymbol{\theta}_c)$ is the probability of $K_s$ given the $CI$ and the cluster parameters $\boldsymbol{\theta}_c$. Since our photometric model takes into account the equal-mass binaries, we include them proportionally to their fraction, ($1-\pi_{CB}$). Thus,

\begin{align}
p(K_s | \boldsymbol{\theta}_c) & =  \int \left[\pi_{CB}\cdot p_{Cs}(K_s| CI, \boldsymbol{\theta}_c) + (1-\pi_{CB})\cdot p_{Bs}(K_s| CI, \boldsymbol{\theta}_c)\right]\nonumber \\& \cdot p_{CI}(CI|\boldsymbol{\theta}_c)\cdot dCI. \nonumber \\
& =   \pi_{CB} \int p_{Cs}(K_s| CI, \boldsymbol{\theta}_c) \cdot p_{CI}(CI|\boldsymbol{\theta}_c) dCI \nonumber \\
&+ (1-\pi_{CB})\int p_{Bs}(K_s| CI, \boldsymbol{\theta}_c) \cdot p_{CI}(CI|\boldsymbol{\theta}_c)\cdot  dCI. \nonumber \\
\end{align}

In this equation, $Cs$ and $Bs$ stand for cluster and equal-mass binaries sequences, respectively. The terms inside the integrals correspond to Equations \ref{eq:lik-seq} and \ref{eq:colordist}. However, since here we focus only on the distribution of $K_s$, we marginalise the rest of the bands. Also, we change the integration limits to those of the truncated colour distribution ($CI_{min}=0.8, CI_{max}=8$). Finally, we obtain

\begin{align}
&p(K_s | \boldsymbol{\theta}_c)  =   \pi_{CB} \int_{CI_{min}}^{CI_{max}}\left[ \left[\sum_{i=1}^5 \pi_{CI,i} \cdot \mathcal{N}_t(CI| \mu_{CI,i},\sigma_{CI,i})\right]\right. \nonumber \\
&\cdot  \left.\int_{\tilde{Y},\tilde{J},\tilde{H}}\mathcal{N}(\{CI,\tilde{Y},\tilde{J},\tilde{H},K_s\}|\boldsymbol{\mathcal{S}}(CI, \boldsymbol{\beta}),\Sigma_{clus})~d\tilde{Y}~d\tilde{J}~d\tilde{H}\right] \cdot dCI \nonumber \\
& + (1-\pi_{CB}) \int_{CI_{min}}^{CI_{max}}\left[\left[\sum_{i=1}^5 \pi_{CI,i} \cdot \mathcal{N}_t(CI| \mu_{CI,i},\sigma_{CI,i})\right]\right.\nonumber\\
&\cdot \left. \int_{\tilde{Y},\tilde{J},\tilde{H}}\mathcal{N}(\{CI,\tilde{Y},\tilde{J},\tilde{H},K_s\}|T_{Bs}(\boldsymbol{\mathcal{S}}(CI, \boldsymbol{\beta})),\Sigma_{clus})~d\tilde{Y}~d\tilde{J}~d\tilde{H}\right]\cdot dCI. \nonumber 
\end{align}

The derivation of the $J$ and $H$ magnitude distributions is similar to the procedure described for $K_s$. We notice that, the derivation of these magnitude distributions takes into account the equal-mass binaries and the systems which could have different mass ratios. Therefore, these distribution are the system magnitude distributions. 

\subsection{The probabilistic graphical model}
\label{sub:graphical-model}
\begin{figure*}[tb]
  \begin{center}
  \resizebox{0.9\linewidth}{!}{\input{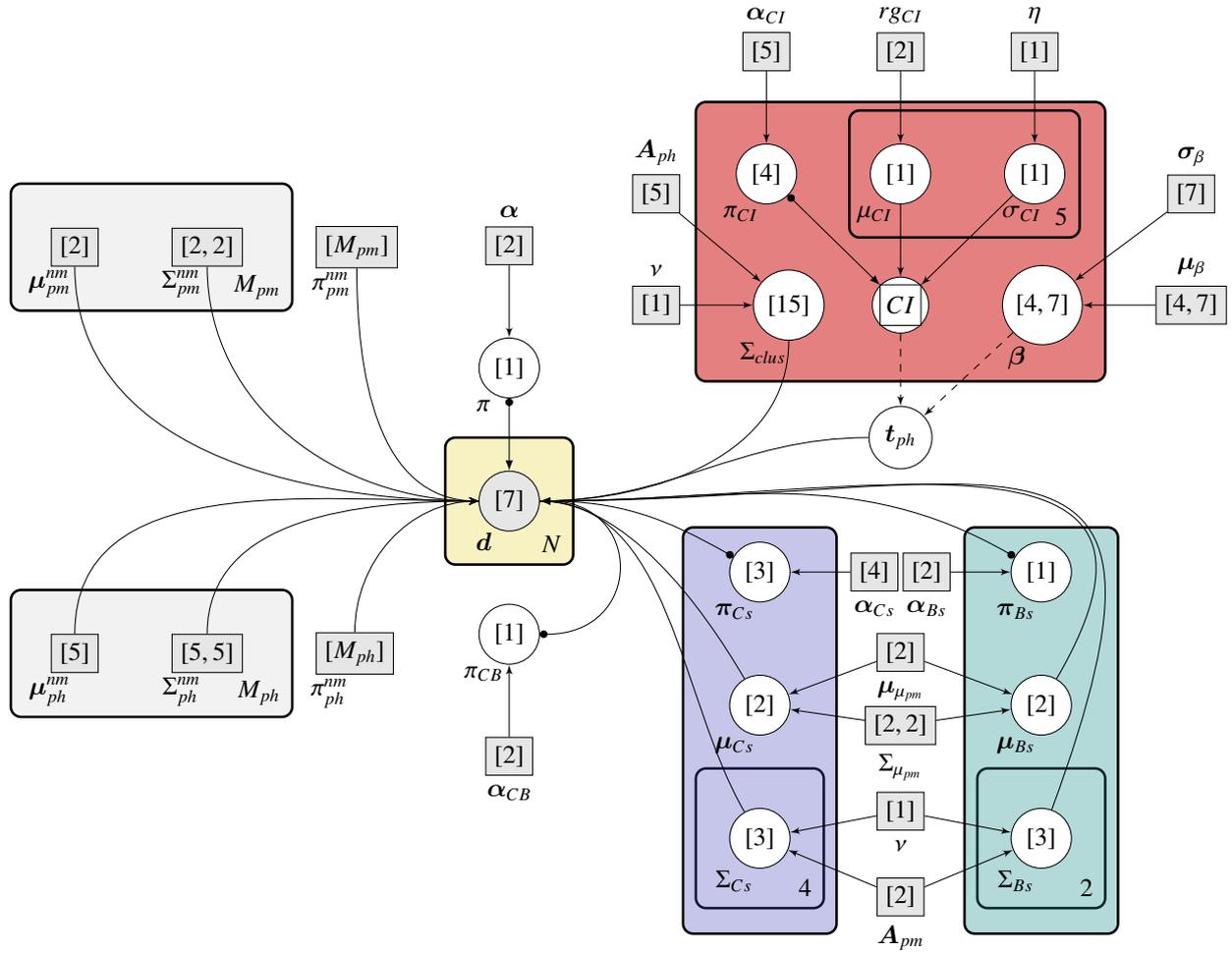}}
  \end{center}
  \caption{Probabilistic graphical model. The left grey plates show the field model. The middle yellow plate shows the node where the likelihood is computed for each datum, $\boldsymbol{d}$. The right plates describe the relations among parameters in the cluster model. The photometric cluster model (red) is on top, while the proper motions cluster (blue) and equal-mass binaries (green) are at the bottom left and right, respectively. See description in the text for more details.}
  \label{fig:HM}
\end{figure*}

A probabilistic graphical model is a graph which expresses the relationships, either deterministic or stochastic, among random variables in a model: parameters and observations. Figs. \ref{fig:HM} shows the probabilistic graphical model of our Hierarchical Bayesian Model. In this figure, the following characteristic apply: i) conditional relations are depicted with arrows, solid when the condition is stochastic (i.e. given by a probability distribution function) and dashed when it is deterministic, ii) random variables are surrounded by circles (also known as nodes) while constants by rectangles; the marginalised parameter ($CI$) is drawn as a square inside a circle, iii)  black dots indicate that categorical variables have been  marginalised, iv) the dimension of constants or independent parameters is written in brackets inside the nodes, v) figures filled with grey indicate that their value is known (e.g. data), and vi) plates join variables which repeat together, the number of repetitions is indicated in one corner.

The left panels of Fig. \ref{fig:HM} represents the set of model parameters that we use to describe the field population. Since these parameters remain fixed throughout the inference (see Sect. \ref{sect:field}), we consider them constants, thus we depict them with grey squares. The right panels show the parameters of the cluster population. The top right and bottom right panels describe the photometric and kinematic models, respectively. The top inner plate inside the photometric panel shows the GMM that we use to describe the $CI$ distribution. The bottom left and right panels show the cluster and equal-mass binaries proper motion models, respectively. The plates inside them designate the covariance matrices of each GMM. Because each gaussian in the mixture shares the mean, it lies outside the plate. Finally, the middle (yellow) plate depicts the comparison between the true quantities and the measured ones. Therefore, it is here were we compute the likelihood of each elements in the data set.

\end{appendices}

\bibliographystyle{aa} \bibliography{olivares}
\end{document}